\documentclass[twocolumn,times,tighten]{aastex63}
\usepackage{color}
\usepackage{enumitem}
\usepackage{hyperref}
\usepackage{verbatim}
\usepackage{amsmath,amstext,mathrsfs}
\usepackage[all]{hypcap} 
\usepackage{afterpage}  
\usepackage{marginnote}
\usepackage{makecell}

\usepackage{ifxetex,ifluatex}
\usepackage{fixltx2e} 
\usepackage{multirow}
\usepackage{natbib}
\usepackage{graphicx}
\defcitealias{LP00}{LP00}

\newcommand{\specialcell}[2][c]{\begin{tabular}[#1]{@{}c@{}}#2\end{tabular}}

\DeclareFontFamily{OMS}{oasy}{\skewchar\font48 }
\DeclareFontShape{OMS}{oasy}{m}{n}{%
     <-5.5> oasy5   <5.5-6.5> oasy6
   <6.5-7.5> oasy7   <7.5-8.5> oasy8
   <8.5-9.5> oasy9   <9.5-> oasy10
   }{}
\DeclareFontShape{OMS}{oasy}{b}{n}{%
    <-6> oabsy5
   <6-8> oabsy7
   <8-> oabsy10
   }{}
\DeclareSymbolFont{oasy}{OMS}{oasy}{m}{n}
\SetSymbolFont{oasy}{bold}{OMS}{oasy}{b}{n}

\DeclareMathSymbol{\smallleftarrow}   {\mathrel}{oasy}{"20}
\DeclareMathSymbol{\smallrightarrow}  {\mathrel}{oasy}{"21}
\DeclareMathSymbol{\smallleftrightarrow}{\mathrel}{oasy}{"24}

\graphicspath{{./}{}}

\shorttitle{Velocity Decomposition Technique: Extracting velocity {  fluctuations} from spectroscopic data}
\shortauthors{Yuen, Ho \& Lazarian}
\received{\today}
\submitjournal{ApJ}

\begin{document}
\title{Technique for separating velocity and density contributions in spectroscopic data \\and its application to studying turbulence and magnetic fields}
\author[0000-0003-1683-9153]{Ka Ho Yuen}
\affiliation{Department of Astronomy, University of Wisconsin-Madison, USA}
\email{kyuen@astro.wisc.edu}

\author[0000-0003-1683-9153]{Ka Wai Ho}
\affiliation{Department of Astronomy, University of Wisconsin-Madison, USA}
\email{kho33@wisc.edu}

\author[0000-0002-7336-6674]{Alex Lazarian}
\affiliation{Department of Astronomy, University of Wisconsin-Madison, USA}
\affiliation{Center for Computation Astrophysics, Flatiron Institute, 162 5th Ave, New York, NY 10010}
\email{alazarian@facstaff.wisc.edu}

\begin{abstract}
Based on the theoretical description of Position-Position-Velocity(PPV) statistics in \cite{LP00}, we introduce a new technique called the Velocity Decomposition Algorithm(VDA) in separating {  the PPV fluctuations arising from velocity and density }fluctuations. Using MHD turbulence simulations, we demonstrate its promise in {  retrieving }the {  velocity fluctuations from PPV cube} in various physical conditions and {  its} prospects in accurately tracing the magnetic field. We find that for localized clouds, the velocity fluctuations are most prominent at the wing part of the spectral line, and they dominate the density fluctuations. The same velocity dominance applies to extended HI regions {  undergoing} galactic {  rotation}. Our numerical experiment demonstrates that velocity channels arising from the cold phase of atomic hydrogen (HI) are still affected by velocity {  fluctuations} {  at} small scales. We apply the VDA to HI GALFA-DR2 data corresponding to the high-velocity cloud HVC186+19-114 and high latitude galactic diffuse HI data. {  Our study confirms the crucial role of velocity {  fluctuations} in explaining why linear structures are observed within PPV cubes. } We discuss the implications of VDA for both magnetic field studies and predicting polarized galactic emission that acts as the foreground for the Cosmic Microwave Background studies. {  Additionally}, we address the controversy related to the filamentary nature of the HI channel maps and explain the importance of velocity {  fluctuations} in the formation of structures in PPV data cubes. VDA will allow astronomers to obtain velocity {  fluctuations} from almost every piece of spectroscopic PPV data and {  allow direct investigations} of the turbulent velocity field in observations.
\end{abstract}


\section{Introduction} \label{sec:intro}

Spectroscopic Doppler-shifted lines carry information about astrophysical turbulence in interstellar media, in particular, neutral hydrogen, ionized and molecular species. The corresponding data for \ion{H}{1} and molecular lines, e.g. , CO, are stored in Position-Position-Velocity (PPV) {  data cubes that contain a wealth of information} for diagnosis of the properties of turbulence in interstellar media. It is thus essential to understand what is contained in the velocity channel data, which is the v-slice of data from PPV cube, i.e. , a slice with a certain velocity $v_0$ and channel width $\Delta v$. 

The theory that relates the statistics of spectroscopic intensity fluctuations in the PPV space and the underlying statistical {  fluctuations} of turbulent velocities and density was developed in \citet{LP00} (henceforth \citetalias{LP00}) and extended in the subsequent theoretical studies \citep{LP04,LP06,LP08,KLP16,KLP17a,KLP17b}. The aforementioned theory describes how the statistics of intensity fluctuations in PPV arise from velocity crowding. The latter we will call {\bf velocity caustics}, although {previously in \citetalias{LP00}}, this term is applicable only in the absence of thermal broadening. Such structures can be seen in PPV cubes obtained with spectroscopic data {in} turbulent regions. The formation of caustics does not require shocks but is a natural result of turbulent motions.

\citetalias{LP00} for the first time {formulated} the statistical description of the PPV statistics and stimulated many further developments both in terms of theory and observational studies (see \citealt{2009fohl.book..357L}). In particular, it predicted the change of the spectral index of the spectrum of the channel maps' emissivity fluctuations with the change of the thickness of the velocity slice and related this change with the change of the relative contributions of density and velocity fluctuations. {The velocity fluctuations are actually the velocity caustics we refer to above and are the center of interest in the series of papers of Lazarian \& Pogosyan}. The quantitative relations derived in \citetalias{LP00} provided the theoretical foundations of the Velocity Channel Analysis (VCA) technique that proved to be a useful tool for {extracting the velocity statistics from turbulent interstellar media}. In particular, the VCA predictions were successfully tested numerically \citep{2009ApJ...693.1074C} and {were} applied to numerous sets of observational \ion{H}{1} and CO data (\citealt{2001ApJ...551L..53S,2006ApJ...653L.125P,Chepurnov2010ExtendingData}, see also a summary in \cite{reply19}, see \S \ref{sec:theory} for a comprehensive review)\footnote{The publicly available VCA software is available at \url{https://github.com/Astroua/TurbuStat}}. The \citetalias{LP00} approach was generalized to address the studies of turbulence anisotropies in \cite{KLP16}.The corresponding study made use of the description of different modes of anisotropic turbulence in \cite{LP12}. {  The aforementioned works open an opportunity in isolating the contributions of fast, slow, and {   Alfven basic MHD modes in observations {\citep{2020NatAs.tmp..174Z}}}.}

It is very important that the aforementioned understanding of the mapping of the anisotropic velocity and density fluctuation into PPV cubes, together with the nature of this anisotropy that follows from the theory of MHD turbulence \citep{GS95} and turbulent reconnection \citep{LV99}, resulted in a radically new technique of tracing magnetic field using PPV cube data. This technique, termed Velocity Gradient Technique, showed promising results in mapping magnetic fields using both HI \citep{YL17a,YL17b,LY18a} as well as in CO and other molecules \citep{LY18a,2019ApJ...873...16H,survey,velac}. The VGT does use the information about the velocity caustics in order to trace the magnetic field. The density fluctuations are sensitive to shocks \citep{YL17b,IGVHRO}. Therefore, it is very advantageous to separate velocity and density contributions in the PPV cube. 

While {  a way for} the statistical separation of the velocity and density contribution in PPV cubes, \citetalias{LP00} did not directly address the actual practical separations of velocity and density contributions to the individual observed fluctuations. The separation of velocity and density fluctuation is fundamentally important for studying turbulence statistics in PPV cubes for the following reasons: (1) {  It enables the study of} studying true velocity statistics from observational data. (2) It allows one to cross-check with the theoretical {  framework} of \citetalias{LP00}. (3) The original development of \citetalias{LP00} dealt with the statistics of the central channels of the spectral line. In this paper, employing the statistical approach in \cite{LP04} we deal with the statistics of the channels that are far away from the spectral line peak, aka. Wing channels, and find the differences from those of the central channels. The separation of velocity fluctuations in PPV cubes will help us in identifying the channels that carry the largest portion of the velocity information. {(4) Based on the theory of PPV statistics, we would like to examine the influence of thermal broadening to turbulent statistics}. The thermal broadening is an essential factor that affects the velocity caustics in the PPV cubes. \citetalias{LP00} showed that the thermal broadening {effect} acts to increase the effective thickness of velocity channels. The thermal broadening {effect} also blurs velocity caustics, decreasing the contribution of the latter to the intensity fluctuations observed within velocity channels. Earlier, a number of approaches for dealing with the problem were discussed: the effective kernel as discussed in \citetalias{LP00}, {  thermal kernel modelling \citep{2006ApJ...638..797D}}, the use of centroids \citep{EL05,KLP17a}, and the thermal deconvolution method \citep{GA}. However, it is hard to connect the \citetalias{LP00} theory to the product of these approaches since fundamentally we do not have access to true velocity caustics nor {  to} their turbulence statistics in observations.

In this paper, we examine a broad range of interstellar conditions to which the \citetalias{LP00} theory is applicable, from molecular species like CO that are considered isothermal to a more complex multi-phase gas like HI, which contains cold, warm, and unstable phases. While the isothermal media is straightforward to analyze, the multi-phase gas requires a more complicated numerical setting and an additional analysis. For instance, in \citetalias{LP00} it was noted that due to thermal instability, cold clumps are expected to be formed within the eddies of turbulent warm HI and to carry the momentum of the warm gas. The non-thermal turbulent velocities of cold HI atoms are larger than the {intrinsic} thermal velocities of {  cold} HI atoms and, as a result, one can assume that caustics of cold clumps in sufficiently thin channel maps are marginally affected by the thermal broadening according to \citetalias{LP00}. Another scenario is the formation of caustics due to the presence of passive tracers in turbulent molecular hydrogen gas $H_2$, e.g., CO, in which the thermal width of the passive tracers is significantly narrower than that of $H_2$ because the former has a larger molecular weight.

In this paper, we do not appeal to passive tracer arguments but address the thermal broadening problem in its most complicated form, i.e.{assuming a single-phase or multiphase} fluid with the thermal linewidths {larger than} the turbulent linewidth. We show that even in this case, our approach for the velocity-density separation provides a way for us to study both velocity turbulence and accurately trace the magnetic field through the caustics structures. In particular, to understand the nature of fluctuations in HI velocity channels, we need to build an appropriate model for CNM/WNM caustics and cross-check with observations. In particular, we would like to answer the following questions:
\begin{enumerate}
  \item Is the concept of density/velocity fluctuations pixel-based, or is it only valid in a statistical sense?
  \item What is the role of velocity caustics in channel maps when the CNM dominates the emission?
  \item What is the relative importance of velocity and density fluctuations in a spectral line's central and wing channels? 
\end{enumerate}
We expect our present study to resolve several {controversial points} regarding the caustics in the literature. For instance, the importance of velocity caustics in creating fluctuations of 21 cm intensity fluctuations in thin channels was questioned in \cite{susan19}. With our new procedure, termed Velocity Decomposition Algorithm (VDA), we can extract velocity caustics in observations and discuss their importance relative to density fluctuations in HI emissions. {Furthermore, discussed in later sections of the paper, the availability of velocity caustics will have a far-reaching impact {\bf allowing new, unexplored sets of data to be easily extracted from nearly every spectroscopic and interferometric data cubes}}.

{This paper is structured as follows. In \S \ref{sec:theory} we first review how theoretically velocity caustics are defined and what we expect in the presence of multiphase media. In \S \ref{sec:la} we discuss a {  new} algorithm for obtaining the velocity caustics from any PPV cubes based on the statistical principle. In \S \ref{sec:sim} we discuss our numerical set-up. In \S \ref{sec:test} we test the method of VDA {  to} both isothermal and multiphase simulations. {  In \S \ref{sec:one_sigma} we discuss a unique property of velocity caustics fluctuations as a function of line-of-sight velocity named {\it "1-$\sigma$ criterion"} based on the \citetalias{LP00} framework.} In \S \ref{sec:observations} we examine {an important observation example, the high-velocity cloud (HVC), that the velocity caustics fluctuations are dominant in the HVC HI velocity channels}. In \S \ref{sec:VCA} we discuss the implication of VDA to the Velocity Channel Analysis method (VCA). In \S \ref{sec:implication} we discuss the implications of VDA to a wide range of studies related to the physics of spectroscopic data and its underlying velocity statistics. In particular, we discuss the potential impact of this paper and compare our results to the recent series of papers \citep{susan19,2019ApJ...886L..13P,kalberla2019,kalberla2020a,kalberla2020b} that {  challenge} the concept of velocity caustics. In \S \ref{sec:discussions} we discuss the prospect of the VDA for future studies. In \S \ref{sec:conclusion} we conclude our paper. The important discovery of the current paper is listed in the flowchart in Fig.\ref{fig:flowchart}.

We also include some essential supplementary materials in the appendices. In Appendix \ref{sec:apb} we discuss how the correlation of density and velocity would impact our result. In Appendix \ref{secap:kernel} we discuss the differences between the thermally broadened channels and intensity maps. In Appendix \ref{secap:origin} we discuss the theoretical meaning of "velocity channel gradients" based on the formulation of PPV statistics. In Appendix \ref{secap:centroid} we discuss how to use VDA in obtaining the constant density velocity centroid. The velocity centroid in constant density form was extensively studied numerically in \cite{2003ApJ...592L..37L,EL05,2015ApJ...814...77E} and also theoretically in \cite{KLP17a}. In Appendix \ref{sc:SVDA} we discuss the formal theoretical construction of VDA and how to make modifications if the assumptions of VDA do not hold. In particular, we discuss an improvement of the VDA algorithm in supersonic turbulence. At last, in Appendix \ref{secap:vda_emissions_absorptions} we discuss the application of VDA to passive tracers.}

\begin{figure*}[th]
  \centering
  \includegraphics[width=0.88\textwidth]{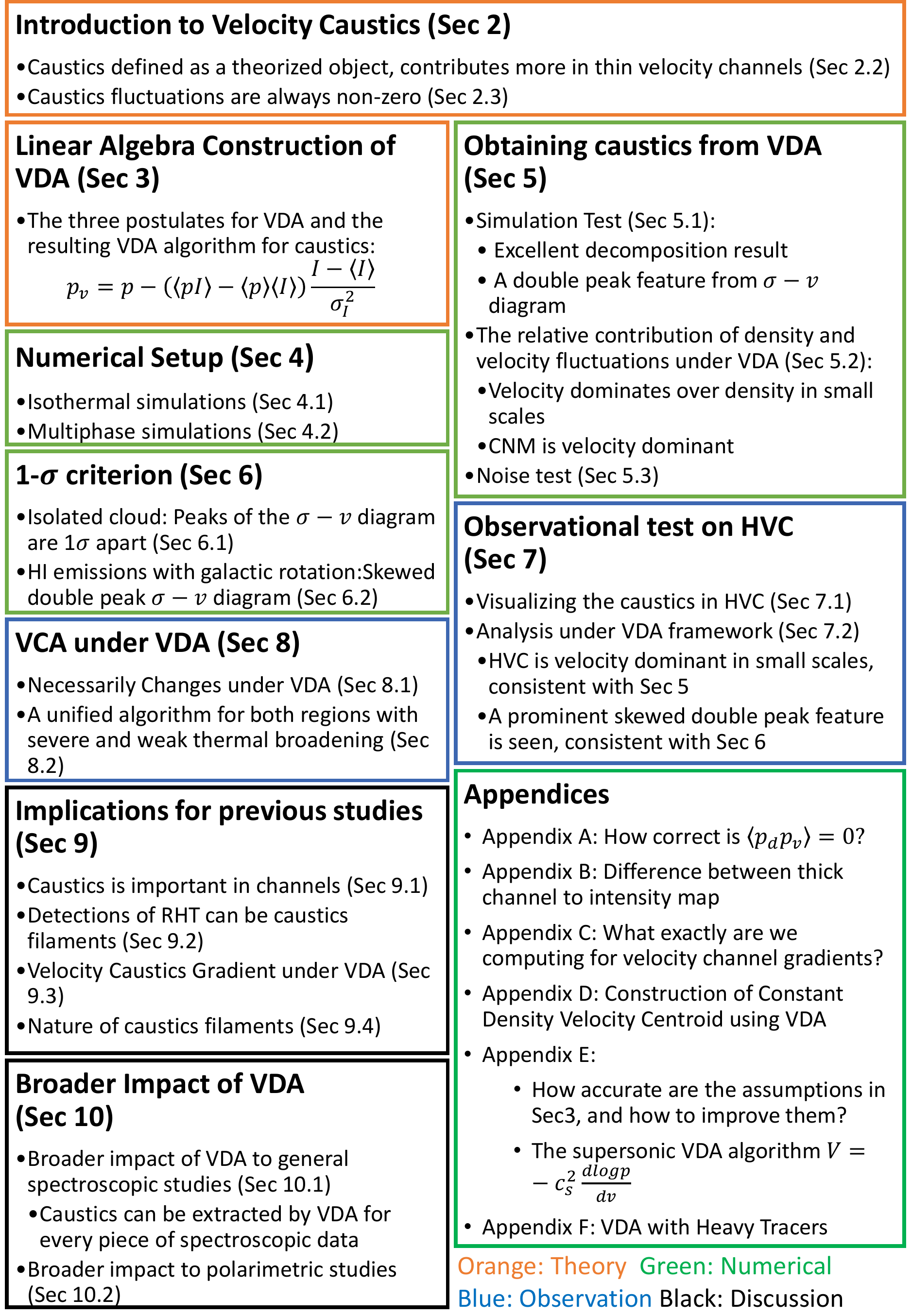}
  \caption{\label{fig:flowchart} A flowchart summarizing the result of the current paper. For readers' easier reference, we classify the sections by their nature by changing the color of the box edge: theory (orange), numerics (green), observations (blue) and discussions (black). }
\end{figure*}

\section{Theoretical considerations {of the PPV statistical theory}} \label{sec:theory}

\begin{figure*}[th]
  \centering
  \includegraphics[width=0.99\textwidth]{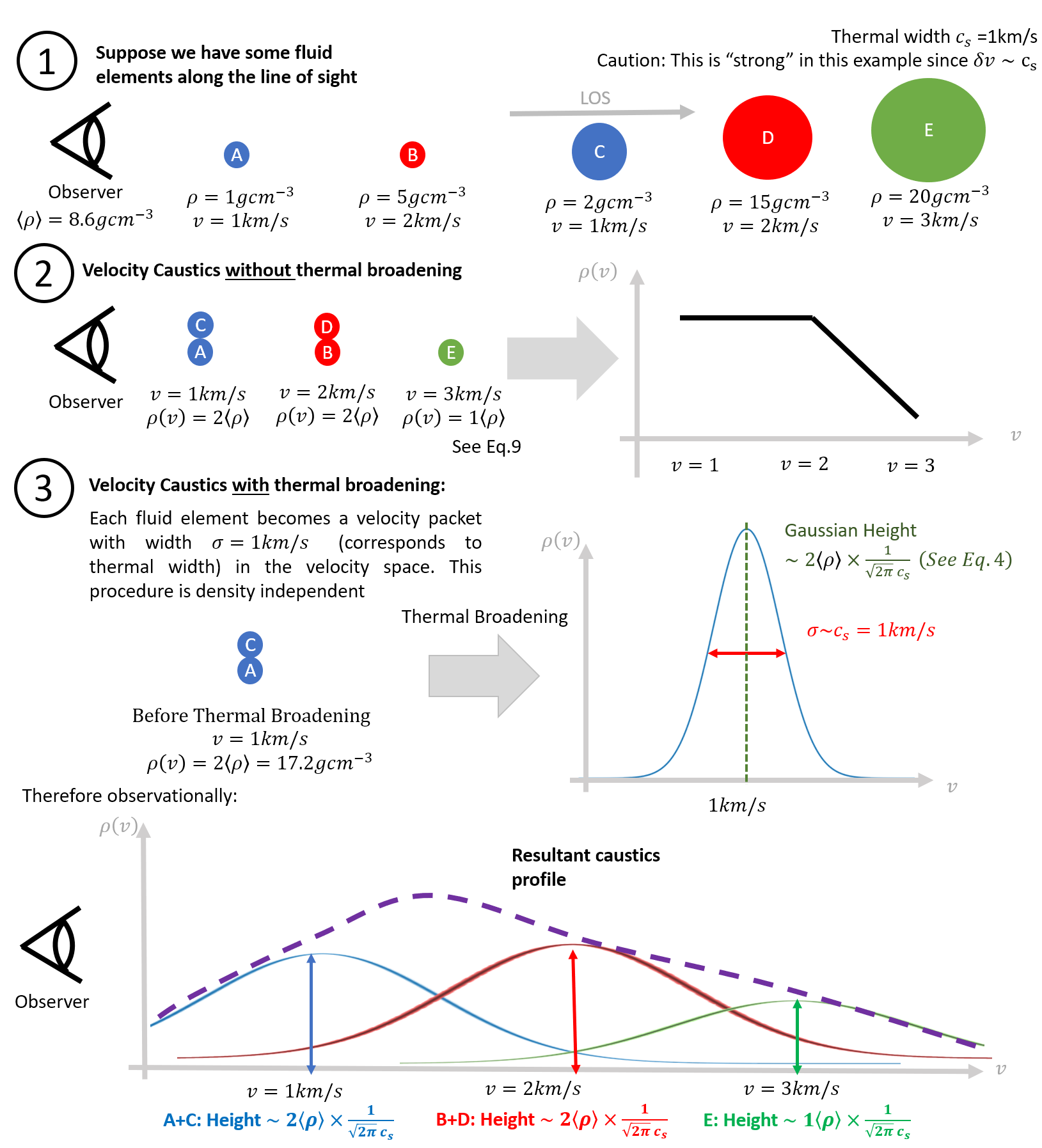}
  \caption{\label{fig:cartoons} A cartoon illustrating how velocity caustics are formed in spectroscopic PPV cube, and the concept of velocity and density fluctuations in PPV data. From the top: Panel (1) shows our example that facilitate the discussion of velocity caustics. Panel (2): We show how the velocity caustics based on the example in (1) look like without thermal broadening. Panel (3): The velocity caustics with thermal broadening, in the view of the $\rho(v)-v$ diagram, where we plot the caustics profile (the purple dash line) as a function of $v$. }
\end{figure*}

\begin{figure*}[th]
  \centering
  \includegraphics[width=0.99\textwidth]{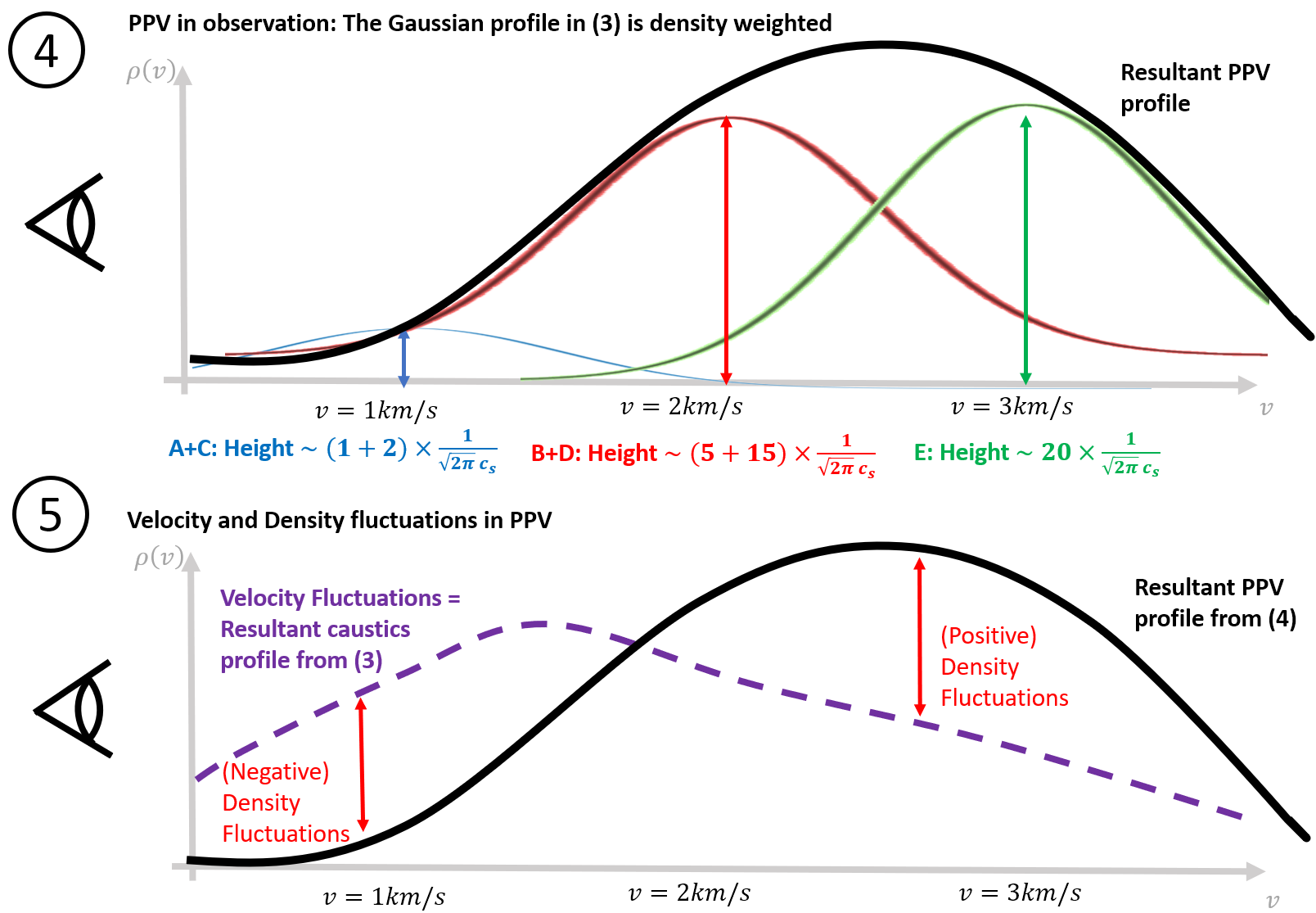}
  \caption{{\bf (continuing Fig.\ref{fig:cartoons})}. Panel (4): The real PPV profile in observations (the black line) according to our example in panel (1). Panel (5): The differences between the caustics profile and the true PPV profile corresponds to the density fluctuation while the caustics profile itself is the velocity fluctuation. }
\end{figure*}

\subsection{The correlation and structure function of turbulent media}

In what follows, we consider turbulent gas {with} temperature $T$ and mean mass of the turbulent fluid $m$. The latter is usually written as $m=\mu m_H$ where $\mu$ is the mean molecular weight, and $m_H$ is the weight of the hydrogen. The thermal broadening of such gas is given by $\beta_T=k_{B}T/m$. The turbulent velocities of gas $v$ in our model can be {  larger or smaller} than $\beta_T$. The intensity fluctuations in the PPV are arising from {both density} and velocity fluctuations, {where the} statistics of {the latter} are characterized by the {  LOS-component of the velocity structure function $D(r)$ (hereafter, denoted as the z-direction}:
\begin{equation}
  D_z({\bf r}=({\bf X},z))=\langle (v_z({\bf r}')-v_z({\bf r}+{\bf r}'))^2\rangle
  \label{eq:v_z}
\end{equation}
where $v_z$ is the z-direction velocity and $D_z$ averages over ${\bf r}'$. Similarly the correlation function $\zeta$ of turbulent density is given by
\begin{equation}
  \zeta_\rho({\bf r})=\langle \rho({\bf r})\rho(r+{\bf r}')\rangle
  \label{eq:cf_rho}
\end{equation}
where $\rho$ is the density. From the theory of MHD turbulence (see \citealt{2019tuma.book.....B}), it follows that the fluctuations of velocity are {generally larger as the scale increases}. {Notice that the correlation/structure functions are connected to the power spectrum under a Fourier transform. For example, One can show that a one dimensional velocity spectrum $E(k)\sim k^{-\alpha}$ is equivalent to a velocity structure function as {\it a scalar distance of r} $D_z(r)\propto r^{\alpha}$ (See \citealt{LP06}). The  value of $\alpha$ would allow one to classify the turbulence system into two regimes. The $\alpha>1$ case is termed "steep spectrum" in \citetalias{LP00} while otherwise termed "shallow spectrum". For instance,}, the Kolmogorov turbulence has $\alpha=5/3$, which is a steep spectrum. The density spectrum can have both steep and shallow spectrum ($\alpha <1$). {  Notice that the shallow density spectrum emerges from either high sonic Mach number turbulence \citep{2010ApJ...720..742K} or from self-gravity (See \citealt{2018MNRAS.477.4951L}). As a remark, the sonic Mach number is defined as $M_s=V/c_s$, where $V$ is the turbulent velocity at the injection scale and $c_s$ is the sound speed.}

{It is worth noting that } both {the velocity structure function and the density correlation }functions are {mostly} anisotropic {
, (\citealt{GS95}, see also \citealt{2019tuma.book.....B} for the condition of anisotropy)} with the {local } magnetic field playing the dominant role for determining the anisotropy of velocities. The latter property is essential for using PPV information for determining the magnetic field direction with the VGT.

\subsection{Thin and Thick velocity channels}

For our study, we have to adopt several concepts from \citetalias{LP00}. In particular, we have to define thick and thin velocity channels. Importantly, the criterion that the channel is \textit{thin} or \textit{thick} {is given by the following criterion}\footnote{  Notice that in the case of no thermal broadening, the thin criterion of Eq.\ref{eq:thin_criterion} collapses to $\Delta v^2 \ll D_z({\bf X}, z=0)$, which is a compassion between the channel width and the turbulence dispersion along the line of sight.}
\begin{equation}
\begin{aligned}
\mathit{thin} &:& \Delta v^2 + 2\beta_T \ll D_z({\bf X}, z=0) \\
\mathit{thick} &:& \Delta v^2 + 2\beta_T \gg D_z({\bf X}, z=0),
\label{eq:thin_criterion}
\end{aligned}
\end{equation}
{where the right hand side of the equation is $D_z({\bf X},z=0)$, the {z-component} velocity structure function,} {  depending} on the {plane-of-sky} (POS) separation ${\bf X}$ .{In other words, in \textit{thick} channels, the velocity effects are integrated out in velocity channels, while in \textit{thin} channels, they are retained.} It was shown in \citetalias{LP00} that the relative contribution of velocity and density changes as the velocity channel gets from thin to thick regime. Also, note that the definition of the thin and thick channels is {  position}-dependent, i.e., it depends on the scale of the fluctuations that we measure. Therefore, the channels can be thin for small-scale fluctuations and remain thick for large scale fluctuations or otherwise. Notice that\citetalias{LP00} demonstrated that effect of the thermal broadening is similar to the increases the effective width of the velocity channel $\Delta v_{effective}$ following Eq.\ref{eq:thin_criterion}:
\begin{equation}
  \Delta v_{effective}^2 \approx \Delta v^2 +\frac{2k_{B}T}{m}
  \label{eq:effective_dv}
\end{equation}
We emphasize that whether a channel is thin or not depends on the planer position ${\bf X}$.  {  Checking the thin criterion in multiphase turbulent environments with a varying temperature needs extra caution since both $D_z$ and $T$ are both functions spatial distance $r=({\bf X},z)$ }

\subsection{Velocity caustics from thin channel maps}
\label{subsec:caustics}

The concept of {\bf velocity caustics} in PPV {is a theoretical object} used in \citetalias{LP00} to denote the effect of velocity crowding due to the turbulent velocities along the line of sight (See Fig.\ref{fig:cartoons} for a cartoon explaining the reasoning of velocity caustics). Due to this effect, the atoms that are at different physical positions along the same line of sight happen to overlap when viewed in the PPV space. A more detailed discussion of the mapping from real to PPV space is provided in \cite{2009fohl.book..357L}. {In the following, we shall first discuss pictorially what {  it meant} by velocity caustics and how it is defined analytically based on \citetalias{LP00}. }

\subsubsection{Physical picture}

{
Suppose we have five fluid elements A-E along the line of sight associated with some specific density and velocity values associated with interstellar turbulence in panel (1) of Fig.\ref{fig:cartoons}. We further assume that this system's thermal width is $1km/s$, which is considered strong here since $\delta v\sim c_s$. In the absence of thermal broadening, the spectral line is simply the histogram of velocities weighted by the mean density $\langle \rho\rangle$. In our example $\langle \rho\rangle=8.6gcm^{-3}$, as we show in panel (2) of Fig.\ref{fig:cartoons}. In the presence of thermal broadening, the contribution of each fluid element to PPV will be Gaussians instead of discrete numbers. For our example here, the caustics density is $2\langle \rho \rangle =17.2 gcm^{-3}$ at $v=1km/s$ like what we show in panel (3) of Fig.\ref{fig:cartoons}. The density profile in the PPV will then be a Gaussian with height $2\langle \rho\rangle/\sqrt{2\pi}c_s$ (See Eq.4) and width $\sigma=1km/s$. These Gaussians (the blue, red, and green curves in the lower part of the panel (3) Fig.\ref{fig:cartoons}) at different velocities will sum up, and the resultant profile (the purple dash curve in (3) of Fig.\ref{fig:cartoons}) corresponds to the velocity fluctuations, aka. Velocity caustics, of this turbulence system. Notice that the caustics profile is independent of the individual density values of the fluid elements. Therefore, this velocity caustics map is uncorrelated to the density field. The true PPV cube has both density and velocity effects, but the construction of the PPV cube from individual fluid elements is similar to the case of velocity caustics as we show in panel (4) of Fig.\ref{fig:cartoons}. The only difference between panel (3) and (4) in Fig.\ref{fig:cartoons} is that, the height of the Gaussians are related to the true densities of the fluid elements. For example, the total density of the fluid elements at $1km/s$ (i.e. A and C) is just $3 gcm^{-3}$ while that for $2km/s$ and $3km/s$ are both $20gcm^{-3}$. The resultant profile generated from the Gaussians will then be the black curve in panel (4) of Fig.\ref{fig:cartoons}. One can see from panel (5) of Fig.\ref{fig:cartoons} that the {  density weighted profiles and the caustics profiles} are very different. In particular, the differences between the two profiles correspond to the contributions of the density fluctuations to PPV. In our example here, we see that the velocity contribution is significantly larger at $1km/s$ relative to the total PPV cube but being smaller at $v=3km/s$. Notice that the density contribution can be negative but not for the velocity contribution. This simple example illustrates how velocity caustics and density fluctuations are defined in the channel maps.} 

One of the most important consequences for the effect of velocity caustics is the creation of PPV intensity structures that is {independent} of the true density structures {  in 3D}. As a result, {  there are non-zero PPV intensity fluctuations in spectroscopic data} even from incompressible turbulence {or in observational maps that are unrelated to velocity fluctuations (e.g. synchrotron intensity fluctuations that {  contain} of density and magnetic field fluctuations)}. {  Even for the case of incompressible turbulence, the thin channel maps still contain intensity fluctuations due to the non-linear mapping of velocity turbulence structures along the line of sight.(See panel (4),(5) of Fig.\ref{fig:cartoons}). This explains why the velocity channel has to be thin in observing the velocity fluctuations in PPV data.} 

\subsubsection{Mathematical formulation}

To formulate the statistical behavior of velocity channels in non-isothermal environment {mathematically}, we need to rewrite the velocity channel formula from \cite{LP04} with explicit {consideration of thermal broadening.} The density in PPV space of emitters with temperature $T$ moving along the line-of-sight with stochastic turbulent velocity $v_{turb}(\bf x)$ and regular coherent velocity $v_{\mathrm{g}}(\bf x)$ is (See \citealt{LP04}):
\begin{equation}
\begin{aligned}
p(\mathbf{X},v_0,\Delta v) &= \int dz \rho(\mathbf{X},z) \left(\frac{m}{2\pi k_BT}\right)^{1/2} \times\\
&\int_{v_0-\Delta v/2}^{v_0+\Delta v/2} dv W(v) e^{-\frac{m(v-v_{turb}(\mathbf{X},z))^2}{2k_{B}T(\mathbf{X},z)}},
\label{eq:rho_PPV}
\end{aligned}
\end{equation}
where sky position is described by 2D vector $\mathbf{X}=(x,y)$ and $z$ is the line-of-sight coordinate and $W(v)$ is a window function given by the instrument. Eq.~(\ref{eq:rho_PPV}) is \textit{exact}, including the case when the temperature of emitters varies in space, $T = T(\mathbf{X})$. One can see that the three {  quantities}, namely, $\rho$, $v_{turb}$ and $T$ enter differently into Eq.~\ref{eq:rho_PPV}. This provides the physical basis for separating these different contributions to the velocity channel maps.

To facilitate the discussions of what is contained in velocity channels, we would use the PPV density:
\begin{align}
\tilde{\rho}({\bf X},v) &= \int dz \rho({\bf X},z) f(v({\bf X},z)),
\end{align} where \(f(v)\) is the distribution function of the turbulent velocities along the z-axis. Note, that $\tilde{\rho}({\bf X},v)$ is the measure directly available from observations. Then Eq.~\ref{eq:rho_PPV} can also be rewritten as
\begin{equation}
\begin{aligned}
p(\mathbf{X},v_0,\Delta v) = \int_{v_0-\Delta v/2}^{v_0+\Delta v/2} dv \tilde{\rho}(\mathbf{X},v) \times\\W(v)\left(\frac{m}{2\pi k_BT(\mathbf{X},v)}\right)^{1/2} e^{-\frac{m(v-v_0)^2}{2k_{B}T(\mathbf{X},v)}},
\label{eq:rhov_PPV}
\end{aligned}
\end{equation}

\citetalias{LP00} quantifies the statistics of intensities observed in velocity channels {through correlation functions} \footnote{Notice that the use of correlation or structure function has to be computed in a large sampling region \citep{2018ApJ...865...54Y}. However, a hole-punching test performed in \cite{2018ApJ...865...54Y} showed that the structure function is insensitive to the percentage of the area with missing data up to 40\%. }. There the two-point correlation function of velocity channel intensity is given by: 
\begin{equation}
\begin{aligned}
 \zeta_I({\bf X},v_0;\Delta v)&\propto \int_{0}^{S} dz \frac{\zeta'_\rho}{(D_z({\bf r})+2\beta_T)^{1/2} }\times\\
 &\int_{v_0-\Delta v/2}^{v_0+\Delta v/2} dv W(v) e^{-\frac{(v-v_0)^2}{D_z({\bf r})+2\beta_T}},
 \label{eq:LP00}
\end{aligned}
\end{equation}
where $\zeta'_\rho = \zeta_\rho/\langle \rho^2\rangle$ {(See Eq.\ref{eq:cf_rho})} is the normalized three-dimensional over-density function, $D_z({\bf r})$ is the velocity structure function projected along the line of sight (Eq.\ref{eq:v_z}), $\beta_T$ is the thermal contribution\footnote{As a remark (1) the thermal contribution can actually be a function of distance $\beta_T=\beta_T(r)$ (2) in the case of non-isothermal case the temperature field can act as the passive scalar and reflect the turbulent velocity statistics (see \citealt{2019tuma.book.....B}). This effect was recently observed in the fluid undergoing thermal instability in \cite{2017NJPh...19f5003K}.} {  and $S$ is the line of sight depth of the turbulent cloud. }

For any density spectral index, the over-density correlation function can be {  modelled as (See Appendix of \citealt{LP06} or \citealt{KLP17a}, and a numerical test in \citealt{2010ApJ...720..742K})}:
\begin{align}
  \zeta'_\rho \propto 1-\text{sign}(\gamma+3)(\frac{r_0}{r})^{|\gamma+3|},
  \label{eq:corr_model}
\end{align}
where $r_0$ is a characteristic length-scale signifying the turbulent correlation length\footnote{There is an implicit condition $r>r_0$ for this model to hold where $r_0$ is treated as a cut-off of the correlation/structure functions. See \cite{LP04}.} {  and $\gamma$ is a modelled parameter denoting the density dependencies to $r$.} By inserting Eq. \ref{eq:corr_model} back to Eq. \ref{eq:LP00}, we get two terms, one arising from the {first term (i.e. the 1)} in Eq. \ref{eq:corr_model}, which provides the pure velocity contribution effect :
\begin{equation}
\begin{aligned}
  {\cal I}_1({\bf X})&=\int_{0}^{S} dz \frac{1}{(D_z({\bf r})+2\beta_T)^{1/2} }\times\\
  &\int_{v_0-\Delta v/2}^{v_0+\Delta v/2} dv W(v) e^{-\frac{(v-v_0)^2}{D_z({\bf r})+2\beta_T}}
  \end{aligned}
  \label{eq:const_d},
\end{equation}
and the {second part of Eq. \ref{eq:corr_model} contains} {mainly density fluctuations}\footnote{It is worth noting that, despite we call ${\cal I}_2$ as the density fluctuations in \citetalias{LP00}, later literature and even this work,$ {\cal I}_2$ actually depends both to density and velocity. }: 
\begin{equation}
\begin{aligned}
   {\cal I}_2({\bf X})&=\int_{0}^{S} dz \frac{-\text{sign}(\gamma+3)(\frac{r_0}{r})^{|\gamma+3|}}{(D_z({\bf r})+2\beta_T)^{1/2} }\times\\
   &\int_{v_0-\Delta v/2}^{v_0+\Delta v/2} dv W(v) e^{-\frac{(v-v_0)^2}{D_z({\bf r})+2\beta_T}}
   \end{aligned}
   \label{eq:orho_d}
\end{equation}
We note, that $D_z(r) \rightarrow 0$ as $r\rightarrow r_0$. Therefore, at small scales $r\sim r_0$ the integral is affected by the density structures {\it weighted by the combined effect from turbulent and thermal velocities}. While dealing with multiphase HI case whose temperature is not constant, we shall assume the HI environment is in the Local Thermal Equilibrium (LTE). This simplifies our theoretical treatment. 

It is important that while the ${\cal I}_1$ presents the pure velocity contribution, {and the density fluctuations dominate in ${\cal I}_2$}. The relative importance of ${\cal I}_1$ and ${\cal I}_2$ depends on the length scale $r$. It is {  useful} to compare the Fourier {  transform} of ${\cal I}_1$ and ${\cal I}_2$, namely $P_v(K)$ and $P_d(K)$ respectively as \citetalias{LP00} did, in characterizing the relative importance of velocity and density fluctuations as a function of wavenumber $K=1/X$\footnote{  We use the capital ${\bf K}$ to refer to the 2D wavevector, and $K=|{\bf K}|$ to be the wavenumber of the 2D vector. Similarly, ${\bf k}$ is the 3D wavevector, and $k=|{\bf k}|$ is the 3D wavenumber. } In fact, as long as the density is not {straightly constant everywhere}, the relative importance of velocity and density fluctuations is always finite {  (i.e. the quantities $P_d$ and $P_v$ do not diverge).} 

Moreover, it is evident to see that the effective width of the velocity channel is critical in determining the relative importance of velocity and density fluctuations. For example, in the case of extremely thin channels ($\Delta v\rightarrow 0$ in Eq. \ref{eq:thin_criterion}), the velocity integral can be approximated as a constant integral,i.e.
\begin{equation}
\int_{v_0-\Delta v/2}^{v_0+\Delta v/2} dv W(v) e^{-\frac{(v-v_0)^2}{D_z({\bf r})+2\beta_T}} \approx \textit{const}
   \label{eq:vintegral}
\end{equation}
in this scenario, both Eq. \ref{eq:const_d} and Eq. \ref{eq:orho_d} will be left with:
\begin{subequations}
\begin{align}
{\cal I}_1\approx\int_{0}^{S} dz \frac{1}{(D_z({\bf r})+2\beta_T)^{1/2} }\\
{\cal I}_2\approx\int_{0}^{S} dz \frac{-\text{sign}(\gamma+3)(\frac{r_0}{r})^{|\gamma+3|}}{(D_z({\bf r})+2\beta_T)^{1/2}}
\end{align}
\label{eq:thin_case}
\end{subequations}
{  Notice that $I_1$ contains no density-related terms. We can see from here that the factor $-\text{sign}(\gamma+3)(\frac{r_0}{r})^{|\gamma+3|}$ is critical in determining the relative importance of velocity and density fluctuations.}

{  The} \citetalias{LP00} study {  results} in the development of the Velocity Channel Analysis (VCA) that employs the difference of how ${\cal I}_1$ and ${\cal I}_2$ depend on the thickness of channel map\footnote{The spectrum of fluctuations along with the velocity coordinate is also affected in different ways by the velocity and density contributions. The latter will bring changes to the Velocity Coordinate Spectrum (VCS) technique (\citealt{LP00,LP06, LP08, 2009ApJ...693.1074C,2015ApJ...810...33C}) that we do not consider in this paper.} for studying velocity and density statistics of the ISM. VCA turned out to be a reliable tool in probing the velocity spectral index from observational spectroscopic maps. Table. \ref{tab:L09} lists the predicted spectral dependencies of VCA according to \cite{LP00,LP04} {  in which the 3D velocity spectral index $m$ is involved in the discussion of VCA}. The velocities and densities' contributions in VCA were separated by changing the thickness of the velocity channels. The technique was successfully tested numerically in \cite{2009ApJ...693.1074C} {  and} was applied to both HI and CO data (see \citealt{2001ApJ...551L..53S}, \citealt{2006ApJ...653L.125P,reply19}). 

\subsection{The need {  for} exact velocity caustics in real observational data}
\label{ssec:need}

{  The issue of thermal broadening is not that severe in the case of molecular tracers  like ${}^{12/13}CO$, {or} in molecular hydrogen H$_2$ according to \citetalias{LP00} since the turbulent environment at such instance is usually isothermal and supersonic.\footnote{Notice that the turbulence in molecular gas can be subsonic, but the thermal broadening $\sim 2k_{B}T/m_{CO}$ may be not important. Naturally, thin slice regime is {available} for CO and heavier molecules {despite the turbulent velocities in H$_2$ are still subsonic (See Appendix \ref{secap:vda_emissions_absorptions}).} As a result, it is rather easy to see a velocity dominant channel, aka velocity caustics, in observation.} As a result, the thin channel criterion (cf Eq.\ref{eq:thin_criterion}) can be easily fulfilled.}

\citetalias{LP00} applied similar considerations to neutral hydrogen media consisting of Warm ($\sim 10^4$K) and Cold ($\sim 10^2$K) components \citep{1990ASPC...12....3M,2005ApJ...630..238D,Draine2011PhysicsMedium}. The turbulence in galactic HI is {  generally} subsonic {  (in high latitudes)} to trans-sonic { (near the inner regions of galactic disks, see \citealt{2006ApJ...638..797D})} in terms of its warm component {and often supersonic for the colder one.}  However, the two components are not independent but interconnected. First of all, they are connected by the magnetic field. Second, the two components are subjected to thermal instability with a significant fraction of matter being in the intermediate formally unstable state \citep{1990ASPC...12....3M,2005ApJ...630..238D,Draine2011PhysicsMedium}. In this situation, it is natural to consider that the condensation of HI into the Cold phase is happening within the flow of the Warm turbulent phase, which means the two phases share the same turbulent velocity. This was the model adopted in \citetalias{LP00} to represent galactic HI. In terms of coupling of motions of the Warm and Cold phase, other authors adopted similar models concerned with aspects of multiphase HI dynamics not related to the applicability of the VCA. For instance, \cite{2009ApJ...704..161I} discussed the condition of CNM formation in WNM collision flow. Furthermore, in \cite{2016ApJ...833...10I} they further characterize the conditions for the filamentary structures to be perpendicular to the magnetic field in multiphase media (See also \citealt{2020MNRAS.497.4196S} for the molecular cloud variant). 

The concept of comoving phases has far-reaching implications to the determinations of thin and thick channels in multiphase neutral hydrogen observations {  since the comoving phase argument is an assumption well taken by theorist \citep{LP00,2009ApJ...704..161I} but never been able to characterize in observation as of our knowledge.} {  If} the two phases have similar velocities, their velocity structure functions $D_z$ are the same. As a result, for a given velocity channel width $\Delta v$ it is entirely possible that the channel itself is thin for CNM but thick for WNM due to the dramatic differences of their temperature by Eq.\ref{eq:thin_criterion}. The resultant velocity channel can be composed of velocity caustics from the colder component with the density fluctuations from the warmer component in multiphase neutral hydrogen media. This coupling of density and velocity structures between different phases was never considered adequately in \citetalias{LP00}. Nevertheless, it was previously impossible to disentangle the density and velocity fluctuations from different phases to our knowledge, making the determination of velocity turbulence statistics very difficult, as reported by some authors in the community.

In this study, we, however, go beyond the particular model adopted in \citetalias{LP00} and formulate a recipe {  for} decoupling the density and velocity fluctuations from observational multiphase data. Our current paper addresses a general and more complex problem of the effect of thermal broadening on the channel map statistics. In other words, we consider the case {where the} intensity fluctuation in PPV {contains} significant thermal broadening and without heavy tracers, which is the worst case from the point of VCA study if turbulence is subsonic. This paper will show how our new procedure of separating velocity and density contributions works in this case. 
\begin{table*}
\centering

\begin{tabular}{| c | c | c |}
\hline
Slice thickness & \specialcell{Shallow 3D density spectrum \\$k^{-3+\gamma}$,$\gamma <0$} & \specialcell{Steep 3D density spectrum\\ $k^{-3+\gamma}$,$\gamma>0$} \\ \hline \hline
Thin slice  & $K^{-3+\gamma+m/2}$ & $K^{-3+m/2}$ \\
Thick slice & $K^{-3+\gamma}$ & $K^{-3-m/2}$\\
Very Thick slice & $K^{-3+\gamma}$ & $K^{-3+\gamma}$\\
\hline
\end{tabular}
\caption{\label{tab:L09} The spectrum asymptotics of velocity channels in the limiting cases, where $\gamma$ is the reduced density power spectrum slope, while $m$ is the velocity structure function index $D_z \sim \langle (v_z({\bf r'}) - v_z({\bf r}+{\bf r'}))^2 \rangle_{\bf r'} \propto r^m$ {(See Eq.\ref{eq:v_z})}.  Note that the definition of the thin and thick channels is $r$-dependent according to Eq. (\ref{eq:thin_criterion}). Extracted from \cite{2009fohl.book..357L}.}
\end{table*}

\begin{table}
\centering
\begin{tabular}{c c c c c}
Model & $M_S$ & $M_A$ & $\beta=2M_A^2/M_S^2$ & Resolution \\ \hline \hline
huge0         & 6.36 & 0.22 & 0.00049 & $1200^3$\\
e5r3          & 0.61 & 0.52 & 1.45 & $1200^3$ \\
{multiphase }      & {1.42} & {0.80} & {0.625} & $480^3$ \\\hline \hline
\end{tabular}
\caption{\label{tab:sim} {Descriptions} of MHD simulation cubes {which some of them have been used in the series of papers about VGT, (\citealt{YL17a,YL17b,LY18a,LY18b}},{Ho et.al in prep}). $M_s$ and $M_A$ are the R.M.S values at each snapshots are taken. }
\end{table}

\begin{table}
\centering
\begin{tabular}{c c c c}
 & WNM & UNM & CNM \\ \hline \hline
Density filling factor & 0.081 & 0.284 & 0.635 \\
Volume filling factor & 0.608 & 0.316 & 0.077 \\\hline \hline
\end{tabular}
\caption{\label{tab:MP_factor} The density filling factor and Volume filling factor for each phase in the multi-phase simulation. {The numerical details will be in Ho et.al in prep}.}
\end{table}

\section{The construction of the Velocity Decomposition Algorithm}

\label{sec:la}

To proceed with the analysis, we describe a mathematical procedure that allows us to separate the velocity and density contributions {in} velocity channels. In what follows, we will term this procedure the Velocity Decomposition Algorithm (VDA). This algorithm uses the statistical properties of fluctuations induced by MHD turbulence within thin velocity channel maps. Here we mostly focus on the subsonic regime, which was the most challenging case within the VCA as it is formulated in \citetalias{LP00} (See \S \ref{ssec:need}). We also test numerically the same approach for the case of supersonic turbulence. Below we lay out the formal deviations and discuss the possible improvements in Appendix \ref{sec:apb} for readers to understand the theoretical foundations of the algorithm. {Moreover, we would also discuss another method of velocity decomposition in Appendix \ref{sc:SVDA}}.

In \citetalias{LP00} it was analytically described how the density and velocity fluctuations contribute to the fluctuation of velocity channels (See \S \ref{sec:one_sigma} for the formal expressions in terms of their respective correlation functions). Formally we can always write the fluctuation of velocity channel intensity as:
\begin{equation}
  p(\mathbf{X},v_0,\Delta v) - \langle p\rangle_{{\bf X}\in A} = p_d(\mathbf{X},v_0,\Delta v) + p_v(\mathbf{X},v_0,\Delta v)
  \label{eq:sp}
\end{equation}
where $\langle p\rangle_{{\bf X}\in A}$ represents the velocity channel averaged over a certain spatial area $A$. The subtraction of the mean value in Eq.\ref{eq:sp} is required as we deal with the fluctuations arising from turbulence. Notice that $p_d$ and $p_v$ are functions of the Plane of Sky (POS) two dimensional vector $\mathbf{X}$, as well as the velocity channel position $v_0$ and channel width $\Delta v$. In what follows, we shall refer to $p_v$, i.e. the velocity contribution to velocity channels, as the velocity caustics contribution.

There are a few properties of the density ($p_d$) and velocity ($p_v$) contributions that we will use below, namely:
\begin{enumerate}
  \item {\bf Orthogonality of $p_d$ and $p_v$ when $M_s\ll 1$}: {  \citetalias{LP00} postulated that the density and the velocity part are statistically uncorrelated for the case of MHD turbulence.} That means that that the average $\langle p_\rho p_v\rangle_A$ should be zero in the case of subsonic turbulence\footnote{From our numerical experiment, $\langle p_d p_v \rangle$ is not always zero for supersonic turbulence (Appendix \ref{sec:apb}). \citetalias{LP00} discussed this situation {  in} their Appendix D and we provide a corresponding deviation in the case of non-orthogonality in Appendix \ref{sc:SVDA}.}.
  \item {\bf $p_v=0$ when $\Delta v\rightarrow \infty$}: The caustics should not be observed when the channel width is large. This is a property {  that} naturally follows from \citetalias{LP00} theory for {  velocity} caustics. The {  increase} of {  the} number of emitters in one channel means the removal of {  emitters} from other channels. Thus the {  caustics fluctuations} must decrease {  when the channel width increases}. 
  \item {\bf $p_d\propto I$ when $M_s \ll 1$}: In the case when the sonic Mach number is low, \citetalias{LP00} {  predicts} that the thermal broadening will significantly increase the effective channel width of the velocity channel. When we deal with subsonic turbulence, the emission from the entire volume arrives to every channel map (if we do not consider the {  galactic} rotation curve). This emission is proportional to the total column density.{  In the case of thick channels, the integration kernel will cover the whole velocity axis.} As a result, the {  thick} velocity channel would then {\it look like } the intensity map (See Appendix \ref{secap:kernel} for the {  technical} reason and the differences of the integration between a thick channel and the intensity map, and Appendix \ref{sc:SVDA} for a method that is independent of $M_s$.). {Moreover, this construction guarantees that $\langle p_d\rangle =\langle p_v \rangle =0$.}
\end{enumerate}

In what follows, we shall restrict our discussion to sub-sonic turbulence and leave the supersonic counterpart in Appendix \ref{sc:SVDA} as the latter case is less influenced by the thermal broadening. Using property 3, we can then write the form $p_d$ for each single channel:
\begin{equation}
  p_d (\Delta v=\infty)\propto I-\langle I\rangle_{{\bf X}\in A}
  \label{eq:postulate}
\end{equation}
where I is the intensity map. From Property 1 we have $\langle p_\rho p_v\rangle_A=0$. That means for any velocity channel $p$ that can be expressed by Eq.\ref{eq:sp}, we can multiply it by $p_d$ and take the areal average:
\begin{equation}
\begin{aligned}
  \langle (p - \langle p\rangle)p_d\rangle &= \langle p_d p_d\rangle + \langle p_v p_d \rangle
\end{aligned}
\end{equation}
which the second term is zero according to Property 1\footnote{In fact the areal average operator is an inner product in linear algebra, and here $p_d$ and $p_v$ "spans" the velocity channel space (as we see in Eq.\ref{eq:sp}). That means if we can write out the "unit vector" of $p_d$ and $p_v$, namely $\hat{p}_d$ and $\hat{p}_v$, we can simply write
\begin{equation*}
  p = A\hat{p}_d + B\hat{p}_v
\end{equation*}
for some A,B. It is straightforward to derive that $\hat{p}_{d,v} = p_{d,v}/\sigma_{p_{d,v}}$. The space span by ($\hat{p}_d$,$\hat{p}_v$) is a complete vector space with the inner product operator be the areal average operator $\langle ... \rangle_A$}. With properties 1 and 3, we can then formally write $p_v$ as
\begin{equation}
  p_v = p - \langle p \rangle - C(I-\langle I\rangle)
  \label{pvp}
\end{equation}
for some constant C that is a function of line of sight velocity $v$. We first multiply two sides of Eq. (\ref{pvp}) by $I-\langle I\rangle$, then take the areal average operator and using the Property 2 we get:
\begin{equation}
  C = \frac{1}{\sigma_I}\left\langle (p-\langle p\rangle )(\frac{I-\langle I\rangle}{\sigma_I})\right\rangle
\end{equation}
where $\sigma_I^2=\langle (I-\langle I\rangle)^2\rangle$ is the dispersion of $I$ {\it within the area A}. Then we can write $p_v$ as :
\begin{equation}
  p_v(\mathbf{X},v_0,\Delta v) = (p-\langle p\rangle) - \left\langle (p-\langle p\rangle )(\frac{I-\langle I\rangle}{\sigma_I})\right\rangle\frac{I-\langle I\rangle}{\sigma_I}
  \label{eq:lineardecomposition}
\end{equation}
Eq.\ref{eq:lineardecomposition} gives an expression of velocity caustics in the case of subsonic media. Notice that Eq.\ref{eq:lineardecomposition} fulfills the following properties
\begin{enumerate}
  \item The sum of $p_v$ across all channels are zero, which we can see from the definition that $\sum_v p(v) = I$, $\sum_v \langle p(v)\rangle=\langle I\rangle$ (since the averaging operator and the summation direction are commutative ) and $\langle (I-\langle I\rangle)^2\rangle=\sigma_I^2$. This is exactly property 2 we listed before.
  \item If the area of averaging $A$ is not statistically large enough (meaning: it does not recover the MHD statistics in this area), then $p_v\rightarrow0$. This property is very important since velocity caustics can only be measured statistically (See \S \ref{sec:one_sigma}). We shall explore numerically below (See Sec.\ref{sec:test} ) on the size of $A$ required in the simulation.
\end{enumerate}
With these properties in mind, we simplify Eq.\ref{eq:lineardecomposition} to:
\begin{equation}
\begin{aligned}
  p_v &= p - \left( \langle pI\rangle-\langle p\rangle\langle I \rangle\right)\frac{I-\langle I\rangle}{\sigma_I^2}\\
  p_d &= p-p_v\\
  &=\left( \langle pI\rangle-\langle p\rangle\langle I \rangle\right)\frac{I-\langle I\rangle}{\sigma_I^2}
\end{aligned}
\label{eq:ld2}
\end{equation}

This constitutes the {\bf foundation of velocity decomposition algorithm}, i.e., the VDA. The suggested procedure is entirely new, and it is not developed based on \citetalias{LP00} theory, but the underlying idea was based on the properties of velocity caustics described in \citetalias{LP00}. We expect the VDA to work in the case of significant thermal broadening, i.e., when the performance of the traditional procedures described in \citetalias{LP00} drops. In particular, for such a significant thermal broadening, both the VCA and the Velocity Channel Gradients (VChGs) introduced in \cite{LY18a} are not expected to work. We shall augment both the VCA (\S \ref{sec:VCA}) and VChGs (\S \ref{subsec:VGT}) with the VDA and test numerically the new techniques below.

\section{Numerical simulations}
\label{sec:sim}

\subsection{Isothermal simulations}

{The isothermal simulations are set up similarly to the simulations in \cite{LY18a}. The 3D simulation cubes are obtained from a single fluid, operator-split, staggered grid MHD Eulerian code ZEUS-MP/3D \citep{2006ApJS..165..188H} to set up a three dimensional, isothermal ($T=10K$), saturated turbulent medium under triply periodic conditions. The turbulence is injected solenoidally {  by Fourier-space forced driving at $k=2$ (See e.g, \cite{1998ApJ...508L..99S}) and used by several authors \citep{CL03,2007ApJ...661..262D,2008ApJ...678L.105D} . The choice of force stirring over the other popular choice of decaying turbulence is because only the former will exhibit the full characteristics of turbulence statistics (e.g, power law, turbulence anisotropy) extended from $k=2$ to {a dissipation scale of $12$ pixels} in a simulation, and matches with what we see in observations (e.g., \citealt{Armstrong1995ElectronMedium, Chepurnov2010ExtendingData}).}

For the particular application of the VDA, we consider two extreme cases that correspond to the highly subsonic and supersonic turbulence, with both being sub-Alfvenic $M_A<1$. The conditions of these simulations are listed in Tab.\ref{tab:sim}. Due to the spatial resolution requirements on both thermal broadening and the VCA method (See \citealt{2009ApJ...693.1074C}), the two simulations are both $1200^3$, such that the inertial range extends to close to two orders of magnitude ($k=2$ {to} $k=100$). The two limiting cases will correspond to the strong thermal limit and the supersonic density crowding limits.}

\subsection{Multiphase simulation}
\begin{figure}[th]
  \centering
  \includegraphics[width=0.49\textwidth]{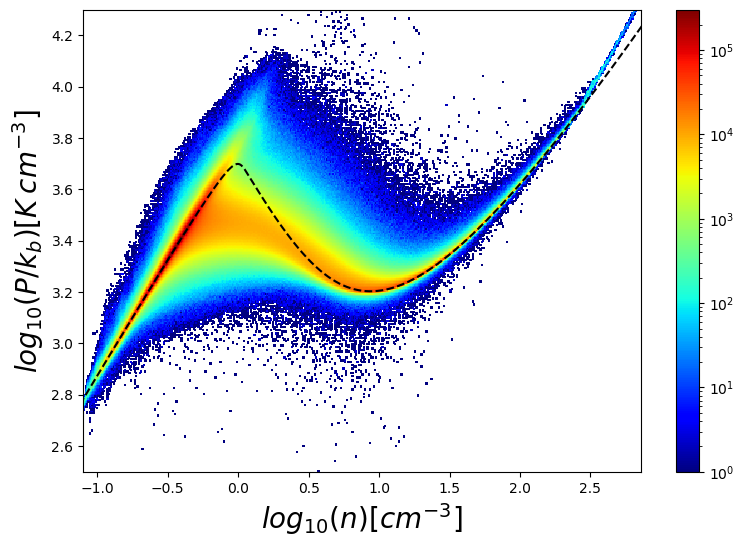}
  \caption{\label{fig:phase_diagram}  A 2D histogram showing the volume-weighted phase diagram of our simulation with the count rendered in log scale. The black dash line indicates the steady equilibrium state of ISM with the cooling function. {The simulation details will be in Ho et. al (in prep).} }
\end{figure}

\begin{figure*}[th]
  \centering
  \includegraphics[width=0.99\textwidth]{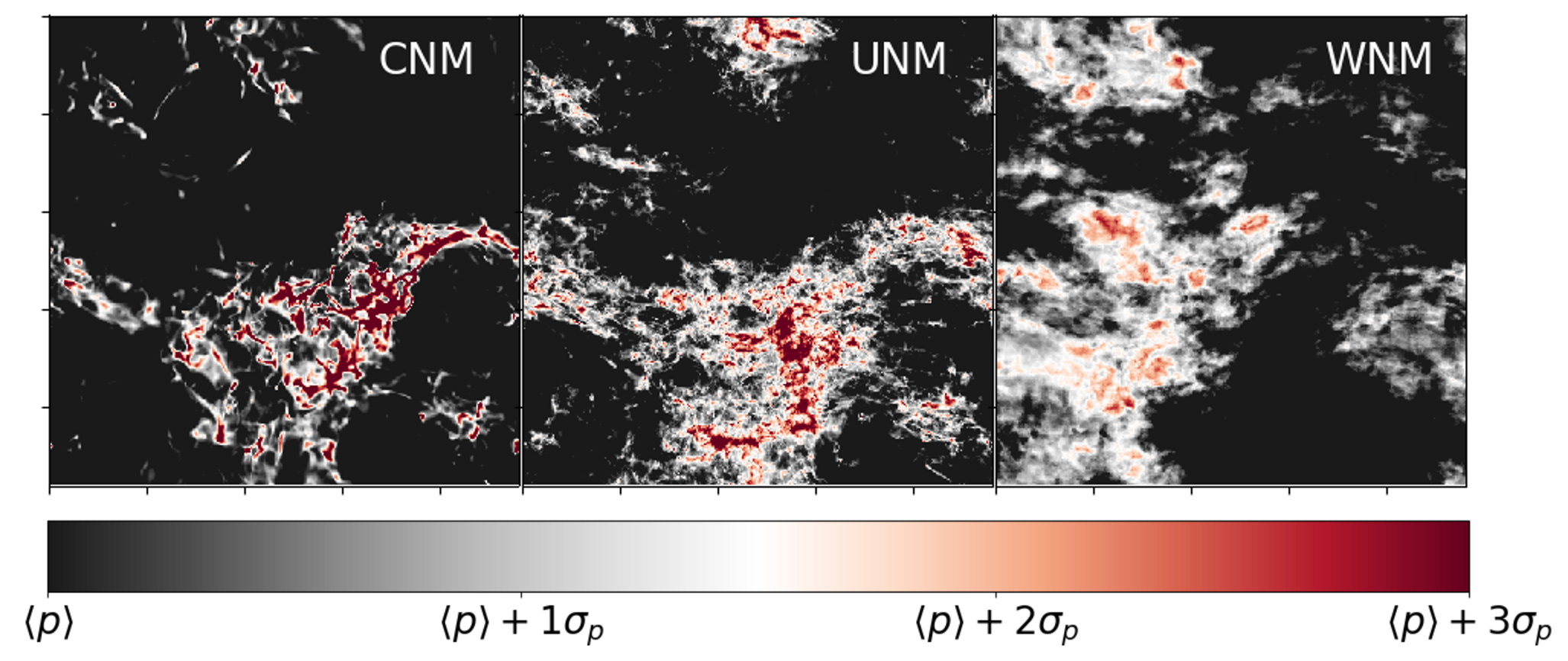}
  \caption{\label{fig:multiphase_illus}  A set of figures showing the center channel intensity structure of cold ($T<200K$, left), unstable ($200K < T < 5250K$, middle) and warm neutral media ($T>5250K$,right ), from our multiphase simulation (See Tab.\ref{tab:sim} for the simulation condition). {The simulation details will be in Ho et. al (in prep).} }
\end{figure*}

{The multiphase simulations are set up with the modern MHD code ATHENA++ \footnote{https://github.com/PrincetonUniversity/athena-public-version/wiki} \citep{2016ApJS..225...22W,2020ApJS..249....4S}. To solve the radiative heating and cooling, we adopt the simplified generalized heat loss function proposed by \citealt{Koyama02}\footnote{  The volume and mass filling fractions depends also on the other sources of heating, such as the local star formation rate, which is not accounted in \cite{Koyama02}. }. We neglect explicit thermal conduction as it cannot be resolved in the resolution used in this simulation. For our purpose here, we only provide one simulation that has a condition similar to the realistic multiphase neutral hydrogen gases with mass/volume fractions {  being} consistent with observations. For the initial state, we set up a 3D periodic cube with the length of 200 pc {and we are assuming the fluid represents the bulk neutral hydrogen in the interstellar media} The mean {number} density {for the neutral hydrogen} set as $n_h = 3 cm^{-3}$ and magnetic field strength {is} $1 \mu G$. The turbulence is {driven at} every iteration step to generate {a saturated} turbulent medium. We take a snapshot at about $110 Myr$. We obtain a classical three-phase turbulence system under this set-up, namely cold, unstable, and warm phase \citep{1977ApJ...218..148M,2012MNRAS.421.3488H}. For the convenience of the discussion, we define two temperature thresholds to differentiate {the} three-phase: {We shall define the gas as the cold phase when the temperature of the gas is below $200 K$ while those} above $5250K$ {as} to warm phase, while the gas in between is the unstable phase. Fig.\ref{fig:phase_diagram} shows volume-weighted phase diagram. Besides, we computed the volume filling fraction and mass {  filling} fraction of each phase at Table \ref{tab:MP_factor}. Those fractions consist of the previous study of the multiphase simulation \citep{2017NJPh...19f5003K,2003ApJ...586..1067H}. In Fig.\ref{fig:multiphase_illus}, we show the channel map at the peak of the spectral line for each phase, which are cold ($T<200K$, left), unstable ($200K < T < 5250K$, middle) and warm neutral media ($T>5250K$, right). We can see from Fig.\ref{fig:multiphase_illus} that structurally the cold and unstable phases are more filamentary than that of the warm phases. More details of the numerical method and the simulations' physical picture are provided in Ho et al. (in prep).}

\section{Numerical testing of the VDA}
\label{sec:test}

{In this section, we numerically test the VDA method using the simulations from \S \ref{sec:sim}. In particular, we test the accuracy of VDA in \S \ref{subsec:testVDA}, using VDA to discuss the important question of whether velocity or density fluctuations are dominating the channel fluctuations in \S \ref{subsec:pd2pv} and discuss the effects of noises to VDA in \S \ref{subsec:noise}. }

\subsection{Accuracy of the VDA in numerical simulations}
\label{subsec:testVDA}
The parameter that influences the thermal broadening strength is the ratio between the line of sight velocity to that of the sonic speed $\delta v_{LOS}/c_s$, which we will refer to be the line-of-sight sonic Mach number $M_{s,LOS}$. If {  the gas} is isothermal {  and isotropic}, we expect $M_{s,LOS}=\delta v_{LOS}/c_s\sim M_s/\sqrt{3}$. {  In the case of anisotropic turbulence, a higher or lower ratio between $M_{s,LOS}$ and $M_s$ could be possible depending on the cloud's inclination and also the beam size}. The ratio between $M_{s,LOS}$ and $M_s$ is expected to be smaller in the case when the turbulence is magnetized with the mean-field being perpendicular to the line of sight. The relation between $M_{s,LOS}$ and $M_s$ as it is possible that the interstellar media is weakly supersonic ($M_s>1$), but its emission/absorption lines are strongly broadened, as the latter is characterized by the condition $M_{s,LOS}<1$ only.

We would first like to visualize the power of VDA using numerical simulations. We compute the realistic thermal broadening synthesis (PPV) based on \cite{reply19} (See \S \ref{sec:one_sigma}) which requires three input: the density $\rho$, the line of sight velocity $v_{LOS}$, and also the sonic speed $c_s$. We compute both the density-weighted PPV cube (i.e the velocity channels) $p({\bf X},v)=ppv(\rho,v,c_s)$ and the constant density PPV cube (i.e. the velocity caustics) $n({\bf X},v)=ppv(\langle \rho\rangle,v,c_s)$. For the density-weighted PPV cube, we compute the $p_v$ term based on Eq.\ref{eq:ld2}. Notice that in isothermal simulations that we employ in this experiment, the sonic speed can be re-scaled with other simulation variables since periodic saturated magnetized turbulence simulations are scale-free (See \citealt{2018ApJ...865...54Y}). That means we can freely adjust the broadening strength by varying $c_s$ in the simulation. We shall utilize this fact in later sections, but here we prepare two simulations with different sonic Mach numbers and perform synthetic observations by using their intrinsic $\delta v_{LOS}/c_s$. 

Fig. \ref{fig:illus} illustrates how the algorithm (Eq.\ref{eq:ld2}) works pictorially (See Appendix \ref{sec:apb} for the discussion on density and velocity correlation) For visualization purpose we only {  display} the relative enhancement/discrepancy of local structures in the channel map according to Eq.\ref{eq:ld2}, Therefore, the color bar of Fig.\ref{fig:illus} has been adjusted to be $[\langle p\rangle,\langle p\rangle+3\sigma_p]$ where $\langle p\rangle$ and $\sigma_p$ here are the mean and standard deviation of each image's intensity. We can see visually from Fig.\ref{fig:illus} that in both subsonic and supersonic cases, their respective decomposed $p_v$ are highly similar to those of the constant density velocity channel (i.e., velocity caustics) proposed in LP00. The result from Fig. \ref{fig:illus}  is auspicious in terms of estimating the statistics of velocity caustics using observationally available parameters $p_v$.

\begin{figure*}[th]
  \centering
  \includegraphics[width=0.98\textwidth]{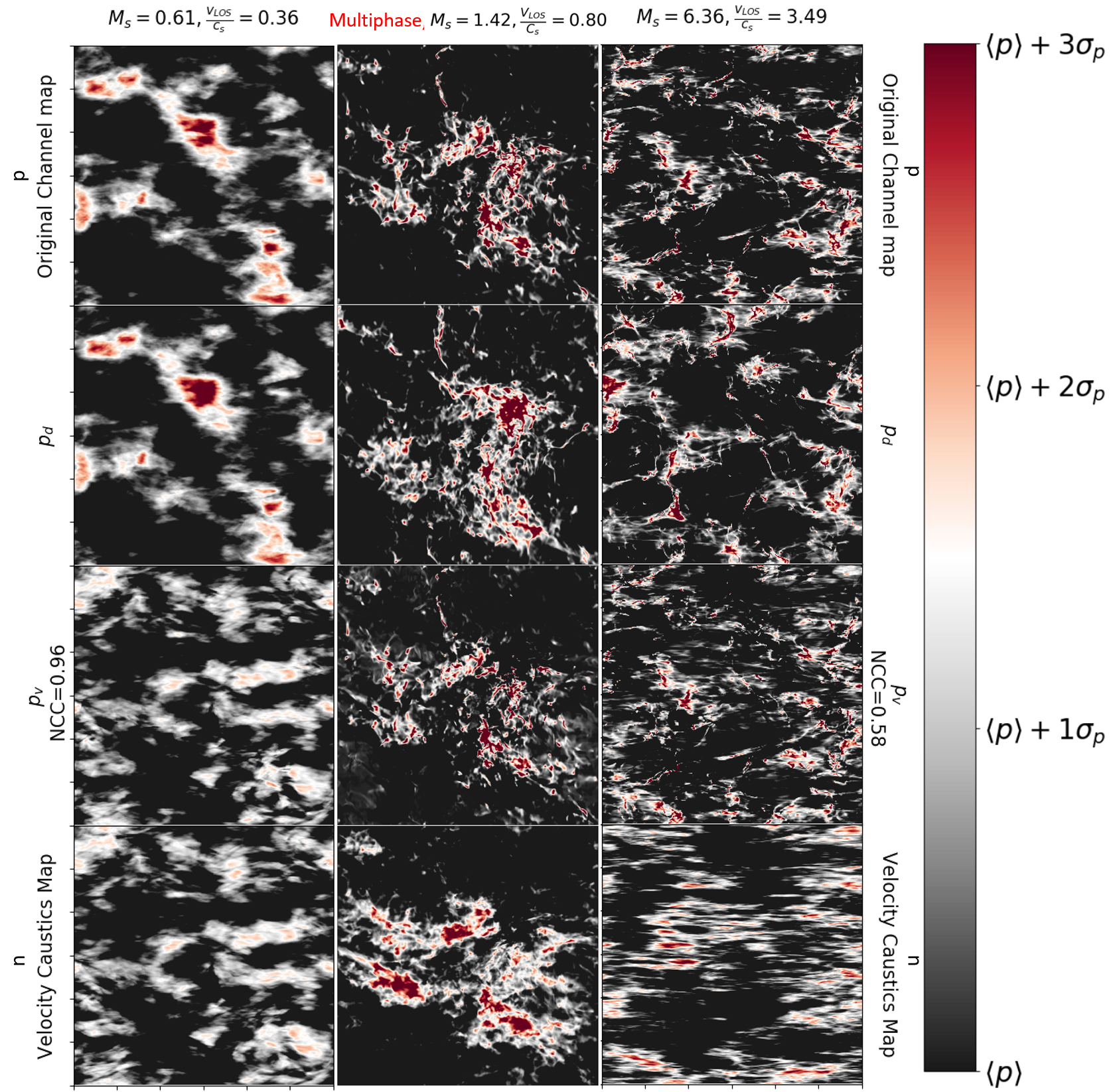}
  \caption{\label{fig:illus} The structure of the velocity channel according to Eq.\ref{eq:ld2} for both subsonic($M_s=0.61$ ,$\delta v_{LOS}/c_s = 0.36$, left column) and supersonic cases ($M_s=6.36$ ,$\delta v_{LOS}/c_s = 1.23$, right column). {We also include the multiphase case for comparison (middle column).} From the top row: Raw density velocity channel at velocity $v_0=0$, 2nd row: the density only velocity channel $p_d$, 3rd row : the decomposed $p_v$ and the respective NCC$(p_v,n)$; bottom row: The true velocity caustics map $n$. The color bar scaling is adjusted such that the minimum scale corresponds to the figure's mean value (black) while the maximum scale corresponds to the mean value plus three standard deviation. }
\end{figure*}

\begin{figure*}[th]
  \centering
  \includegraphics[width=0.98\textwidth]{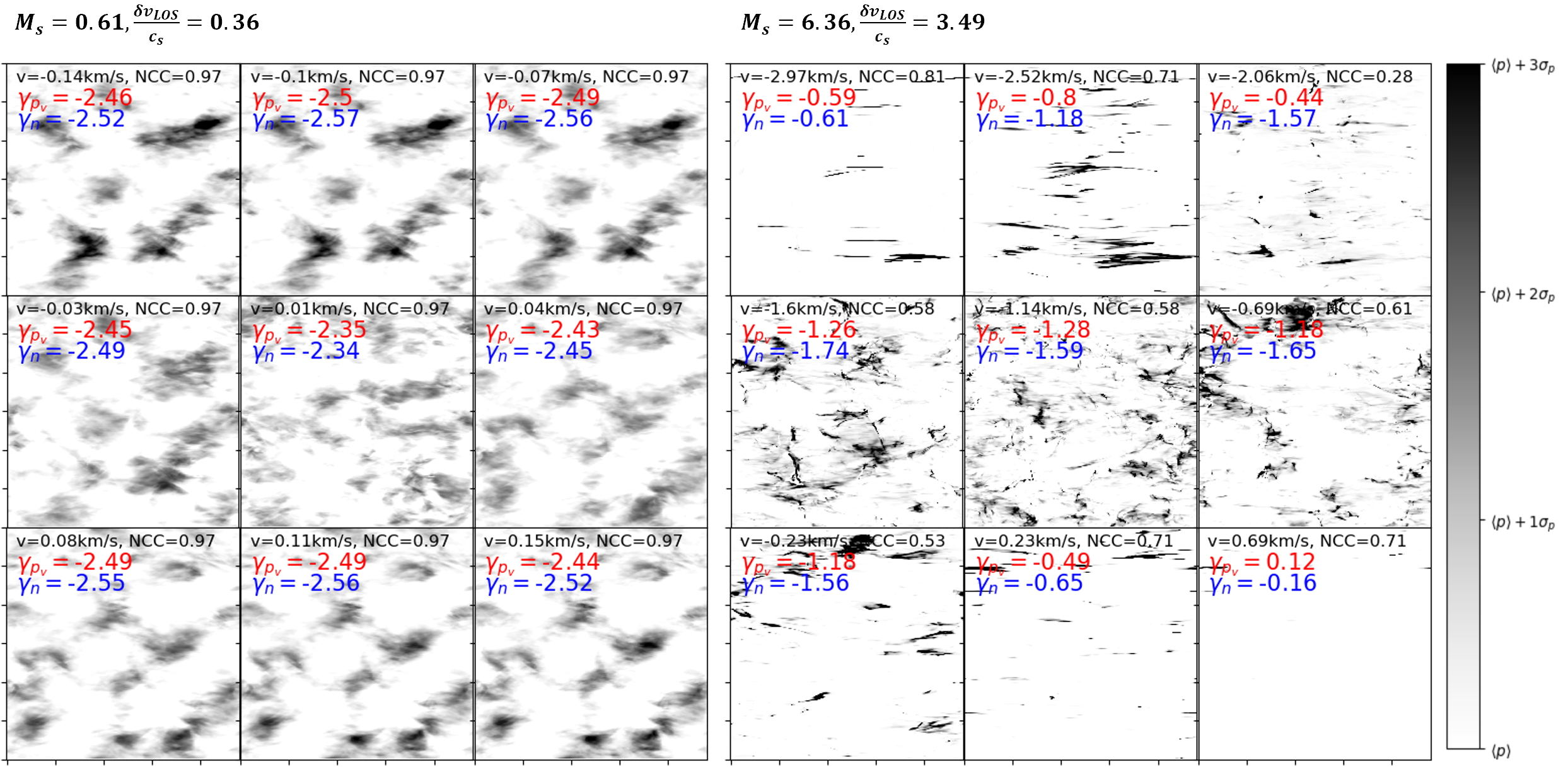}
  \caption{\label{fig:sim_illustration} (Left) An illustration of the decomposition algorithm (Eq.\ref{eq:ld2}) on the different density-weighted velocity channels synthesized from a numerical simulation "e5r3" ($M_s\sim 0.61$). We compute the NCC {between $p_v$ (from Eq.\ref{eq:ld2}) to the real constant density PPV channel $n$, i.e. $NCC(n,p_v)$}. We see that {the numerical value of NCC is }close to 1 for all 9 cases. Notably, the decomposed power spectral slope for each $p_v$ based on Eq.\ref{eq:ld2} (red) is almost the same as that of the constant density PPV channel (blue). The color bar scaling is adjusted such that the minimum scale corresponds to the figure's mean value (white) while the maximum scale corresponds to the mean value plus three standard deviation. (Right) A similar figure {with the same method applied to the} supersonic simulation "huge-0" ($M_s\sim 6.36$). }
\end{figure*}

From here we define the Normalized Covariance Coefficient (NCC, \citealt{reply19}) which is defined as :
\begin{equation}
  NCC(A,B) = \frac{\langle (A-\langle A\rangle)(B-\langle B\rangle)\rangle}{\sigma_A\sigma_B}
  \label{eq:NCC}
\end{equation}
{  where A and B are generic terms that refer to two distinct images.} Notice $NCC(A,B)\in[-1,1]$. The use of NCC would allow us to determine whether the two 2D maps are correlated. If $NCC=1$, then the two maps $A, B$ are statistically perfectly correlated. If $NCC=-1$, then the two maps are statistically anti-correlated. If $NCC=0$, then the two maps are statistically uncorrelated. For our purpose, we would like to see whether $NCC(n,p_v)$ would be close to 1. Since we are unable to obtain {  the true velocity caustics map} $n({\bf X},v)$ from observations, if we can obtain $NCC(n,p_v)\sim 1$ for most of the velocity values, then we have a reliable way in reconstructing $n({\bf X},v)$ from observations.

We are going to {apply VDA} with {simulations} (See Table\ref{tab:sim}) {to see if the channels after VDA do} indeed return the {correct caustics map}. We use a simulation "e5r3" that has $M_s\sim 0.61$. On the top left of Fig. \ref{fig:sim_illustration} we show the results of the NCC correlation when applied to the numerical data from "e5r3" using the simulations' intrinsic thermal broadening strength. {To quantify the effectiveness of VDA, we compute the NCC between the true caustics map $n$ and the VDA-decomposed caustics map $p_v$.} Our results demonstrate that the decomposition works well for channels in different velocity position $v$. We see that these nine randomly drawn velocity channels have their NCC values close to 1. This indicates that the decomposition algorithm (Eq.\ref{eq:ld2}) works very well numerically for channels in different velocity position. Moreover, we compute the power spectral slopes for both $p_v$ (red text in Fig. \ref{fig:sim_illustration}) and $n$ for each v (blue text in Fig. \ref{fig:sim_illustration}). We can see that the computed spectral slopes are very similar, with a deviation of less than $10\%$ is observed. We can conclude that the decomposed $p_v$ captures most of the statistical features of the velocity caustics for every velocity channel in the case of subsonic turbulence. 

{  The excellent performance of VDA suggests that the velocity caustics \footnote{  Notice that while the density term $p_d$ do still contain a small contribution of velocity, the $p_d$ term in subsonic turbulence is structurally similar to the intensity map. Yet we are aware that in supersonic turbulence the claim that $p_d$ contains purely density contribution does not hold, and thus we address in Appendix \ref{sc:SVDA} on how to tackle this problem.} can be obtained from observations, which is important for both the statistical and structural studies in PPV cube.} In particular, the agreement of spectral slopes between $n$ and $p_v$ is {  also} very important as the VCA (\citetalias{LP00}, \citealt{2009ApJ...693.1074C}) and VCS technique (\citealt{LP00,LP06, LP08, 2009ApJ...693.1074C,2015ApJ...810...33C}) can be applied much more easily in the case when we can extract the caustics. Also, with the caustics, we can perform the real velocity gradients according to the recipe of \cite{YL17a} (See also \citealt{LY18a}) as the formulation of VGT is based on the dominance of statistics of velocity fluctuations in velocity channels\footnote{The magnetic field tracing is also possible if density fluctuations are passively induced by velocity fluctuations. This is the case of the passive scalar approximation for the description of density statistics (see \citealt{2019tuma.book.....B}). In this situation, the Intensity Gradient tracing is feasible \citep{YL17b,IGVHRO}. However, the accuracy of the passive density approximation fails as shocks are formed in supersonic turbulence.} 

We can also apply the same method to supersonic simulations. On the right of Fig. \ref{fig:sim_illustration} we show the structure of velocity channels for the case when we have supersonic simulations "huge-0" which has an intrinsic $M_s \sim 6.36$. We can see that comparatively, the VDA performs not as good in the supersonic cases as in the subsonic cases, particularly at the center channel $v=-1.14 km/s$ where the spectral peak lies. While wing channels generally have a better NCC than the center channels, the spectral slopes of $p_v$ in the wings are generally steeper. {  The reason why VDA works for supersonic turbulence is rather simple: the $p_d$ term now does not collect as much information as it is in the subsonic counterpart (See, e.g. Appendix \ref{secap:kernel}) due to thermal broadening. Interestingly enough, when the VDA algorithm applies to supersonic simulations, we see from the right of Fig. \ref{fig:sim_illustration} that VDA can recover the caustics reasonably well in terms of NCC, in particularly, in wing channels. We have to emphasize that the drop of the NCC} would not be a problem for VDA as (1) highly supersonic simulations do not suffer from the thermal problem as indicated by the success of the VCA method in supersonic molecular clouds (See \citealt{2001ApJ...551L..53S,2006ApJ...653L.125P}, see also Appendix \ref{sc:SVDA}). (2) In multiphase media, the supersonic regions, likely to be cold neutral media, occupies only a few percent in terms of volume fraction (See Tab.\ref{tab:MP_factor}). By performing proper Gaussian decomposition along the line of sight, we can ignore those regions and focus on the underlying warm neutral media statistics. 

\begin{figure}[th]
  \centering
  \includegraphics[width=0.49\textwidth]{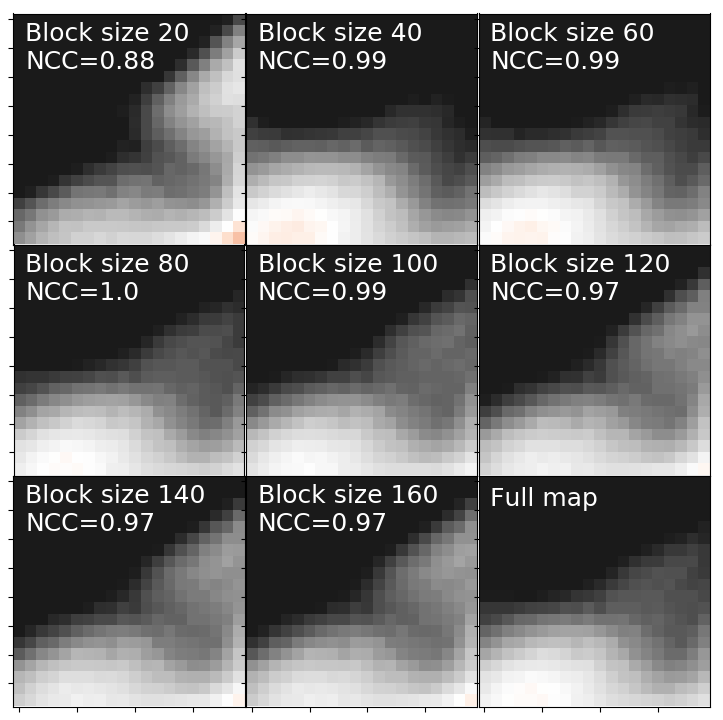}
  \caption{\label{fig:blocksize} {A set of figures showing how the different size of block before the application of Eq.\ref{eq:ld2}} would change the resultant structures and the NCC value (Eq.\ref{eq:NCC}).}
\end{figure}

One might wonder the sensitivity of the VDA algorithm to the size of the area that we are taking average with, {since the algorithm we proposed in Eq.\ref{eq:ld2} depends on the statistical area that is used for the VDA computation.}  Therefore, we perform a simple test to see what is the smallest area required for the VDA to work. To start with we first sub-sample the PPV cube in a size of $N\times N\times N_v$ where $N$ is the block size and $N_v$ is the number of channels along the line of sight. We prepared 8 samples that with $N=20,40,60,80,100,120,140,160$. Notice the original PPV cube has $N_0=1200$. {For each $N$ and also $N_0$, we extract the PPV cube $N\times N\times N_v$ and compute VDA on this smaller cube.} After VDA, we extract the same region and plot their structure as in Fig.\ref{fig:blocksize}. To compare with we also show the same area when we have $N_0=1200$, and compute the NCC (Eq.\ref{eq:NCC}) between the sample map and the original map. We can see visually from Fig.\ref{fig:blocksize} that only for the case of $N=20$ was the structure of the $p_v$ map behaving differently to the other maps. This is reflected by the drop of the NCC value in the top-left panel of Fig.\ref{fig:blocksize}. As for other panels, the NCC values are almost equal to 1, indicating the VDA technique is robust even when one samples a tiny area\footnote{  Another test that might influence the performance of VDA is to change the telescope beam size of the synthetic map. We report that the VDA method under the change of beam size has no change in terms of its performance even if we degrade the resolution for 60 times. }. 

\begin{figure}[th]
  \centering
  \includegraphics[width=0.49\textwidth,height=0.40\textwidth]{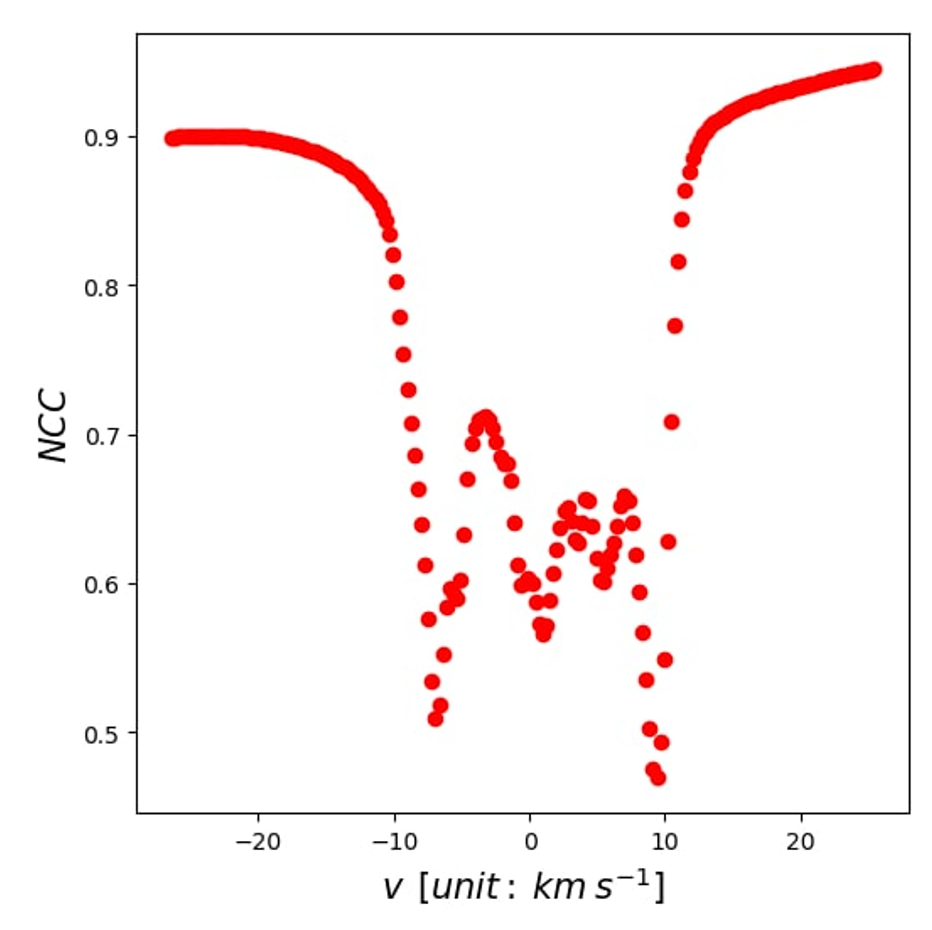}
  \caption{\label{fig:multiphase_NCC_v}  A figure showing the variation of $NCC(p_v,n)$ for the multiphase simulation {(See Tab\ref{tab:sim})} as a function of velocity $v$. }
\end{figure}

{Readers might also wonder how the same experiments we made above will behave for multiphase media. The multiphase media is a more complicated case since it involves both the subsonic warm phases and the supersonic cold phases. The former will perform precisely the same as the subsonic limit in this section, while the latter will be similar to this section's supersonic counterpart. We show the visual decomposition results in the middle column of Fig.\ref{fig:illus}. The NCC would therefore be a function of the relative fractions between the cold and warm parts. However, since we learned from Fig.\ref{fig:sim_illustration} that the NCC and both subsonic and supersonic cases are relatively high ($> 0.5$), we expect that the performance of VDA in multiphase media will perform reasonably well both in the center channel and also the wing channels. More precisely, we expect that the center channel, which is more CNM-crowded, has a lower NCC while the wing channel has a higher NCC. Indeed, the NCC across synthetic channels from multiphase media is relatively high, as we can see from Fig.\ref{fig:multiphase_NCC_v}, suggesting that our method also works for multiphase media.}

\subsection{Relative contribution of density and velocity {fluctuations}}
\label{subsec:pd2pv}
\subsubsection{Isothermal simulation test}
\label{sssec:isothermal}
With the decomposition algorithm, it is much easier to compare the relative contribution of density and velocities in each velocity channel in observations. \citetalias{LP00} study contains a description of how the spectra of density and velocity affect the spectrum of the velocity channels and specifically discusses the case of shallow density spectrum ({See \S \ref{sec:one_sigma}, Tab.\ref{tab:L09}})\footnote{It is well known (See, e.g., \citealt{2007ApJ...658..423K}) that the density spectrum becomes shallow in high Mach number systems. } It was noted that in the latter case, it is unavoidable to have both density and velocity contributions at small scales intensities in thin velocity channels, while velocity {fluctuations dominate over that of density} in the case of steep density spectrum (see Table\ref{tab:L09}). { Therefore, we would develop a systematic method in quantifying the relative contribution of density and velocity fluctuations in observation using Eq.\ref{eq:ld2}.}

\begin{figure*}[th]
  \centering
  \includegraphics[height=0.49\textwidth,width=0.98\textwidth]{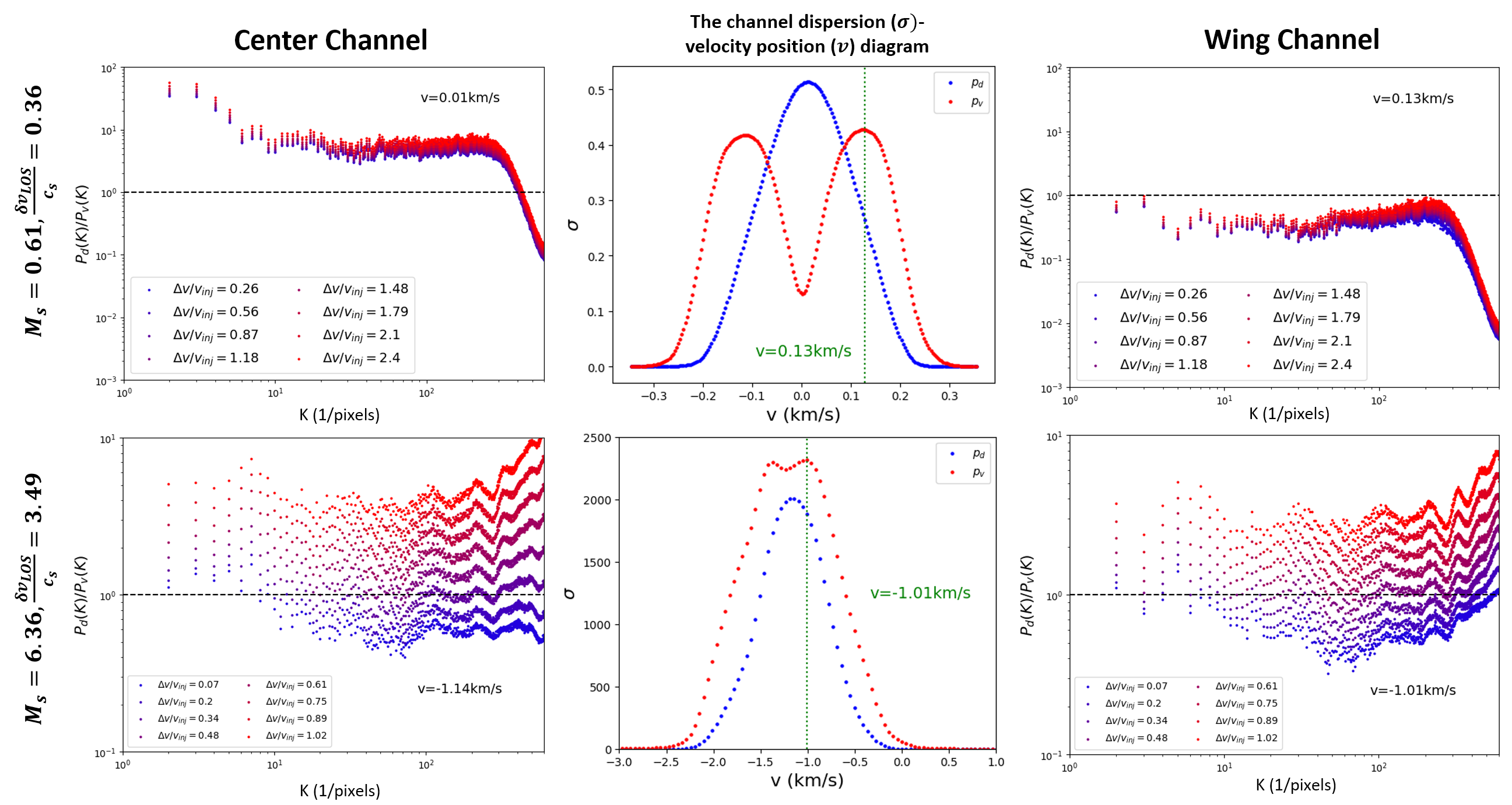}
  \caption{\label{fig:pd2pv_huge} A set of figures showing how the spectral ratio $P_d/P_v$ behave as a function of 2D wavenumber $K$ in our two simulations. We first plot the channel dispersion ($\sigma$)-velocity position ($v$) diagram (center column) for both density only channel $p_d$ and velocity only channel $p_v$, and we notice that the maximum dispersion of $p_v$ for both cases are not at the center of the spectral line. We then compute the $P_d/P_v$ for the center (left column) and wing channel (right column) with maximum $\sigma$ in the $\sigma$-v diagram for both subsonic (top row) and supersonic cases (lower row) as a function of the normalized channel width $\Delta v/\delta v_{inj}$. The horizontal dash line indicates that the relative contribution of density and velocity in the scale $K$ equals to each other.}
\end{figure*}

To start with, we prepare two sets of simulations that are supersonic and subsonic based on Table \ref{tab:sim}. With the simulations, we produce the thermal broadened velocity channels and proceed with the decomposition method as we suggested in \S \ref{sec:la}. We then compute the power spectra for density-only $p_d$ and velocity-only channel $p_v$ for different channel width $\Delta v$ and different $v$ and plot their ratio as a function of wavenumber $K\sim 1/R$ (see Fig.\ref{fig:pd2pv_huge}). Here we normalize the velocity channel width with the injection velocity $v_{inj}$ {  the r.m.s. velocity of injected turbulent motions.}. Notice that the channel width that we measure here should be {checked with Eq.\ref{eq:thin_criterion}}. We can see from Fig.\ref{fig:pd2pv_huge} that the features predicted by \citetalias{LP00} are observed in the case of thermal broadening. 

Before proceeding with the analysis of the power spectra ratio $P_d(K)/P_v(K)$ in Fig.\ref{fig:pd2pv_huge}, we would first like to see how the simple dispersion of $p_d$ and $p_v$ look like as a function of $v$ for both cases. In the center column of Fig. \ref{fig:pd2pv_huge} we plot the channel dispersion ($\sigma$)-velocity position ($v$) diagram for both density only channel $p_d$ and velocity only channel $p_v$. If $p_d$ and $p_v$ truly represents the fluctuations of density and velocity in velocity channels, the $\sigma-v$ diagram will then tell us whether a particular channel is velocity or density dominant. While it is expected that the dispersion of $p_d$ would be largest near the peak of the spectral line, it is surprising for us to see that there is a two-peak distribution for the dispersion of $p_v$ as a function of $v$. The maximum dispersion of $p_v$ for both subsonic and supersonic cases is not at the center of the spectral line. {Moreover,} the local minimum of the dispersion of $p_v$ is located at the center of the spectral line. More importantly, if we consider the ratio of dispersions between $p_d$ and $p_v$, this value will be straightly smaller than 1 in the case of supersonic media. For the subsonic case, the ratio of dispersion will only be smaller than one when we are inspecting the wing channels. 

Similar conclusions can be drawn when one computes the power spectra ratio $P_d(K)/P_v(K)$. In the case of high sonic Mach number (lower row of Fig.\ref{fig:pd2pv_huge}), we can see that in the case of thin velocity channels both at the center (lower left of Fig.\ref{fig:pd2pv_huge}) and the wings (lower right of Fig.\ref{fig:pd2pv_huge}) the relative contribution $P_d(K)/P_v(K)$ is generally smaller than one when the normalized channel width is smaller than $0.4$ in small scales ($k>30$). In the other case ($\Delta v/v_{inj}>0.4$) we can see that small scale relative contribution $P_d(K)/P_v(K)$ is larger than one. This experiment indicates that in the case of supersonic turbulence, we can achieve velocity dominance by (1) reduce the width of velocity channels (2) filter large scale contribution away to retain small scales that have velocity contributions dominate over that of densities. These effects correspond to the predictions in LP00 where the dominance of velocity fluctuations was predicted for thin velocity channels for asymptotically large $k$. This effect of decreasing the fluctuations of channel intensity arising from velocity fluctuations with the decrease of $\Delta v$ corresponds to the prediction from LP00, where it was shown that thermal velocity acts to increase the effective width of the channel maps. This effect naturally decreases the effects of velocity caustics induced by turbulence.

A more non-trivial case is the subsonic limit which we are showing the behavior of $P_d(K)/P_v(K)$ as a function of $K$ and $\Delta v/v_{inj}$ on the top row of Fig.\ref{fig:pd2pv_huge}. We can see that the relative ratio of the power spectra has a significantly different behavior compared to the supersonic case. The most important thing to notice is that, when we are looking at the center channel (top left of Fig.\ref{fig:pd2pv_huge}), the ratio $P_d(K)/P_v(K)$ stay larger than $1$ for almost all length scales for all possible choices of channel width that we selected. This agrees with the top central panel of Fig.\ref{fig:pd2pv_huge} that the velocity caustics fluctuations attain a minimum at the center channel. However, as we {move} from the center to the wings, we can see that velocity fluctuations start to {dominates} over density fluctuations. The top right of Fig.\ref{fig:pd2pv_huge} shows the spectral ratio $P_d(K)/P_v(K)$ at one of the local maxima in the $\sigma-v$ diagram. We can see that the ratio is straightly smaller than 1, indicating the velocity caustics, which is nicely traced by $p_v$ as indicated by Fig.\ref{fig:sim_illustration} by the NCC value, actually dominate over the density counterpart in the wing channel. As a side note, we do not observe any changes of the spectral ratio $P_d(K)/P_v(K)$ as a function of the channel width, opposed to \citetalias{LP00}'s point of view. We also report that the spectral ratio $P_d(K)/P_v(K)$ will decrease further if we move away from the spectral peaks.

{For techniques like VCA, VCS, VGT, or RHT, it is not necessary to discuss the relative contribution of density and velocity in $p_v$ in every channel in detail despite we now can trace back the caustics in every channel (See Eq.\ref{eq:ld2}). }However, some modifications have to be done to these techniques for them to be applicable in subsonic media (See \S \ref{sec:VCA},\ref{subsec:VGT}) as {the statistics of velocity caustics is far more complicated than what \citetalias{LP00} studied}. Also, in realistic turbulence systems, the injection scale is separated from the smallest observable scale much further than in numerical simulations. Therefore, we do not expect the first few points with small $K$ in both panels of Fig.\ref{fig:pd2pv_huge} will play an important role in discussing the dominance of either density or velocity fluctuations in velocity channels. 

\subsubsection{Multi-phase media test}
\label{sssec:multiphase}

The same analysis framework can be extended to cases in multiphase simulation. Here we use our multiphase simulation (see \S\ref{sec:sim} ) and perform the same analysis as in the previous section. For our current purpose, this simulation is statistically supersonic and contains three phases with the correct volume and mass fraction consistent with observation. Here we define the cold phase as $T<200K$ and the warm phase to be $T>5250K$.

\begin{figure*}[th]
  \centering
  \includegraphics[width=0.98\textwidth]{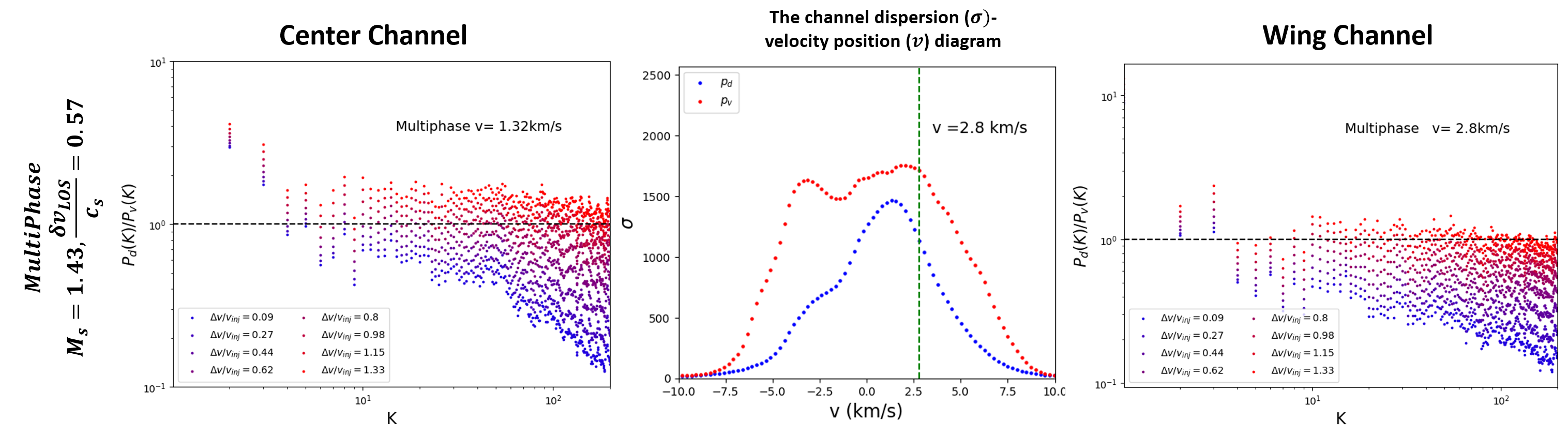}
  \caption{\label{fig:pd2pv_multiphase} A set of figures showing how the spectral ratio $P_d/P_v$ behave as a function of 2D wavenumber $K$ in the multiphase simulation similar to what we did in Fig.\ref{fig:pd2pv_huge}: We first plot the channel dispersion ($\sigma$)-velocity position ($v$) diagram (center column) for both density only channel $p_d$ and velocity only channel $p_v$, and we notice that the maximum dispersion of $p_v$ for both cases are not at the center of the spectral line. We then compute the $P_d/P_v$ for the center (left column) and wing channel (right column) with maximum $\sigma$ in the $\sigma$-v diagram.}
\end{figure*}

The result is shown in Fig.\ref{fig:pd2pv_multiphase}. We can see that the spectral ratio $P_d/P_v$ is sensitive to the channel width, indicating the supersonic nature of the channel map in multiple areas. Moreover, we observe that the spectral ratio is significantly below 1 (notice the y-axis of Fig.\ref{fig:pd2pv_multiphase} is in log-scale), {which} indicates that in the case of sub-Alfvenic multiphase media, it is possible to have velocity {fluctuations } to be dominant despite that the cold neutral media in this simulation {has a higher mass fraction (See Tab.\ref{tab:MP_factor})} and being highly supersonic.

We can further analyze the multiphase data by separating the phases using the two density threshold as defined {above}. Fig.\ref{fig:pd2pv_multiphase_XNM} shows the $\sigma-v$ diagram and the spectra ratio $P_d/P_v$ figures for both center channel and wing channel for the cold, warm and unstable phase gas.
\subsubsection{Similarities of the results of isothermal and multi-phase media cases}

\begin{figure*}[th]
  \centering
  \includegraphics[width=0.98\textwidth]{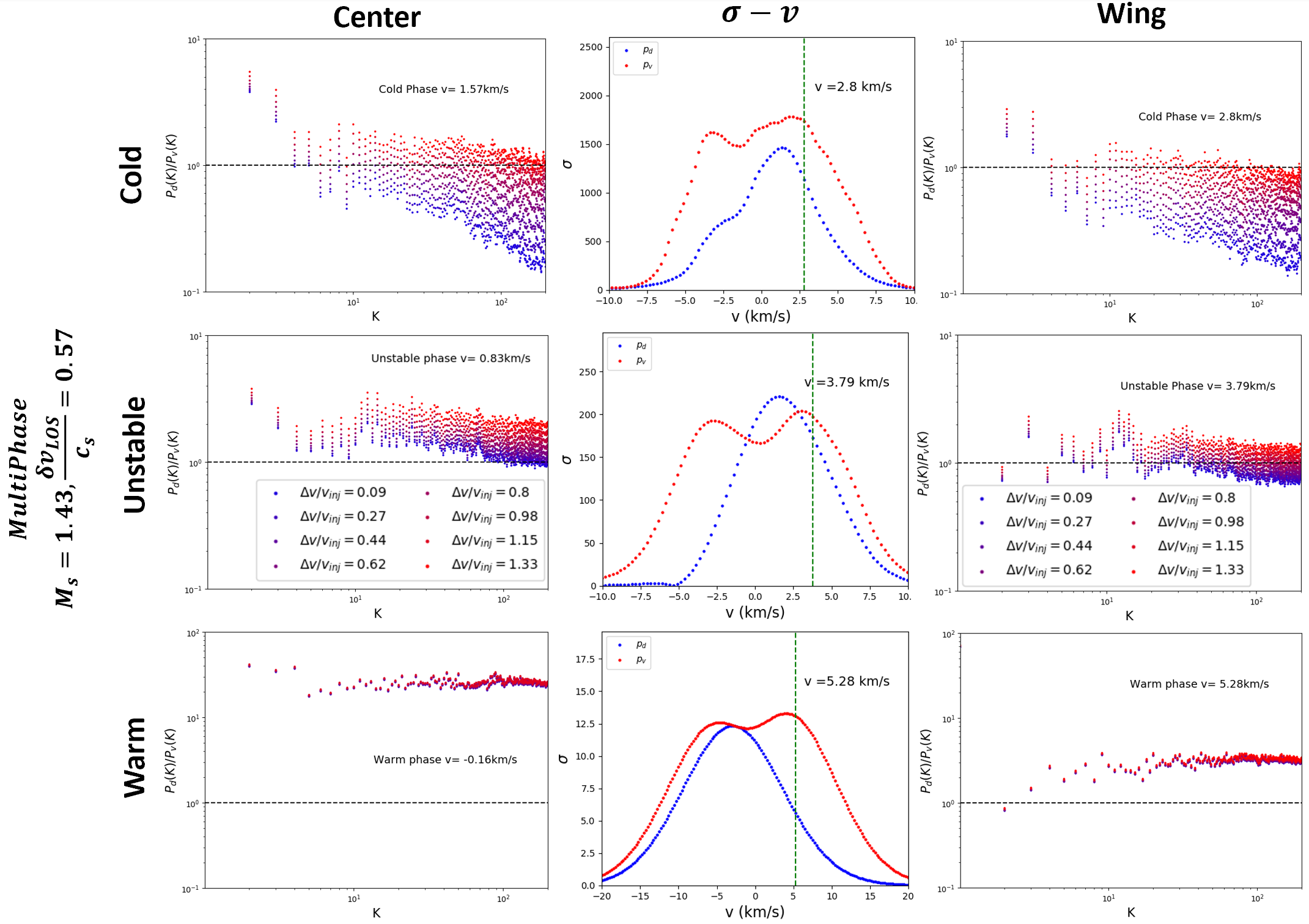}
  \caption{\label{fig:pd2pv_multiphase_XNM} A set of figures showing how the spectral ratio $P_d/P_v$ behave as a function of 2D wavenumber $K$ in the multiphase simulation {\it for each phase} similar to what we did in Fig.\ref{fig:pd2pv_huge} and Fig.\ref{fig:pd2pv_multiphase}. From the top: Cold, middle: unstable, bottom: warm neutral media. The choice of the channel widths are the same for all 6 cases (See the legend of the middle row).}
\end{figure*}

{
In \S \ref{sssec:isothermal} and \S \ref{sssec:multiphase} we see some consistent behavior on both the $\sigma-v$ diagram and the power spectra ratio $P_d/P_v$. We can see from Fig. \ref{fig:pd2pv_huge} and \ref{fig:pd2pv_multiphase_XNM} that (1) There is always a clear double peak seen for the $\sigma-v$ diagram for $p_v$, which is true for both isothermal and multiphase simulations. (2) The power spectra ratio $P_d/P_v$ is generally smaller than 1 when $\delta v/c_s>1$. This is true also for CNM since $c_{s,CNM}$ is usually very small. (3)The power spectra ratio $P_d/P_v$ varies much more for supersonic isothermal turbulence/CNM than the subsonic isothermal turbulence/WNM. For instance, one can see from top row of Fig.\ref{fig:pd2pv_huge} that the $P_d/P_v$ simply does not change as a function of $\Delta v$. That happens similarly for the WNM in the lower row of Fig.\ref{fig:pd2pv_huge}. Conversely the trend of $P_d/P_v$ as a function of $\Delta v$ is basically the same for the supersonic isothermal turbulence (lower row of Fig.\ref{fig:pd2pv_huge}) and both CMW and WNM (top and middle row of Fig. \ref{fig:pd2pv_multiphase_XNM}).} From here we see that the {phenomena (double peak, spectral ratio)} that occur in the isothermal case are also occurring in the multiphase limit. 

{The similarities between the statistical properties of velocity caustics in isothermal and multiphase media make the study of the statistical properties of velocity caustics in multiphase media much easier since we can reduce the multiphase simulations a superposition of isothermal simulations with different $M_s$. In particular, we can mimic the properties {of} velocity caustics by using supersonic isothermal simulations to be CNM while that of the subsonic one to be the WNM. 

These similarities can be utilized to analyze the dynamics of CNM and WNM in observations. Notice that cold neutral media could be formed within the collision flow of warm neutral media \citep{2009ApJ...704..161I}, i.e. {  any compression occurring in the WNM will lead to the formation of a CNM phase.}. It is rather natural to guess that the ambient WNM is moving the resultant CNM. If such a guess is correct, then the velocity caustics of CNM should also be embedded inside WNM, which can be easily checked from observation using multi-gaussian decomposition and VDA. Here we shall postpone }the full study of multiphase media {and its statistical analysis} elsewhere (Ho et. al in prep).

\subsection{Effects of noise to VDA}
\label{subsec:noise}
\begin{figure}[th]
  \centering
  \includegraphics[width=0.49\textwidth]{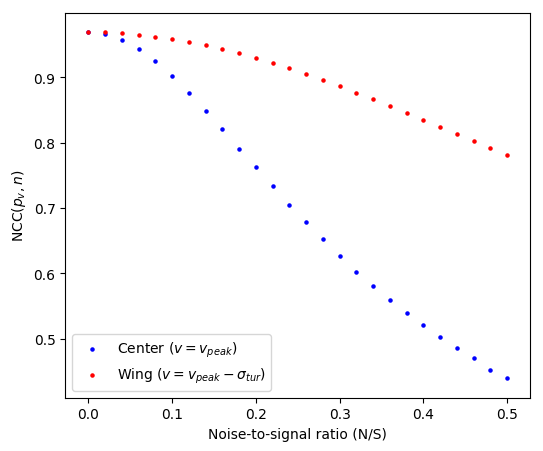}
  \caption{\label{fig:noise} A figure showing how the NCC behaves when we compare the realistic caustics $n$ to the VDA-decomposed $p_v$ from the center (blue) and wing channels (red) as a function of the noise-to-signal level (N/S) in the simulation e5r3. Here we select the wing channel position to be {$v=v_{peak}-0.5\Delta v_{effective}$ (See \S \ref{sec:one_sigma})}.}.
\end{figure}

The VDA is affected by the observational noise, and therefore we would like to see how accurate our prediction of the caustics using VDA would be when we add white noise to the channels. To start with we select two channels from the subsonic simulation "e5r3", one of which is the center channel $p(v=v_{peak})$ while the other is the wing channel {at} $0.5\Delta v_{effective}$ away from the peak position $p(v=v_{peak}-0.5\Delta v_{effective})$. We add white noise with certain amplitudes to these channels and then perform the decomposition algorithm based on the noise-added map with respect to the noise-added intensity map. We then compare the decomposed $p_v$ map from these two cases to the true velocity caustics map obtained by setting $\rho=\text{const}$ in Eq.\ref{eq:rho_PPV}. We use the NCC function (Eq.\ref{eq:NCC}) to characterize how similar the decomposed map $p_v$ and the true caustics map $n$ are as a function of the noise-to-signal ratio. Fig.\ref{fig:noise} shows the results for both center and wing channels. We can see that the $p_v$ from wing channels is less influenced by noise than center channels compared to true caustics. In particular, when the noise to signal ratio is $0.5$, the wing channel $p_v$ is still having an NCC value of $\sim 0.8$ while that of the center channel drops below $0.5$. Therefore, practically it is easier to extract the wing channel caustics in observations in thermally subdominant media.

\section{The standard deviations of velocity {caustics} as a function of velocity: The 1-$\sigma$ criterion}
\label{sec:one_sigma}

{
As we can see from the previous section (\S \ref{sec:test}, Fig.\ref{fig:pd2pv_huge}, Fig. \ref{fig:pd2pv_multiphase}, Fig.\ref{fig:pd2pv_multiphase_XNM}), the dispersion of velocity caustics as a function of $v$ {exhibits} a double peak shape but that does not happen for that of density fluctuations, which something that was not {expected} in the original theory in \citetalias{LP00} and this suggests that the maximal velocity fluctuations do not occur at the center of the spectral line, {but rather some velocity position away from the spectral peak.} The {locations of the double peaks are} not random and can be explained by the theory of \citeauthor{LP04} (\citeyear{LP04}, also D.Pogosyan, private communication). Here we will show where the peaks locate for GS95-type turbulence {using numerical simulations}, but we shall defer the rigorous theoretical treatment {that applies to general turbulence system} in our forthcoming paper. We shall analyze the unique phenomenon of velocity caustics numerically, but the formal analytical studies will be presented elsewhere.

In the following, we shall separate the discussion on two of the most important cases in observations: localized emission regions that correspond to molecular clouds (\S \ref{subsec:localized}), and regions with the line of sight differential velocity that corresponds to general HI emissions (\S \ref{subsec:shear}). In particular, we shall derive a relation called {\bf The 1-$\sigma$ criterion} that allows us to find the maximal velocity caustics fluctuations in observations in spectral line PPV cube.

\subsection{The 1-$\sigma$ criterion in localized emission regions}
\label{subsec:localized}
{
From our previous discussion, we see that {\it in terms of the behavior of velocity caustics}, the multiphase simulations can be treated as a linear combination of supersonic and subsonic isothermal simulations\footnote{{  Notice that for multiphase media there is a continuum of temperature variation even a clear two-phase separation happens. To realistically model the situation in terms of velocity caustics, there is a need to superpose multiple data with temperature T. }}. {  To simplify the analysis}, we can analyze the behavior of $\sigma-v$ curve {using} our isothermal simulations as a function of $v_{los}/c_s$. To test how $\sigma-v$ changes as a function of $v_{los}/c_s$, we change the value of $c_s$ in our simulation "e5r3" ($M_s=0.63$, See Tab.\ref{tab:sim}). Fig.\ref{fig:betatoppv} shows that peak difference $|v_{left peak}-v_{right peak}|$ is a clear function of $v_{los}/c_s$. The double peak feature of the $\sigma-v$ diagram is shifted outward as the thermal broadening strength increases (i.e. $v_{los}/c_s$ decreases). 

Based on the observations from Fig.\ref{fig:betatoppv}, we can measure how the distance of the two peaks $|v_{left peak}-v_{right peak}|$ in the $\sigma-v$ diagram for $p_v$ {varies} as a function of effective channel width $\Delta v_{effective}\sim\Delta v\sqrt{1+(c_s/\Delta v)^2}$ in Fig.\ref{fig:one_sigma}. We see that for sub-thermal cases there is a rather simple linear relation:
\begin{equation}
\begin{aligned}
|v_{left peak}-v_{right peak}| \approx (1.09\pm 0.02)\Delta v_{effective} \\+ (-0.013 \pm 0.002)
\end{aligned}
\label{eq:one-sigma}
\end{equation}
with $95\%$ confidence. Notice that the intercept is almost negligible compared to the choice of values we used for the $\Delta v_{eff}$. Also, one can estimate the relation of $|v_{left peak}-v_{right peak}|/\Delta v_{effective}$ using the theory of \citetalias{LP00}. Theoretically from \citeauthor{LP04} (\citeyear{LP04}, also D.Pogosyan, private communication) suggests that $|v_{left peak}-v_{right peak}|\sim 0.95$. {Since these special velocity locations (Eq.\ref{eq:one-sigma}) are exactly the locations where we have the maximal velocity caustics fluctuations, }we would then refer {Eq.\ref{eq:one-sigma}} to be the {\bf 1-$\sigma$ criterion}.}

As we emphasized earlier, we do not see a similar phenomenon in the {$\sigma-v$ diagram for $p_d$} as shown in previous examples (See Figs.\ref{fig:pd2pv_huge},\ref{fig:pd2pv_multiphase},\ref{fig:pd2pv_multiphase_XNM}). The dramatic differences for the $\sigma-v$ diagrams $p_v$ and $p_d$ {have very important consequences }in discussing the dominance of velocity/density contribution in velocity channels raised by the recent debate \cite{susan19,2019ApJ...886L..13P,kalberla2019,kalberla2020a,kalberla2020b}. Notice that we {\bf do not} consider only warm phase, but the two phases with cold phase that is moved together with the warm one in the galactic disk, at least. We consider in this paper the most challenging setting for the VCA situation (see \S\ref{subsec:VCA} for a further discussion). 

In the series of papers, it is claimed that in the density contribution {\it generally} dominates velocity in {\it every} velocity channels. However, as we can see in Fig.\ref{fig:betatoppv}, the velocity fluctuations are maximal from the center channel (i.e., $v=0$), especially when the thermal broadening strength is large. In the latter case, it is expected according to \citetalias{LP00} to have the suppression of velocity fluctuation at the center channels, but not at wing channels. Therefore, even in this extreme case, the density contributions are not dominant for every channel (see an example in Fig \ref{fig:RHT}). Naturally, in the case of multi-phase media, the thermal broadening effects present a more complex pattern compared to that in Fig.\ref{fig:betatoppv}. Nevertheless, {the double-peak pattern of the velocity contribution is a substantial effect that is important to consider when any judgment on the contributions to the PPV intensities is made. }

The current study combined with the previous work on thermal deconvolution \citep{GA} also sheds lights on applying the VCA techniques to velocity channels heavily dominated by density and affected by thermal effects. We shall discuss this possibility in \S \ref{sec:VCA}.

\begin{figure}[th]
  \centering
  \includegraphics[width=0.49\textwidth]{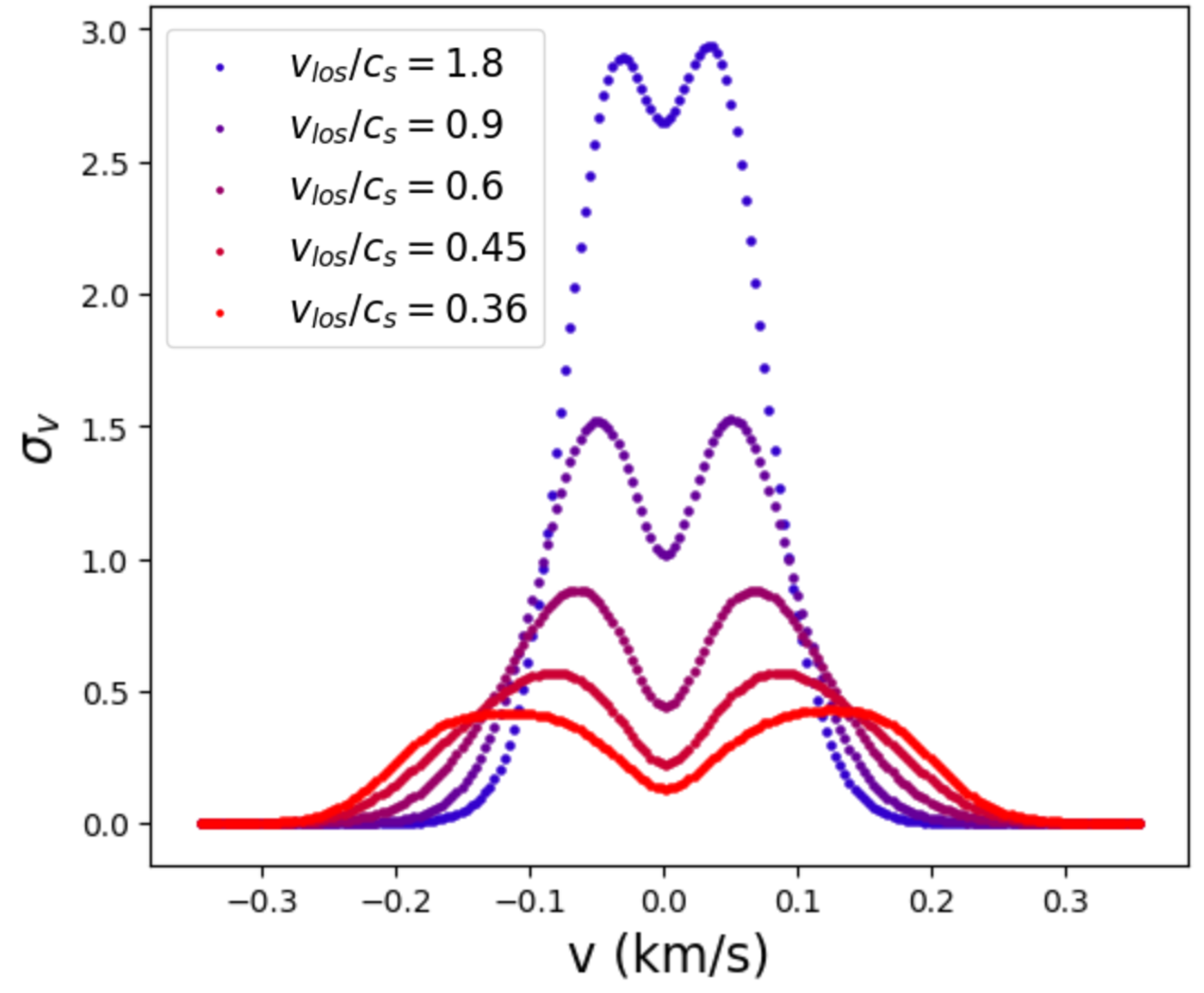}
  \caption{\label{fig:betatoppv}  A figure showing how the variation of velocity caustics dispersion $\sigma$ as a function of velocity position for a variety of choices of $v_{los}/c_s$ by varying $c_s$. }
\end{figure}

\begin{figure}[th]
  \centering
  \includegraphics[width=0.49\textwidth]{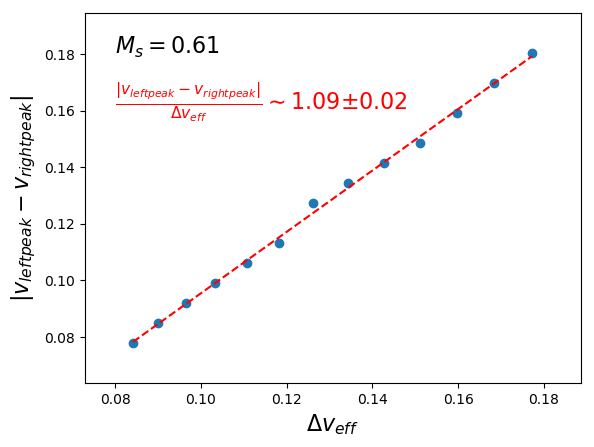}
  \caption{\label{fig:one_sigma}  A figure showing how the differences of the peak positions on the left of Fig.\ref{fig:betatoppv} varies as a function of $\Delta v_{effective}$ (Eq.\ref{eq:effective_dv}). We perform the least squared fit with $95\%$ confidence and obtain a slope of 1.09. }
\end{figure}

}
\subsection{The 1-$\sigma$ criterion in the presence of galactic rotation curve}
\label{subsec:shear}

{For observations of galactic disk neutral hydrogen data, the presence of galactic rotation cannot be neglected. It was shown in \citetalias{LP00} how the galactic rotation affects the velocity channel/caustics map. In terms of the present study, this will impact the double peak feature that we discussed above. In this subsection, we discuss this effect using synthetic observations.

Here we consider a linear shear along the line of sight. For a given 3D numerical data of line-of-sight depth $L$, we add a linear velocity field on top of the original z-component velocity field $v_z$ with {turbulent} dispersion $\delta v_{los}$ in the simulation:
\begin{equation}
  v_{new}(z) = v_z(z) + C\delta v_{los} \frac{z}{L}
  \label{eq:v_add}
\end{equation}
for some constant $C$. Here we shall write $v_{shear} = C\delta v_{los}$. We then synthesize the velocity channel map using $\rho({\bf X},v)=ppv(\rho,v_{new},c_s)$. The caustics map can be obtained using our recipe in \S \ref{sec:la}.

Fig.\ref{fig:betatoppv_shear} shows how the addition of the line of sight velocity shear will change the $\sigma-v$ diagram for velocity caustics. For the illustrative purpose, we keep only three curves that correspond to $C=0,1,2$ in Fig.\ref{fig:betatoppv_shear} to illustrate the effects: (1) The whole double-peak structure is shifted to the right according to the magnitude of $C$. (2) The peak on the right-hand side of the $\sigma-v$ curve is increased with respect to the value of $C$, but that change is minimal for the peak's left-hand side. (3) Most importantly, we do not notice any statistically {  significant} changes to the value of the velocity position differences $|v_{left peak}-v_{right peak}|$ as a function of $C$. 

\begin{figure}[th]
  \centering
  \includegraphics[width=0.49\textwidth]{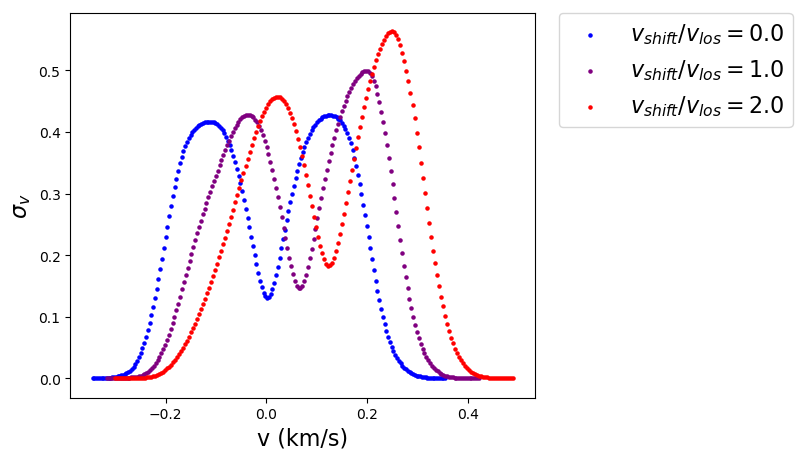}
  \caption{\label{fig:betatoppv_shear}  A figure showing the shape of the $\sigma-v$ diagram for $p_v$ when we add some linear velocity component on top of the turbulence velocity according to Eq.\ref{eq:v_add} in synthetic observations. }
\end{figure}

These results are of vital importance. First of all, in disk HI emission, both turbulent and shear velocity contributions are present. While the shear contributions are discussed in \citetalias{LP00}, the impact of the shear is not discussed in the context of the velocity caustics dispersion. We can see here that in the presence of velocity shear along the line of sight, the heights of the peaks are different. Therefore, it is expected to see unbalanced peaks in the $\sigma-v$ diagram for velocity caustics in HI observations. More importantly, by measuring the differences in the height of the peaks, we can measure the local velocity shear strength. Having the latter value for HI clouds would be advantageous for decoupling the turbulent and shear contributions in observations. 

Readers should notice that the notation of the wings and channels are ambiguous in the presence of galactic rotation. Moreover, in the presence of the galactic rotation curve, the introduction of galactic shear $v_{shear}$ is usually smaller than the intrinsic turbulent velocity value (i.e., $C<1$). Under this condition, the double peak feature would not be distorted very much. Very importantly, the VDA technique will still work regardless of the shape of the $1-\sigma$ double-peak feature. The VDA and the unique double-peak feature will allow us to determine the statistics of the velocity field in observations (See \S \ref{sec:VCA}).}

\section{Observational Application to HVC} 
\label{sec:observations}

{With our establishment from the previous sections, {  we apply our technique to observations.} We select a unique object, the High-Velocity Cloud HVC186+19-114, which is available in GALFA  \citep{stail} and carries a relatively simple phase structure: A cold core plus a warm envelope \citep{2006ApJ...653.1210S}. This simple structure allows us to easily discuss CNM and WNM behavior without worrying about much of the phase exchanges in general HI emissions. }

The observational data is obtained from the GALFA-DR2 \citep{stail} survey which deals with a wide range of observational data of neutral hydrogen 21-cm emission with full ranges of RA and $0^o-34^o$ for DEC. The pixelized resolution is $1'$ (FWHM $\sim 4'$), and the pixelized velocity channel resolution is $0.18km/s$, which is more than enough in resolving CNM with the temperature at the order of $1K$. In that extreme of temperature, the CNM is believed to be absorption-dominant. We would only expect the coldest neutral media observable to have their temperature at $\sim 100 K$, which corresponds to a {  sound speed} of $\sim 0.6 km/s$. The high velocity cloud HVC186+19-114 is located at RA = $108.6^o - 109.8^o$ and DEC $31.0^o-32.4^o$, spanning a total angular area of $1.2^o\times1.4^o$. 

\subsection{Visualizing the caustics from HVC}
\label{subsec:obsdecompo}

\begin{figure}[h]
  \centering
  \includegraphics[width=0.49\textwidth]{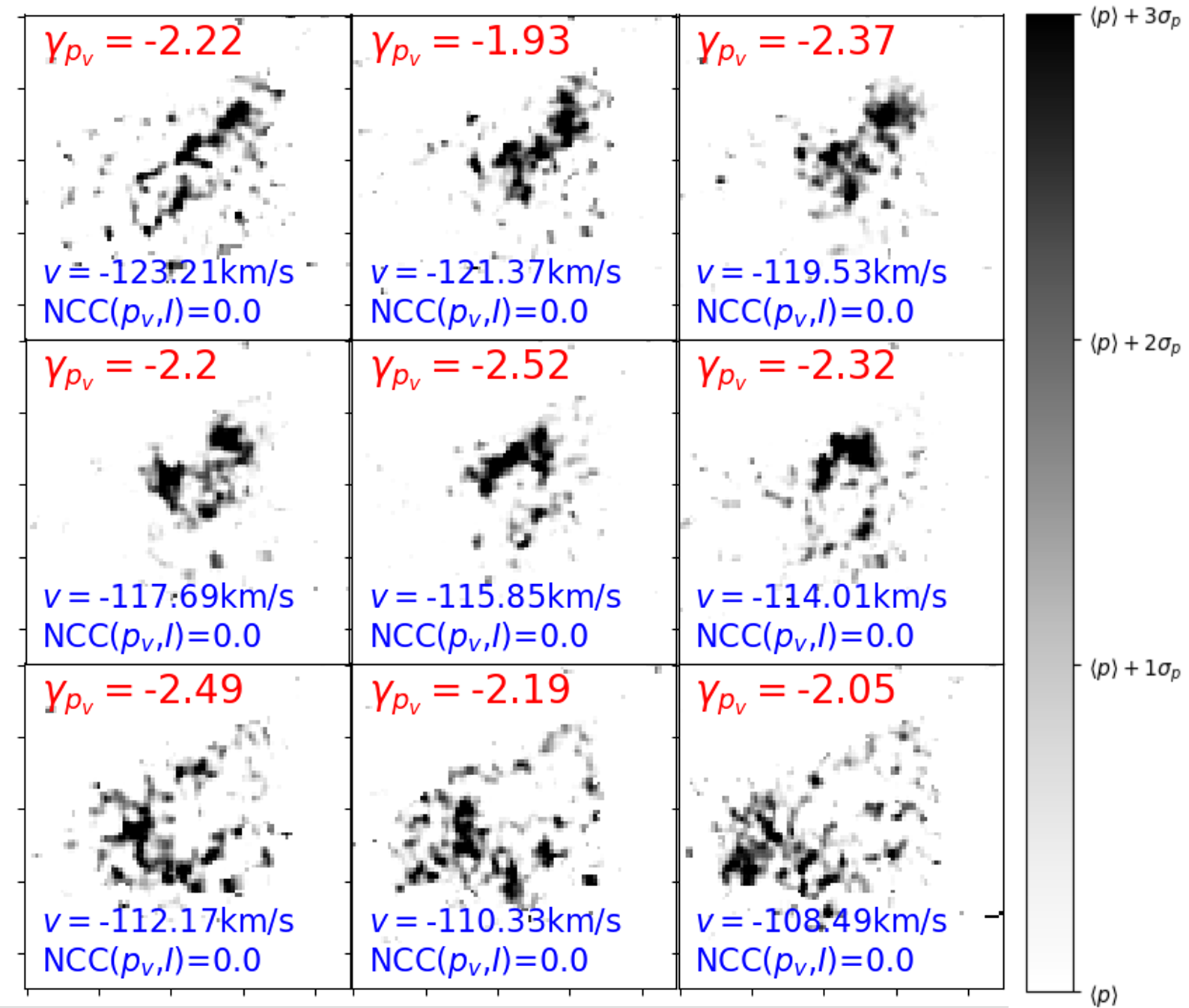}
  \caption{\label{fig:HVC_decomposition} The intensity fluctuations arising from velocity caustics from Eq.\ref{eq:ld2} on multiple velocity channel maps from HVC. The red text in each panel shows the power spectral slopes, while the NCC in blue is computed by $NCC(p_v,I)$.}
\end{figure}

To study the velocity structure of this high-velocity cloud, we follow \S \ref{sec:la} and apply the VDA to the observational data. This allows us to explore how the properties of velocity caustics arising from the high-velocity cloud. Fig.\ref{fig:HVC_decomposition} shows the velocity caustics fluctuations we extracted following \S \ref{sec:la} with a velocity channel width of $\Delta v=1km/s$, which extends from $v=-125 km/s$ to $v=-108km/s$. We include the spectral slopes of each caustics channel $p_v$ {  in} red in Fig.\ref{fig:HVC_decomposition}, and we write $NCC(p_v,I)$ {  in} blue in Fig.\ref{fig:HVC_decomposition}. We can see that the velocity caustics are indeed uncorrelated to the column density map. Moreover, the velocity caustics fluctuations can be spatially different from the channel intensity structure (see \citealt{2006ApJ...653.1210S} ). The rich velocity caustics information that is decomposed by the algorithm {in \S \ref{sec:la}} provides an additional tool for observers to determine the turbulence properties in neutral hydrogen structures (see \S \ref{sec:discussions}). { \cite{2006ApJ...653.1210S} suggests the maximum temperature of CNM is about 350-1000K. If taking the lower limit, that suggests that the CNM in HVC might be supersonic. }

\subsection{Analysis under VDA}
\label{subsec:multigaussian}

\begin{figure}[h]
  \centering
  \includegraphics[width=0.49\textwidth]{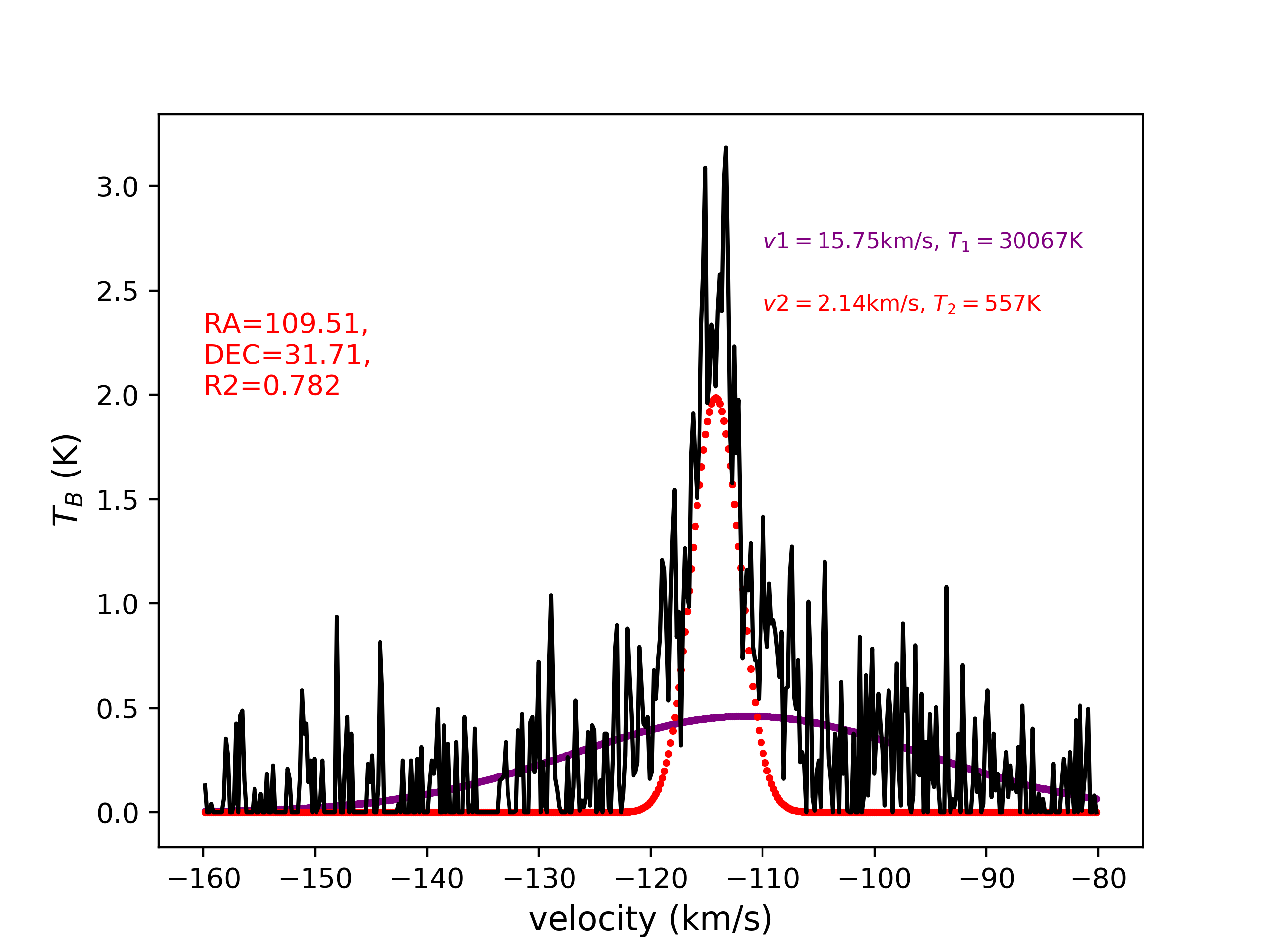}
  \caption{\label{fig:HVC_singlechannel} {One of the example of extremely narrow velocity profiles found }in HVC 186+19-114 from GALFA-DR2, which we are picking the same RA-DEC as in \cite{2006ApJ...653.1210S} (in units of degrees). We use the same two Gaussian profiles to fit the spectral line and we see that the narrower one has a $\sigma=2.14km/s$, while the wider one has a $\sigma=15.75km/s$. The fitting coefficient of determination $R^2$ here is 0.782. Notice that the CNM here might be supersonic.
 }
\end{figure}

\begin{figure*}[th]
  \centering
  \includegraphics[width=0.98\textwidth]{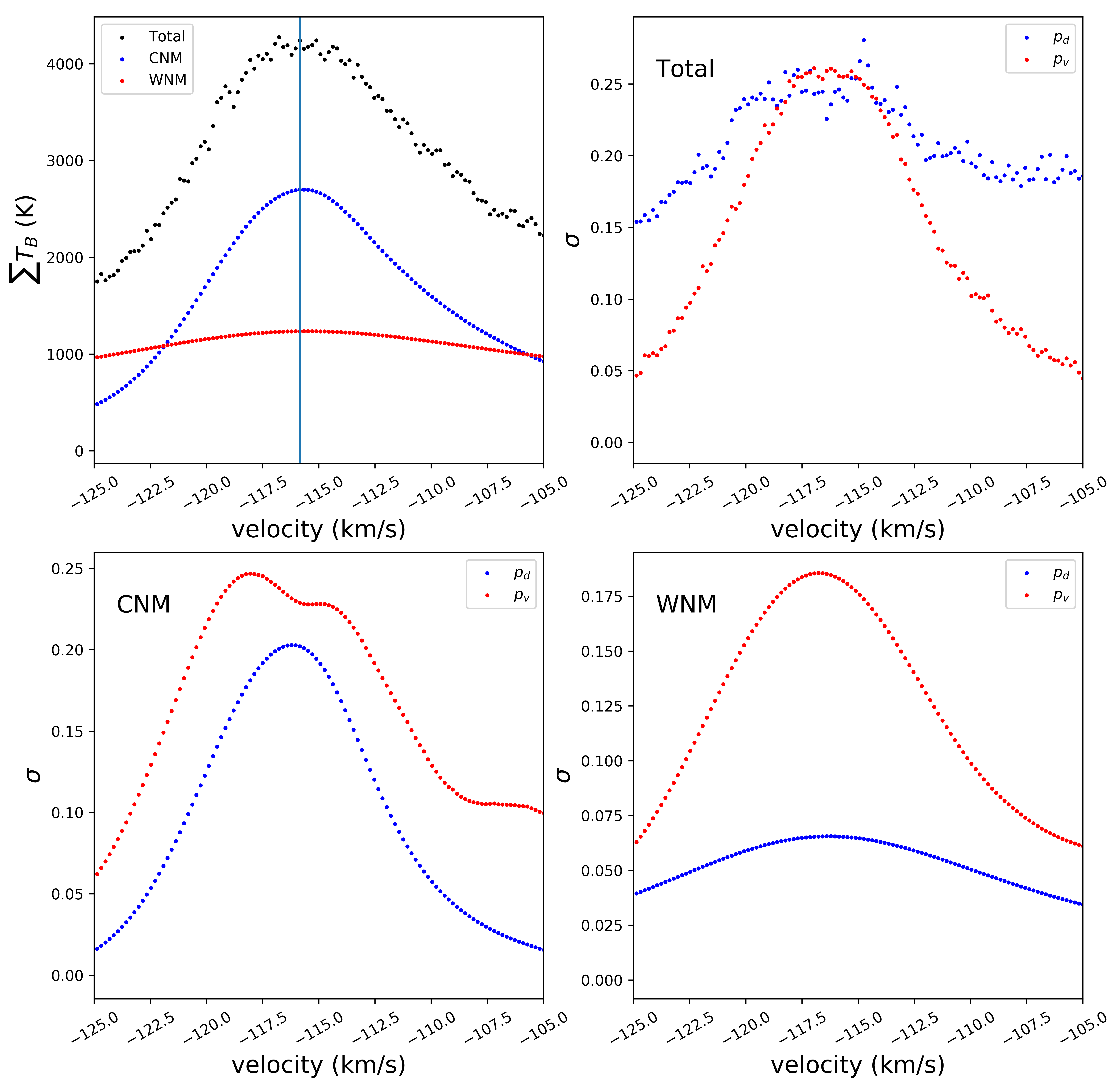}
  \caption{\label{fig:HVC_Gaussian} (Top left) A figure showing the multi-Gaussian decomposition of the {\it total} spectral lines of HVC, {which defines the CNM and WNM in the study of HVC in this paper}. We also compute the standard deviation of each velocity channel as a function of velocity for the total PPV (top left), CNM (lower left) and WNM (lower right) channel by the decomposition method we proposed in Eq.\ref{eq:ld2}. We remove the two scales $k=1,2$ during the calculation of $\sigma-v$ diagram (See Fig.\ref{fig:pd2pv_HVC}).}
\end{figure*}

We would perform the multi-Gaussian decomposition \citep{2007A&A...466..555H} that allows one to separate the cold and warm part of neutral media in GALFA data in the area of interest. Fig. \ref{fig:HVC_singlechannel} shows an illustration of the two-Gaussian fitting of a single velocity spectral line. We can see that the fitting algorithm can recognize two components. This feature is generally true for the spectral lines within the area of interest. Therefore, we {  can group} the fitting Gaussians that have narrower width to be the cold neutral media while those with wider width to be the warm neutral media in our study, i.e. if the velocity channel $p(x,y,v)$ could be fitted with two Gaussian profiles similar to \cite{2006ApJ...653.1210S}. {  The choice of Gaussian over a more theoretically found Voigt profile is that: First of all the majority of the profile decomposition algorithms (e.g. \citealt{2007A&A...466..555H}) are based on Gaussian models. Second of all, our numerical testing shows that the composite Voigt profiles fits the profile in similar goodness compared to that of Gaussian. At last, there is no profile dependencies of the VDA method, as long as the thermal broadening effect is active and the three conditions are \S \ref{sec:la} are fulfilled.}
The resultant spectral line for the whole region is shown in the top-left of Fig.\ref{fig:HVC_Gaussian}. We can see that the cold neutral media contribute {approximately twice} the emission strength compared to the warm neutral media at the center of the spectral line. 

Since this HVC contains a relatively simple {physical} structure, it would be interesting to see how the CNM and WNM velocity caustics behave statistically. We first plot the $\sigma-v$ diagram that corresponds to the PPV cube of HVC in the top-left of Fig. \ref{fig:HVC_Gaussian} as a function of the line of sight velocity.. After we distinguish the CNM and WNM using Gaussian decomposition, we plot the $\sigma-v$ diagrams for CNM and WNM in the lower left and lower right of Fig. \ref{fig:HVC_Gaussian}. {Here, we remove the $k=1,2$ contributions from the decomposed $p_d$, $p_v$ map as they usually do not give insight about the relative dominance of density and velocity fluctuations. Moreover, we zoom into the {velocity} ranges where we see the whole CNM profile.} We can see that, while in the case of total PPV (top left of Fig.\ref{fig:HVC_Gaussian}) the density contribution is {comparable} that of the velocity contribution in the core part of the spectral line, the latter is {not} negligible. In particular, in the case of the WNM (lower right of Fig.\ref{fig:HVC_Gaussian}) the velocity caustics fluctuations {\bf totally} dominates over the density fluctuations. Even for the case of CNM (lower left of Fig.\ref{fig:HVC_Gaussian}), the velocity fluctuations for a large part of the spectral lines is significantly higher than that of the densities, e.g., at $v\sim-110 km/s$. This shows that velocity caustics can be the dominant fluctuations of intensity observed in velocity channels, e.g., in the wing part of the velocity spectral line or warm neutral media. Even in the case of CNM, the velocity fluctuations are not negligible. More importantly, we see the predicted double-peak feature as we predicted from \S\ref{sec:one_sigma}. Notice that the skewness of the $\sigma-v$ diagram suggests that the HVC should have a non-zero line of sight bulk velocity component.

\begin{figure*}[th]
  \centering
  \includegraphics[height=0.40\textwidth,width=0.98\textwidth]{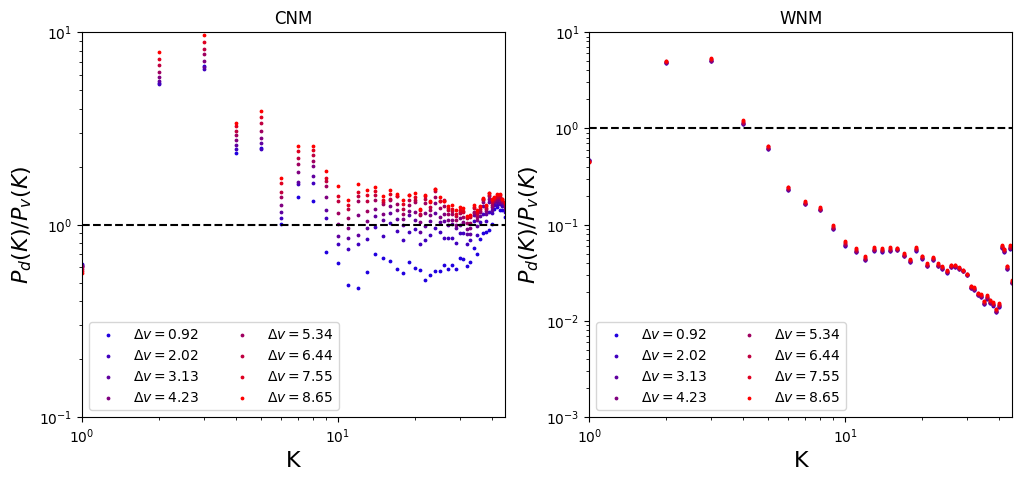}
  \caption{\label{fig:pd2pv_HVC} Two figures showing how the ratio of the power spectra $P_d(K)/P_v(K)$ at the spectral peak as a function of planar inverse length $K$ for the CNM (left) and WNM part (right) of the HVC cloud in different channel width (in km/s) {with the center velocity $v=115.85km/s$. Channel width unit: km/s} .}
\end{figure*}

\begin{figure}[ht]
  \centering
  \includegraphics[width=0.48\textwidth]{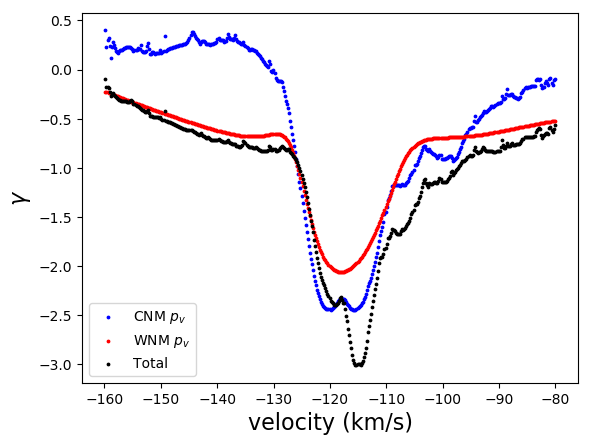}
  \caption{\label{fig:HVC_Spectralslope} A figure showing the spectral slope variation as a function of channel position for both CNM caustics (blue) and WNM caustics (red). We can see that the velocity caustics contribute to the spectral slope of the observed velocity channels {\it without decomposition} (black).}
\end{figure}

{
{  We can} address our findings from the perspective of the theory of intensity fluctuation in PPV space in \citep{LP00,LP04}. First of all, Fig. \ref{fig:HVC_Gaussian} tells us that the turbulent motions associated with the CNM are velocity dominant and the CNM emission dominates the intensity of emission in the velocity range $\approx 123 $ to $110$ km/s. This means that for the corresponding velocity channels, the traditional VCA should be applicable. This also invalidates the arguments in \cite{susan19} that the VCA cannot apply to the multi-phase HI due to strong thermal broadening. We see that this type of problem can occur only beyond the velocity range as mentioned before. However, we should mention that the spatial resolution of the HVC data is relatively {  poor}, that it is likely that the spectral slope estimation will have a large error.}

We can further analyze the relative contribution of the density and velocity fluctuations in velocity channels using the $P_d(K)/P_v(K)$ parameter discussed \S \ref{sec:la}. Fig.\ref{fig:pd2pv_HVC} shows $P_d(K)/P_v(K)$ for both CNM and WNM with variations of the channel widths. We can see very clearly from both Fig.\ref{fig:pd2pv_HVC} that the small scale structures of both CNM and WNM have $P_d(K)/P_v(K)<1$, meaning that velocity fluctuations are dominant at small scales. In particular, for CNM, we see an average $P_d(K)/P_v(K)\approx 1/2$ in small scale., While for WNM, this ratio can extend to the order of $10^{-2}$. This shows a clear example that, even CNM exists extensively in velocity channels {in terms of mass fraction,} the caustics fluctuations associated with CNM are still stronger than its density fluctuation. {The variation of the power spectra ratio $P_d(K)/P_v(K)$ for HVC's CNM is consistent with that of our simulation result (Fig.\ref{fig:pd2pv_multiphase_XNM}) that (1) the relative dominance of density and velocity is indeed a function of $\Delta v$ (2) in small $\Delta v$ velocity fluctuations dominates over the density fluctuations. For HVC's WNM we {observe that} its power spectra ratio $P_d(K)/P_v(K)$ is smaller than 1 {at} almost all scale. This can be seen usually in channels that are very far away from the spectral peak (See Fig.\ref{fig:pd2pv_huge},\ref{fig:pd2pv_multiphase}), or the density fluctuations are simply too small in contributing the $P_d(K)/P_v(K)$, i.e. the incompressible limit. Nevertheless, for HVC we see an observational example that both CNM and WNM are velocity dominant in small scales at the center velocity channel in the thin channel limit. }{We can now have a detailed caustics map for each channel of a physical object that the prediction of the PPV statistical theory \citepalias{LP00} can be applied readily without worrying the density collusion in channel maps. }

We can further examine the changes of spectral slopes as a function of the line of sight velocity for the decomposed channels, which is shown in Fig.\ref{fig:HVC_Spectralslope}, we can see that the total velocity channel power spectral slope (black line of Fig.\ref{fig:HVC_Spectralslope}) shares the core part of that of the CNM $p_v$ (blue of Fig.\ref{fig:HVC_Spectralslope}) and the wing part of WNM $p_v$ (red line of Fig.\ref{fig:HVC_Spectralslope}). This shows that velocity caustics do indeed dominate the intensity fluctuation in thin velocity channels.

\section{Synergy of VCA and VDA}
\label{sec:VCA}
With the availability of velocity caustics from VDA, it is natural to ask what {  are the} implications for the techniques that depend on the theoretical formulation of \citetalias{LP00}. In particular, several questions can be asked (1) With the knowledge that the velocity caustics are dominant on wing channels, {  should we give less importance to the centers of the velocity channel?} (2) How to {  use VCA} in heavily thermally broadened channels?

\subsection{The {  necessary} changes to the VCA method}
\label{subsec:VCA}
The VCA method relies on the difference of the spectral slopes in thin and thick channels to predict the 3D velocity spectral index (See \citealt{2009fohl.book..357L}, Tab.\ref{tab:L09}). The success of applying VCA in observation in {  the} SMC (\citealt{2001ApJ...551L..53S}\footnote{We also show numerically with the high-resolution data that the SMC has $P_d/P_v\sim 1/5$ for most K using the newest observational data, Yuen et al. in prep}) and Perseus \citep{2006ApJ...653L.125P} do make VCA seem like a universal method in obtaining the 3D velocity spectral index. However, there is no change of spectral index between the thin and thick velocity channel in the case of strong thermal broadening. Hence the VCA method is not applicable in subsonic media. {  The new developments presented in this paper }allows one to extract the caustics from subsonic media, which allows one to compare the spectral index between the thin channel caustics and the intensity map to obtain the 3D velocity spectral index. {  Below} we show two examples on how {  we can} estimate the velocity spectral index for a case with the spectral ratio $P_d/P_v<1$ and $P_d/P_v>1$.

First, we have to summarize what changes we need for the VCA method to work in subsonic media. First of all, the caustics statistics should be used in substituting the velocity channel in computing the thin channel spectral slope (Table \ref{tab:L09}). The value of $m$ as required by the VCA method would be the difference between the spectral index between the thin channel caustics and the thick velocity channel. It is evident that we only have the contribution from velocity fluctuations in the case of constant density. Using the VCA method's formula becomes straightforward without any worry about the spectral distortions {  arising} from the density fluctuations. One can also compare the results of the 3D velocity spectral index with that of the constant density velocity centroid (See Appendix \S \ref{secap:centroid}) as the two numbers should be {the same in the} ideal case. Second, since the peak fluctuations of caustics are not at the center of the spectral line, but $\sim \pm 0.5\Delta v_{effective}$ away from it {since the peaks are symmetric about the center of the spectral line} (\S \ref{sec:one_sigma}), all channels should be computed with the center channel put at $|v-v_{peak}|\sim \pm 0.5\Delta v_{effective}$ {where $v_{peak}$ is the spectral peak.} In other words, we would consider the spectral index of the velocity channel, for example, take $v=0.5\Delta v_{effective}$ :
\begin{equation}
\begin{aligned}
  &p({\bf X},v_0\sim 0.5\Delta v_{effective},\Delta v)\\
  \propto &\int_{0.5\Delta v_{effective}-\Delta v/2}^{0.5\Delta v_{effective}+\Delta v/2} \rho(v) W(v)\exp\left(-\frac{(v-v_{central})^2}{2c_s^2}\right)
\end{aligned}
\end{equation}
as a function of velocity channel width $\Delta v$. {In simple words, we are going to consider the differences of the spectral slopes between the column density map and the velocity caustics from thin wing channels at $|v-v_{peak}|\sim \pm 0.5\Delta v_{effective}$ for the application of VCA.}

Below, we show two GALFA examples from \cite{reply19}, one of which we knew that VCA works in its original form, while the other example was identified as a heavily thermally broadened map. We shall illustrate how the above new procedure combined with VDA works for these channels.

\subsection{Applying VDA to GALFA data}
\subsubsection*{Subdominant thermal broadening}
\begin{figure*}[th]
  \centering
  \includegraphics[width=0.98\textwidth]{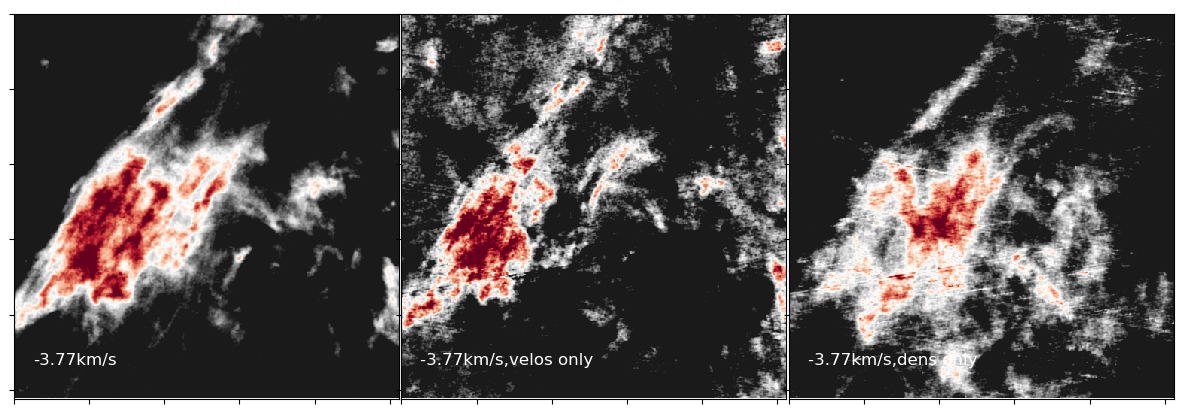}
  \includegraphics[width=0.98\textwidth]{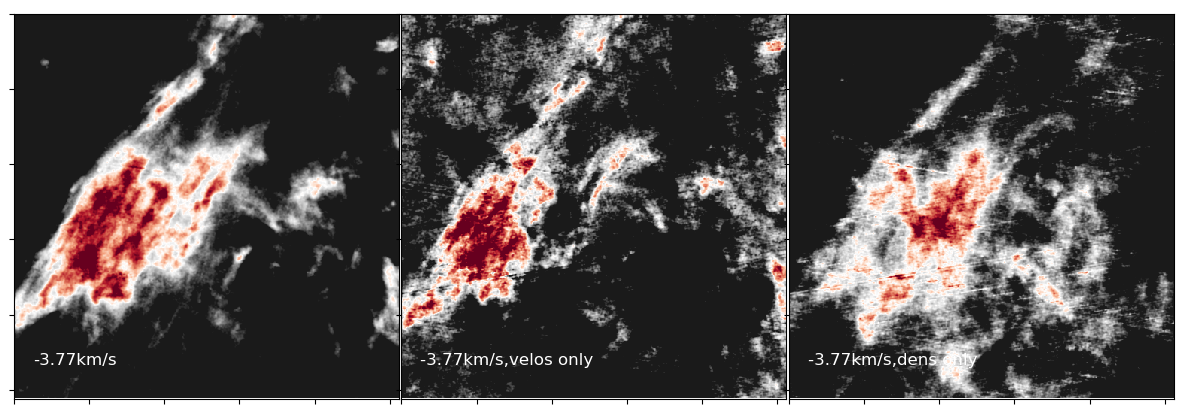}
  \caption{\label{fig:GALFA00400_1035} Two set of figures showing the structure of {raw and decomposed }velocity channels at two selected velocities $-3.77 km/s$ (upper row) and $-9.29 km/s$ (lower row) {for the GALFA data centered at RA=4.00 and DEC=10.35}. {  The latter is one of the positions fulfilling the $1-\sigma$ condition for this piece of data.} From the left: Total intensities of velocity channel; Middle: $p_v$, Right: $p_d$. {  The figures' colorbars are set to be $[\langle p\rangle,\langle p\rangle+3\sigma_p]$.  } }
\end{figure*}

We shall first use the GALFA data that we successfully applied VCA in \cite{reply19}. The region centered at $RA=4^o$ and $DEC=10.35^o$, spanned for an $8^o \times 8^o$ region, is a representative region that has some VCA statistics as discussed in \cite{reply19}. We would first like to see how VDA works for the channels. Fig. \ref{fig:GALFA00400_1035} shows the channels' structures and their decomposition under VDA for two selected velocities.  The two velocities are selected since they are the velocities that produce peaks of the $\sigma_{p_d}$ and $\sigma_{p_v}$ respectively. {  In particular, the velocity position $-9.29 km/s$ fulfills the $1-\sigma$ criterion.}

\begin{figure*}[th]
  \centering
  \includegraphics[width=0.49\textwidth]{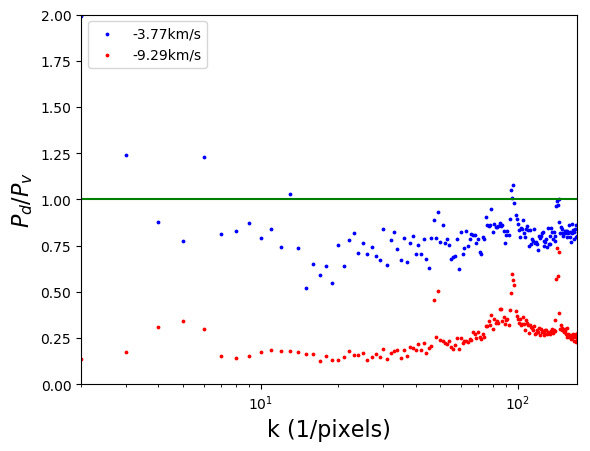}
  \includegraphics[width=0.49\textwidth]{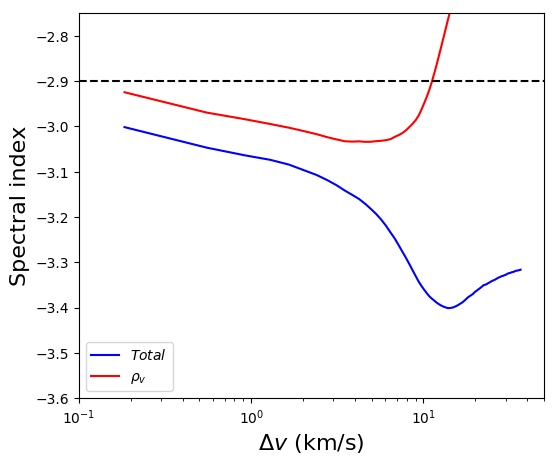}
  \caption{\label{fig:GALFA00400_1035_VCA} (Left) A figure showing the variation of $P_d/P_v$ as a function of $k$ for both selected velocities {for the GALFA data centered at $RA=4^o$ and $DEC=10.35^o$}. (Right) A figure showing the variation of $\gamma$ as a function of $\Delta v$ when we set $v_0=-9.29 km/s$.}
\end{figure*}

We can quantify the results using the $P_d/P_v$ diagram  for easier visualization. On the left of Fig \ref{fig:GALFA00400_1035_VCA} we see the variation of $P_d/P_v$ as a function of $k$ for both velocities. We see that for the case of $v=-3.77 km/s$ the $P_d/P_v$ factor {  does not deviated much} from unity, which signifies that the density and velocity contribution are in "equi-partition" in this selected velocity channel. However, for the case of $v=-9.29 km/s$ we see that the $P_d/P_v$ factor is mostly smaller than 1. Notice that $v=-9.29 km/s$ is the location where we have $\sigma_{p_v}$ attains its maximum, Therefore, according to \S \ref{subsec:VCA} we should use it as the center channel in computing VCA. On the right of Fig \ref{fig:GALFA00400_1035_VCA} we see the variations of the spectral indices for both $p_v$ and $p=p_d+p_v$ as a function of $\Delta v$ by putting $v_0=-9.29 km/s$. We estimate $m \sim 3.4-2.9 \sim 0.5$ for this region, indicating that the 3D velocity spectral index to be $E(k) \sim k^{-3.5}$.

\subsubsection*{Regions with dominant thermal broadening}
\label{sssec:d}

\begin{figure*}[th]
  \centering
  \includegraphics[width=0.98\textwidth]{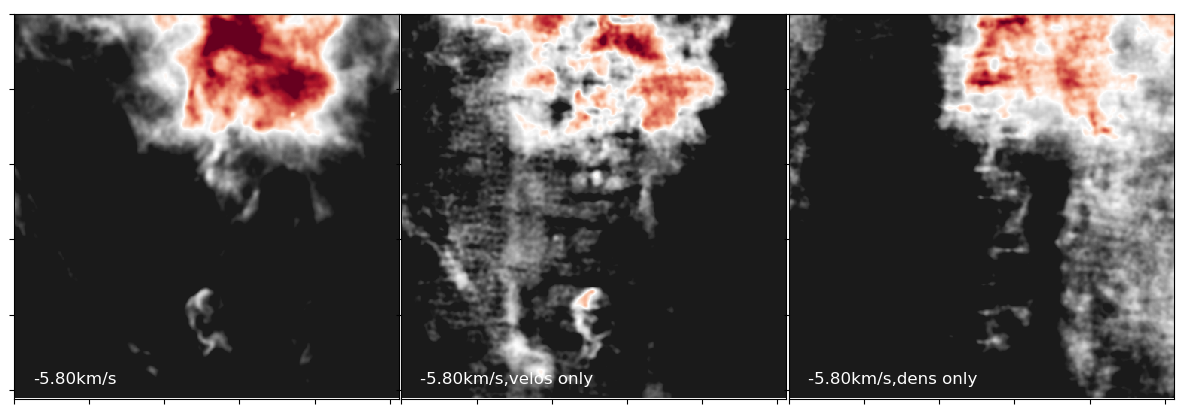}
  \includegraphics[width=0.49\textwidth]{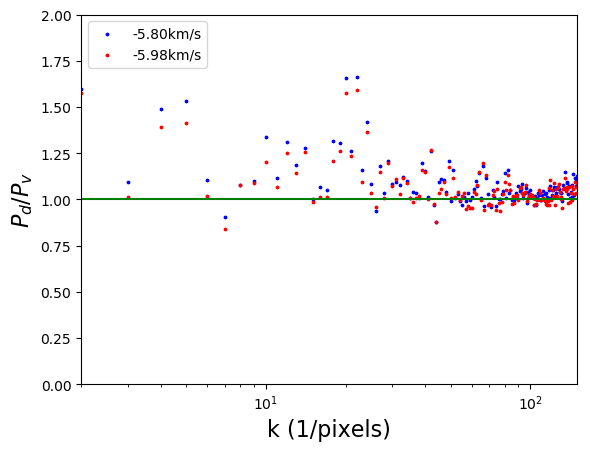}
  \includegraphics[width=0.49\textwidth]{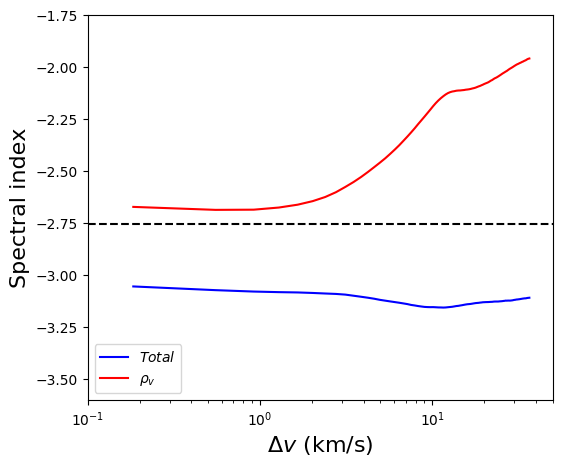}
  \caption{\label{fig:GALFA22800_1835} (Top) A set of figures showing the structure of the channel and its VDA decomposition {for the GALFA data centered at $RA=228.00^o$ and $DEC=18.35^o$}. From the left: Total intensities of velocity channel; Middle: $p_v$, Right: $p_d$. (Bottom left) A figure showing the variation of $P_d/P_v$ as a function of $k$ for two selected velocities. (Bottom Right) A figure showing the variation of $\gamma$ as a function of $\Delta v$. }
\end{figure*}

{  It is interesting to see how we could apply VCA to regions that are suffered from strong thermal broadening.} Note that those regions corresponded to N/A entry in \cite{reply19}. For these regions, the change of the channel width does not change the spectral slope.  The region centered at $RA=228^o$ and $DEC=18.35^o$, spanned for an $8^o \times 8^o$ region, is a good candidate since we can see visually from the top of Fig.\ref{fig:GALFA22800_1835} that the $p_v$ map does not look to be very different from that of the $p_d$ map, which is a bad sign of equipartition of velocity and density contribution in this particular velocity channel. We can see from $P_d/P_v$ curve in the bottom left of Fig.\ref{fig:GALFA22800_1835} that it is unity for most of the scales. As for the VCA method, while we see that there is no variation of the spectral slope as a function of $\Delta v$ for the total velocity channel (blue curve in the bottom right of Fig.\ref{fig:GALFA22800_1835}), there is a significant variation for that of $p_v$\footnote{Notice that according to our discussion in \S \ref{sec:la}, the sum of $p_v$ across channels is 0. Since the $\gamma$ for a nearly constant map is $0$, the thick channel limit of $p_v$ must have a flatter slope compared to the thin channel $p_v$. As a result, the trend of $\gamma$ for $p_v$ might seem different to some readers compared to what \citetalias{LP00} predicted: In thin channels, the spectral slope of $p_v$ is steeper, but that should eventually decrease as the channel width increase. Yet \citetalias{LP00} was actually comparing the spectral slope between the thin channel $p_v$ and the thick channel $p=p_d+p_v$. Therefore, we perform the same in this section.}. Using the spectral index values of thin channel $p_v$ and thick channel $p$, we estimated that $m\sim 3.1-2.7 \sim 0.4$, which corresponds to a 3D velocity spectrum of $E(k)\sim k^{-3.4}$. We note that this value of $m$ cannot be found if one only has the information on the blue curve in the bottom right of Fig.\ref{fig:GALFA22800_1835}. {Only in the case} when we remove the density contamination, we can study the channel's velocity statistics. {Formally, the spectrum is a little bit shallower than the Kolmogorov one (2/3), which is likely because we did not perform a thermal deconvolution in the decomposed velocity channel, or there is a possible self-absorption in these data. Not to mention we did not estimate the noise here since we want to illustrate the power of the VDA for VCA studies. However, the result of this analysis should not be underestimated since it is simply impossible to apply VCA for regions with severe thermal broadening. }{Indeed, the VDA amplifies the capabilities of the VCA significantly. The contribution of velocity caustics arising from the WNM is suppressed in the traditional VCA approach due to the thermal broadening. Combining the VCA and the VDA, we can study both velocity turbulence in the CMN and in the WNM and, therefore, we can compare the properties of turbulence in different phases of HI. This is very advantageous for understanding the dynamics of the multi-phase medium. We will provide the corresponding analysis in our later publications.
}

\section{Implications for previous studies}
\label{sec:implication}

{With our results in the previous sections, it is worth discussing how important velocity caustics are for some previous research. We would discuss some of the important research directions and discuss how the caustics extracted by VDA would change these directions. We shall first discuss the importance of caustics to channel studies in \S \ref{subsec:filaments}. The RHT-related studies will be discussed in \S \ref{subsec:RHT}. We shall discuss how would the VDA fundamentally changes the current recipe of VGT in \S \ref{subsec:VGT}. At last, we shall discuss the nature of HI filaments under the analysis framework of VDA in \S \ref{subsec:nature}. }

\subsection{The importance of velocity caustics in velocity channels}
\label{subsec:filaments}
{  
The theory of PPV statistics \citepalias{LP00} has been verified by numerical simulations and observations. One of the most important predictions in \citetalias{LP00} is the change of velocity channel power spectral slope as a function of velocity channel width, which has been confirmed observationally in {  the} SMC \citep{2001ApJ...551L..53S} and also numerically in MHD simulations \citep{2003ApJ...592L..37L}. The method of obtaining the velocity spectral index from the difference of power spectral slopes of thin and thick velocity channels is later tailored as VCA in \cite{LP04}. Since then, the development of VCA has been extended to emission and absorption lines {for} different tracers \citep{LP04,LP06,LP08}. More importantly, the VCA method is being further improved numerically in \cite{2009ApJ...693.1074C} and the absorption case was tested by \cite{2013ApJ...770..141B}. The VCA is primarily tested observationally for molecular tracers \citep{2006ApJ...653L.125P} and multiple HI cubes that are not severely affected by thermal broadening \citep{reply19}.

Aside from the spectral slopes differences as a function of channel width, \citetalias{LP00} and the subsequent works \citep{LP04,LP06,LP08,2009ApJ...693.1074C} also predict that the anisotropy of velocity channels are associated with the velocity caustics fluctuations. The velocity caustics are both filamentary (see \S \ref{subsec:RHT}, \ref{subsec:nature})\footnote{The formation of "fibrous filaments" arising from velocity fluctuations, i.e., from velocity caustics in PPV space, was reported in simulations by \cite{2018MNRAS.479.1722C}, in agreement with the \citetalias{LP00} predictions. } and independent of the density fluctuations (See Fig.\ref{fig:cartoons}). In this paper, we clarify the concept of velocity caustics and show that it can be observed and fulfills the properties that are predicted by \citetalias{LP00}.

Nevertheless, the importance of velocity caustics was challenged in \cite{susan19} where it was claimed that \citetalias{LP00} is not applicable to multiphase HI, and the filaments in channel maps were mostly density filaments aligned with the magnetic field. Our initial response to \cite{susan19} was made public in \cite{reply19}. With the multiphase simulations and the tools developed in this paper, we are in a position to provide a quantitative response to the critiques by \cite{susan19} and subsequent publications supporting the pure density explanation of 21 cm emission filaments (see \cite{kalberla2019}). A key criticism of \citetalias{LP00} from \cite{susan19} is related to the increased density contribution due to thermal effects. To address criticism such as these, a tool that allows us to disassociate and quantify the relative density and velocity fluctuations in velocity channels is required.}

It is for this purpose that we present the VDA method. The results provided in this paper demonstrate that VDA can answer many of the concerns from the critics of \citetalias{LP00} approach to HI. First of all, the caustics are the most important theoretical case that is studied in \citetalias{LP00} and subsequent works and has been known to follow Tab.\ref{tab:L09} nicely. The extraction of velocity caustics would allow one to determine the 3D velocity spectral slopes based on the caustics map derived from Eq.\ref{eq:ld2}. Second, obtaining the exact contribution of caustic structures enables us to settle the recent debate on whether the velocity channels are dominated by densities or velocities \citep{susan19,2019ApJ...886L..13P,reply19, kalberla2019,kalberla2020a,kalberla2020b}. {From the examples in \S \ref{sec:observations},\ref{sec:VCA} \& \ref{subsec:RHT}, we see that {\bf the statement that "HI velocity channels are exclusively dominated by CNM density fluctuations" is incorrect}}. Note that even in the regions that we identified as "density dominated", ignoring the velocity fluctuations would be inaccurate. While the total intensity fluctuations can arise from densities on large scales, the small scale intensity fluctuations are still dominated by velocity caustics {(See Fig. \ref{fig:pd2pv_huge},\ref{fig:pd2pv_multiphase_XNM},\ref{fig:pd2pv_HVC})}, as opposed to \cite{susan19}'s empirical claim. Furthermore, we can see from the upper row of Fig.\ref{fig:pd2pv_multiphase_XNM} that only when one considers a rather thick channel limit ($\Delta v/v_{inj}> 1.3$) will the spectra ratio ($P_d/P_v$ in Fig.\ref{fig:pd2pv_multiphase_XNM}) stays above unity, which suggests that \citetalias{LP00} is applicable to HI media. 

\subsection{Implication to Rolling Hough Transform and tracing of cold neutral media based on linear structures}
\label{subsec:RHT}

A series of papers \citep{2019ApJ...886L..13P,kalberla2019,kalberla2020a,kalberla2020b} studied the correlation of the enhancement of intensities in HI velocity channels to different observables in order to infer the distribution of cold neutral media on the sky. These works are based on the argument from \cite{susan19} that the features extracted by the unsharp mask (USM), which is the first step of the Rolling Hough Transform (RHT, \citealt{Clark14}), are mostly {dominated by density fluctuations} and reflected cold neutral media structures. In particular, they used a non-normalized version of the NCC (see Eq.\ref{eq:NCC}, \citealt{reply19}) and compared the value of {  NCC} between channel maps and dust column density map to a curve erroneously attributed to \citetalias{LP00} to claim that the velocity channels are density dominated regardless of the channel width. Based on these sorts of arguments, {\cite{susan19} concluded that with the section title 3.3 in \cite{susan19}: {\it"Thin-channel HI Intensity Structures Are Not Velocity caustics"}, and also in their main text {\it "Evidently, not only are thin-channel HI intensity structures not dominated by velocity caustics, but there is no measurable contribution to the HI intensity from velocity caustics at all "}\footnote{\cite{susan19}, Sec 3.3,p.9, last line}, suggesting that velocity caustics are completely absent in HI observation. } 

The above conclusion contradicts our testing in this paper. In fact, our analysis in this paper reveals that each channel contains both density and velocity fluctuations for each phase (See. Fig.\ref{fig:pd2pv_multiphase_XNM}). In particular, previous sections show that (1) velocity fluctuations tend to contribute more when we are observing the wing channels (2) the dominance of density fluctuations in velocity channels occur mostly in the subsonic environment when $|v-v_{peak}|<0.5\Delta v_{effective}$ (See \S \ref{sec:one_sigma}). It is not hard to imagine that the USM or RHT of the $p_v$ map, which is purely velocity features, would still exhibit correlation to cold neutral media. Moreover, it is also possible that the structure of the cold neutral media in channel maps is inherited from the velocity fluctuations. We want to illustrate this effect using the region employed in \cite{Clark15} to show that the velocity channels do contain velocity caustics according to our analysis method in \S \ref{sec:la} and further show that caustics are dominant in a number of channels. 

\begin{figure}[th]
  \centering
  \includegraphics[width=0.49\textwidth]{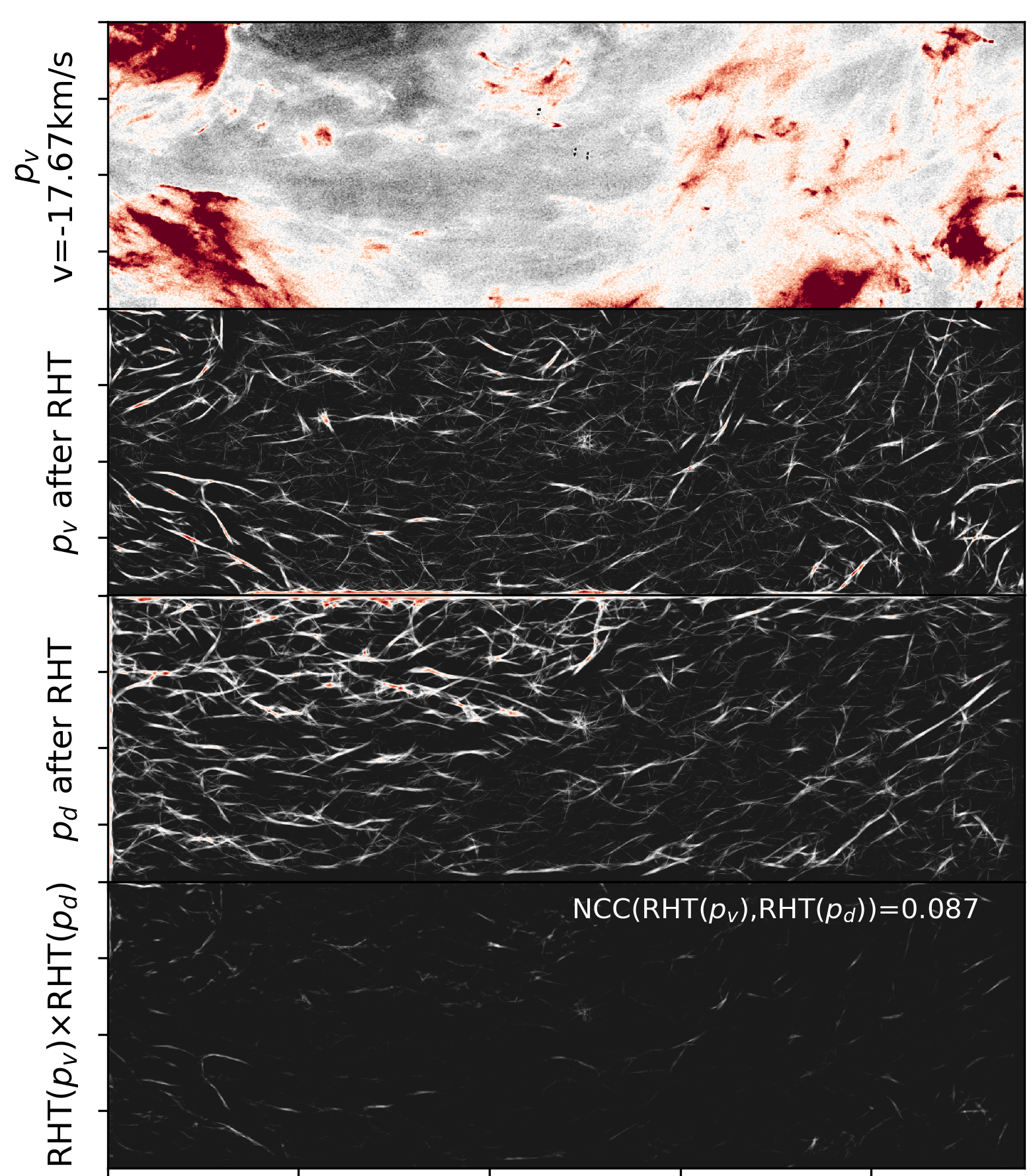}
  \caption{\label{fig:RHT} A set of figures showing how a selected wing channel ($v=-17.67 km/s$, $\Delta v = 2.94 km/s$) from the region in \cite{Clark15} would look like after VDA, and the output of RHT from VDA maps. From the top panel: The $p_v$ map after VDA. 2nd and 3rd panels from the top: The RHT results of $p_v$ and $p_d$ respectively, scaled to $[0,1]$. The lower panel. The product of the RHT results of $p_d$ and $p_v$, scaled to $[0,1]$. The normalized correlation coefficient (Eq.\ref{eq:NCC}, notice $NCC\in[-1,1]$) is 0.087, which means the RHT results of $p_d$ and $p_v$ are basically uncorrelated.
 }
\end{figure}

To start with, we select the regions that have the most "RHT-fibres" (the linear structures that are detected according to RHT) in the region used in \citeauthor{Clark15} (\citeyear{Clark15}, Top panel of Fig.\ref{fig:RHT}). We then perform VDA in the region and compute RHT for both $p_d$ and $p_v$. The RHT output indicates the location of the linear structures in the $p_{d,v}$ maps with the intensity map being rescaled to $[0,1]$. The intensity value of the RHT output indicates the pixel's probability of being a part of the linear feature in the neighboring region. We can see visually from the middle panels of Fig.\ref{fig:RHT} that both $p_d$ and $p_v$ maps exhibit a variety of linear structures. To compare their location, we can multiply the RHT output of $p_d$ and $p_v$, whose intensity value is also in the range of $[0,1]$. The lower panel of Fig.\ref{fig:RHT} shows the multiplied output. We can see that the linear features from $p_d$ and $p_v$ are uncorrelated to each other, which is consistent with the NCC value ($NCC \sim 0.087$, i.e the two maps are statistically not correlated, see Eq. \ref{eq:NCC}) of the RHT output of these two maps. More importantly, it is clear that velocity caustics contain linear features that are identified as filaments by RHT. The appearance of these features follows from MHD turbulence theory and the theory of space-velocity mapping in \citetalias{LP00}. {\it Therefore, it is wrong to disregard the effects of velocity fluctuations in those fibers identified by the RHT.}

\begin{figure}[th]
  \centering
  \includegraphics[width=0.49\textwidth]{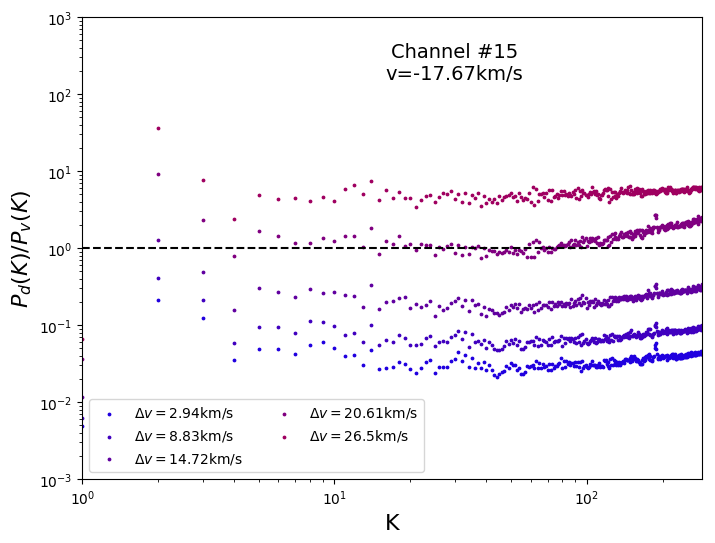}
  \caption{\label{fig:stail_pdpv} The spectra ratio $P_d/P_v$ at the wing channel from the HI data from \cite{Clark15} at $v=-17.67 km/s$ and $\Delta v = 2.94 km/s$. We can see that as the channel width goes smaller, the spectra ratio quickly drops below one for most $K$. }
\end{figure}

It is then important to see how the spectra ratio ($P_d/P_v$, see \S \ref{sec:test}) would behave as a function of the channel width. The channel width in this data is considered to be "thin" according to \citetalias{LP00} (See Eq.\ref{eq:thin_criterion}). Fig.\ref{fig:stail_pdpv} shows that the small scale $P_d/P_v$ ratio is indeed smaller than one when the channel is thin. This shows that velocity fluctuations dominate over that of density in small scales for this particular channel. This result can be easily generalized to different regions. In fact, from our analysis, there are 34 channels out of 41 being velocity-dominant. Furthermore, we would like to estimate the importance of velocity caustics in all channels by computing the weighted percentage along the line of sight: We first compute the importance of velocity fluctuations for each velocity channel by $\sigma_{p_v}/(\sigma_{p_v}+\sigma_{p_d})$ and then multiply this quantity to the spectral line $N(v)$. That is to say, the weighted percentage is given by:
\begin{equation*}
  \text{Weighted Percentage}=\frac{1}{\sum_v N(v)}\sum_v\frac{N(v)\sigma_{p_v}(v)}{\sigma_{p_v}(v)+\sigma_{p_d}(v)}
\end{equation*}
From the above equation, we see that {more than} 50\% of the total pixels in all of the channels in the regions are {velocity fluctuations} even with the density effects fully accounted. This shows that disregarding the effect of velocity caustics for the formation of the structures in thin channels is incorrect. The structures of velocity caustics are linear, filamentary, and expected to align with the magnetic field. We also note that the degree of alignment of the velocity filaments arising from velocity fluctuations with the polarization is higher than for those from the density filaments. This corresponds well to the theory's expectation as the velocity fluctuations in MHD turbulence trace magnetic field direction better than those of density (see \S \ref{subsec:VGT}). However, we believe that the VGT, especially under the modification of VDA (see \S \ref{subsec:VGT}), is a more reliable way of tracing magnetic fields than the currently available methods which utilize the filamentary nature of velocity channel maps.

\subsection{Implications to the Velocity Gradient Technique}
\label{subsec:VGT}

The velocity gradient technique, which is based on the modern MHD theory \citep{GS95,LV99} and the statistics of spectroscopic channel maps \citep{LP00,LP04}, can successfully trace magnetic fields in various astrophysical environments. {However, the velocity channel gradients (VChGs) are founded on the assumption that velocity caustics dominate the fluctuations of the velocity channels in MHD turbulence \citep{LY18a}. In this paper, we show that this argument is not entirely correct. In fact, from our previous sections (e.g., \S \ref{sec:test}), we see that there can be velocity channels that are density dominated. Despite that fact, \cite{LY18a} showed in a variety of MHD simulations with different sonic and Alfvenic Mach numbers the gradient technique is shown to be applicable to HI media. It is then natural to ask: Why do the channel gradients in \cite{LY18a} correctly trace the magnetic field directions? How would the VDA {  developed} in this paper improve the accuracy of the velocity channel gradient in observation? Very importantly, we will have a new quantity under the framework of VGT called {\bf velocity caustics gradient} that would follow the assumptions of \cite{LY18a} nicely, and we expect that to be} more reliable and accurate in tracing the magnetic field.

\subsubsection{Improving magnetic field tracing with velocity channel maps}

The numerical tests \citep{YL17a,YL17b,LY18a} of the VGT cover a large parameter space with various $M_s$ and $M_A$, and are performed with various synthetic observations (see \citealt{2019ApJ...873...16H}). It has been shown repeatedly {  with different numerical setups} that the gradients of different observables (e.g., column density, velocity centroids, velocity channels) are tracing the {\it block-averaged} magnetic field directions as long the \cite{YL17a} Gaussianity is satisfied. That means when one could fit a Gaussian function \footnote{For large $M_A$ cases the fitting should be a von-Mise function.} to the gradient orientation histogram. In this scenario, the peak of the Gaussian fitting function returns the local magnetic field directions. This procedure is termed block-averaging in \cite{YL17a} {and has been the core of the gradient technique despite other procedures or improvements are introduced} (e.g., \citealt{PCA,2020MNRAS.496.2868L}, Ho \& Lazarian 2020). Aside from some special situations, e.g., regions of the gravitational collapse (see, e.g., \citealt{YL17b,IGVHRO}), velocity gradient directions after block averaging are statistically perpendicular to the local block-averaged magnetic field directions. The block-averaging criterion has been extended by \cite{2020MNRAS.496.2868L} under theoretical considerations such that it should be a special Lorentzian-sinusoidal function that fits the gradient orientation histogram. The success of the VGT is further confirmed by the numerous application examples available in different astrophysical environments (See e.g. \citealt{YL17a,LY18a,survey,velac}). {On the contrary, while t}he density field can mimic the statistics of velocities for low $M_s$, it significantly deviates from the {velocity statistics} in high $M_s$ flows {or} in more complicated physical settings \citep{YL17b,LY18a}. Those include, for instance, {multiphase} fluids with thermal instability. {However, previous work shows that even with the density collusion in velocity channels, the velocity channel gradients still trace the magnetic field reliably \citep{LY18a}}.

Simultaneously, the thermal line width of turbulent gas presents a {prominent} complication {in terms of the tracing of the magnetic field as the channel map is affected by the morphological changes of density structure.}\footnote{The importance of such structures gets more prominent for high sonic Mach numbers $M_s$. However, for $M_s\gg 1$, the importance of thermal broadening decreases.} According to the \citetalias{LP00} theory, an increase of the thermal line width will also increase the effective thickness of a velocity channel (See Eq.\ref{eq:effective_dv}). {Moreover, as we see in this paper (See e.g., Fig.\ref{fig:pd2pv_huge}), the contribution of density fluctuations increase in proportion to the thermal line width. Under a strong broadening scenario, the velocity channel will be effectively thicker, which complicates the interpretation of velocity channel gradients in terms of velocity gradients using the \cite{LY18a} recipe. } Earlier, this problem was circumvented by using heavier tracers, e.g., CO, with reduced thermal width, or appealing to the physical model that cold HI clumps move together with the warm HI. With VDA, the problem of {density interference due to thermal broadening} is solved as we can compute the gradients of the decomposed caustics $p_v$ (See Eq.\ref{eq:ld2}) from observational data. 

{We want to stress that}, while some channels are density dominated, the VDA technique provides the pure caustics structures for {\it each} velocity channel and thus the underlying structures will reflect the velocity statistics {as} predicted by \citetalias{LP00} and numerically tested in \cite{LY18a}. Our exploration of velocity channel based on VDA shows that even in the case of subsonic turbulence, the velocity caustics contribute to a significant amount of fluctuations of intensities in velocity channels (\S \ref{sec:test}) which is confirmed by both numerical {(\S \ref{sec:test})} and observational data {(\S \ref{sec:observations},\ref{sec:VCA})}. Elsewhere, we shall discuss how to use the velocity channel data more productively by combining the VGT with the VDA approach.

\subsubsection{Dependencies of caustics gradients on physical conditions and MHD modes}

In fact, when turbulence is highly subsonic, the following is expected by the MHD theory: (1) Alfven mode velocity fluctuations are anisotropic along with the local magnetic field directions \citep{GS95}; (2) the slow mode velocity {fluctuations} are moved by Alfvenic modes as a passive scalar. Therefore, mimic the statistics of Alfven velocity fluctuations; (3) the fast {mode} velocity fluctuations are isotropic (see \citealt{CL03}) with {the relative energy in fast mode being} small compared to the Alfven and slow mode combined \citep{CL02,CL03}. {Notice that the slow mode velocity statistics is slaved by the Alfven mode}. {Under this condition, the velocity fluctuations in channel maps, which mostly coming from Alfven and slow modes, are expected to be anisotropic to the local magnetic field.}

{The statistics of density fluctuations is far more complicated than that of velocity fluctuations \citep{2005ApJ...624L..93B,2007ApJ...658..423K}. In subsonic turbulence, Alfven mode does not contribute to density fluctuations. The majority of the density fluctuations arise mostly from slow modes and they follow the same Alfvenic turbulence scaling as the velocity fluctuations. This also follows from our previous} numerical experiments in \cite{LY18a}, {where} the orientations of the density and velocity {gradients} are {shown to be similar, i.e. perpendicular to the local magnetic field }\footnote{It is worth noting that the gradients of a certain observable will only be able to trace magnetic field when (1) the power spectral slope of that observable to be steeper than $-1$; (2) being anisotropic along the magnetic field (See \citealt{LY18b}, Appendix \ref{secap:origin}). The MHD turbulence (\cite{GS95}, see also \citealt{2019tuma.book.....B}) falls into the requirement, but the gradient technique will also apply to other kinds of the cascade as long as the \cite{LY18b} criterion is satisfied.} in the case of {a mildly subsonic} environment. {In fact,} it is shown numerically (Ho \& Lazarian 2020b) that in the {subsonic} limit, the {velocity channel} gradients perform exceptionally well regardless of the relative contributions of density and velocity fluctuations in the channel maps. Moreover, both of these fluctuations are anisotropic along with the local magnetic field directions. {Yet in supersonic cases, the density fluctuations show rather non-trivial behavior. It was shown in \cite{2005ApJ...624L..93B} that, while the density field below a certain threshold obeys the Alfvenic turbulence statistics, a small fraction of volume filled by high-density structures will have a shallow spectrum and very different statistical properties arising from shocks. Those high-density structures tend to align perpendicularly to magnetic field \cite{YL17b}, which result in a decrease of the accuracy for the channel gradients approach \citep{LY18a}.\footnote{We note that the response of density fluctuations to the presence of magnetic field can be used to get properties of magnetized media, see \citep{IGVHRO}. Using the approaches developed with the velocity gradients, it is also possible to trace magnetic fields \citep{YL17b,IGVHRO}. Notice that density gradients are usually not reliable tracers of the magnetic field in supersonic interstellar media \citep{YL17b}. They, however, can be used in combination with velocity gradients to study both magnetic fields and shocks (see \citealt{IGVHRO}, also \S \ref{sec:934}).} }. {Under the framework of VDA, the observers can deal with} the real velocity fluctuations {from spectroscopic data, allowing the observers to not considering the statistical behavior of density field in MHD turbulence.}

\subsubsection{Caustics gradient dispersion as a probe of media magnetization}

The VGT in its present formulation goes beyond tracing the magnetic field. For example, {the dispersion of the channel gradients} can be used to study magnetization of the media \citep{LYH18}. Notice that the {gradient dispersion} statistics of density and velocity in MHD turbulence could be different in different modes even if their gradient orientations are {statistically} the same. This {leads} to an essential question on whether the magnetization estimation technique \citep{LYH18} is applicable if we have a different weight of density and velocity contribution in velocity channels, as there might be concerns on whether the fitting result from \cite{LYH18} will be applicable to caustics map decomposed from VDA. 

Nevertheless, as the orientation of the velocity caustics gradients is defined purely from velocity statistics (See \S\ref{secap:origin}), we do not expect the orientation-related statistics (e.g. \citealt{LYH18,curvature}) would change dramatically. We also reported numerically that the dispersion of density-weighted channel gradients is not much different from that of the caustics gradients. Thus we believe that the fitting result from \cite{LYH18} should be applicable, but a more careful study should be performed in the near future. Similarly, the improvement techniques (moving average from \citealt{LY18a}, PCA from \citealt{PCA}, spectral filtering from \citealt{Letal17}) will all have to be tested accordingly in the case of caustics map. {   It would also be interesting to see how the mode decomposition technique can be synergetically used with the gradients of velocity caustics in observations (See \citealt{CL03,2020NatAs.tmp..174Z}).}

\subsubsection{Shock identification with the VDA}
\label{sec:934}

The introduction of the VDA technique also allows us to proceed with an accurate shock detection algorithm based on the properties of velocity caustics gradient. It is discussed in \cite{YL17b} and subsequently in later works (\citealt{LY18a,IGVHRO}) that the gradients of densities will be perpendicular to that of velocities in the region of shocks. While in the raw velocity channels we cannot separate density and velocity channels, it was argued in \cite{YL17b} that thin channel gradients would be a good estimate of the caustics map. Therefore, by comparing the gradients of thin and thick channels, one might locate shocks. However, as we see from this paper, the aforementioned method may sometimes be problematic since the gradients of the velocity channel are linear combinations of density and velocity fluctuation. Moreover, the density gradients are now parallel to the magnetic field statistically. A numerical example can be given using the results from \S \ref{sec:test}: Even for the thinnest channel we have at the spectral peak in the lower-left panel of Fig\ref{fig:pd2pv_huge} ($\Delta v/v_{inj} \sim 0.07$), the relative contribution of density to velocity in the center channel is still $\sim 0.7$. That means the gradients of the center channel {\it in the vicinity of the shocks} in Fig.\ref{fig:pd2pv_huge} in the thin channel limit would be comprised of $\sim 41\%$ of the gradients that are density like being parallel to the magnetic field, while $\sim 57\%$ of the gradients that are velocity like will then be perpendicular to the magnetic field. With VDA, we can separate the channel maps into $p_d$ and $p_v$ and consider their gradients' relative orientations in the vicinity of shock regions. We shall refer to the formulation in the forthcoming paper.

\subsubsection{Caustics gradients under self-gravity}

A similar argument also applies when we are using gradients to probe the self-gravitating regions. It was suggested from \cite{YL17b} that the gradients of both density-like and velocity-like features will {\it gradually}\footnote{The word "gradually" is important here since it is not a one-time process for the gradient vectors to flip their directions. Instead, as discussed in \cite{YL17b}, the relative orientation of both intensity and velocity centroid gradients will change from perpendicular to parallel according to the stage of collapse, with the latter being a slower process. } rotate 90$^o$. With VDA, we can see how the velocity caustics behave in the presence of gravity. According to the description in \cite{YL17b}, we believe that the caustics gradients should be least affected by gravity. However, when the gravitational force takes over, the caustics gradients will eventually turn 90$^o$. This change happens as the acceleration induced by gravity gets larger than the acceleration induced by turbulence.  Both the relative orientation between intensity gradient and $p_v$ gradient and that between $p_v$ and polarization will change. This is similar to the use of the velocity gradients (see \citealt{YL17b}), but using $p_v$, we expect to determine the collapse regions significantly better. We shall discuss the corresponding algorithm in the forthcoming paper.

\subsubsection{Caustics gradients in multiphase HI emissions}

The ability to separate the gradients arising from velocities and densities would change the ways of applying VGT in multiphase media. It was shown in Fig.\ref{fig:pd2pv_multiphase_XNM} that all of the channels for the cold and unstable neutral media are velocity dominant, while for the warm neutral media, the velocity channels are velocity dominant only in the wings. When they are combined and observed spectroscopically, we see that only the supersonic, cold/unstable media features are displayed in both the $\sigma-v$ and spectra ratio $P_d/P_v$ diagrams. We can see that the peak location of $\sigma_{p_v}-v$ in Fig.\ref{fig:pd2pv_multiphase} is consistent to that of cold neutral media in Fig.\ref{fig:pd2pv_multiphase_XNM}. Similarly, the $P_d/P_v$ diagrams in both Fig.\ref{fig:pd2pv_multiphase} and \ref{fig:pd2pv_multiphase_XNM} are sensitive to channel width, which is a signature of supersonic system. As we show {in Fig.\ref{fig:pd2pv_multiphase_XNM}}, the numerical simulations suggest that the caustics arising from the CNM clumps dominate {that} from the other phases in multiphase media. {It is worth noting that, due to the large temperature differences and the definition of the sonic Mach number, the turbulence flow can be supersonic for CNM but subsonic for WNM. Again, we emphasize that the thermal problem depends on the parameter $v_{los}/c_{s}$, not the true sonic Mach number. For a two-phase model, there will be at least two numbers $v_{los}/c_{cnm}$, $v_{los}/c_{wnm}$ that will decide whether the velocity channel will be density like or velocity like, where $c_{cnm},c_{wnm}$ is the localized sonic Mach number for CNM and WNM. Readers should be cautious on the dramatic differences of the thermal properties for CNM and WNM and the respective impact on the channel maps (See \S \ref{sssec:multiphase}). }

The VDA approach brings the gradient {technique} to a new level in applications to HI media. This paper shows that the VDA resolves the problem of separating the velocity and density contributions in the situation where the thermal broadening dominates the turbulent contribution. As a result, the VDA makes the caustics gradients applicable in the most unfavorable scenario described in \cite{susan19}. In particular, our practical {applications} of the VDA to the {HI} data in \S \ref{sec:observations} \& \ref{sec:VCA} demonstrates that using caustics gradients in multiphase HI we can, first of all, (a) prove the applicability of caustics gradients in its application to most of the galactic HI, (b) increase the accuracy of both techniques in probing turbulence and magnetic field, (c) obtain reliable results for the regions where the application of the original versions of techniques may be problematic. (d) gives the fundamental reason why small scale gradients are more accurate than that of large scales due to the spectral behavior we show numerically in Fig.\ref{fig:pd2pv_huge},\ref{fig:pd2pv_multiphase_XNM} and observationally in Fig.\ref{fig:pd2pv_HVC}.

{The channel gradients }are being developed based on the better understanding of the position-velocity mapping of turbulent motions \citepalias{LP00}. We {expect} that the VDA {would be} a way of significantly improving their performance {due to the removal of density collusion in channel maps}. The synergy with other approaches presents {a plausible way in} further improving the ways of studying turbulence and magnetic fields. We discussed above the application of the phase decomposition technique \cite{kalberla2019} in combination with the VCA. We believe that this approach {could} be promising together with the caustics gradient {and expect that this synergistic approach } can provide insight into how different phases interact with the magnetic field. Besides, while the classical VCA (\S \ref{sec:VCA}) focuses on obtaining turbulent properties of the cold medium, with the VDA, turbulence in both cold and warm media can be studied, and its properties can be compared. This is very important for the understanding of the dynamics of the multiphase media.

\subsubsection{The use of caustics gradients for predicting CMB polarized foreground}

The VDA, together with VGT, also enables us to study the foreground magnetic field, which would be beneficial in cosmological studies. The VDA is a technique applicable to any spectroscopic data. The application of it to HI will have significant consequences for foreground polarization studies. Indeed, dust polarized emission is an essential component to be removed in CMB polarization studies aimed at detecting the enigmatic gravitational waves in the early universe. 

{As we see from \S \ref{subsec:RHT}, the HI fibres that are extracted by RHT and the related techniques could be either CNM density structures or velocity caustics. While \cite{Clark15} did show that the fibres traced by RHT have a statistical correlation to the local magnetic field on the plane of the sky, whether the correlation is related to density, velocity, or both are subjected to question. As we discussed in the paper, the VDA, with its ability to identify velocity fluctuations, mitigates the effects of density fluctuations on the VGT. As we also discussed earlier, the latter fluctuations are worse aligned with respect to the magnetic field compared to the velocity fluctuations. In other words, on the basis of our study, we conclude that the HI density elongated structures, the orientation of which was viewed in \cite{Clark15} as the way of tracing the magnetic field and the foreground polarization, are in fact, the impediment for the accurate predicting the foreground CMB polarization. A more precise prediction of the CMB foreground polarization, is expected when the contribution of these density filaments is filtered out with the VDA.The corresponding study will be presented elsewhere.
}

\subsection{Implications for studying the nature of the HI velocity channel structures}
\label{subsec:nature}

The current study of the properties of the PPV statistics is not limited to specific astrophysical settings. At the same time, the present study is timely due to a number of reasons. First of all, the intensity fluctuations within HI channel maps were an issue of recent intensive debates. \cite{susan19} suggested that the structures of thin channels, both at the spectral peak and in wing channels, are mostly associated with the density enhancement of cold neutral media. Their argument is based on: (1) in low sonic Mach number cases, and thermal broadening would suppress the contribution of velocity fluctuations in velocity channels (2) in high sonic Mach number cases, the high-density enhancements due to compression dominate the fluctuations of velocity channels (3) When using high-pass filters on velocity channels, it is likely to detect channel fluctuations that are associated with the density enhancements (4) those channel fluctuations are shown to be correlated with a number of alternative observational signatures that are directly or indirectly related to cold neutral media. Several follow-up works support the argument from \cite{susan19} and claim that the observed velocity channel intensity arises exclusively from cold neutral media density fluctuations \citep{susan19,2019ApJ...886L..13P,kalberla2019,kalberla2020a,kalberla2020b}.

These debates have far-reaching implications since the existence of caustics is the foundation of many statistical techniques that are shown to be applicable in observation. For instance, some of the papers (see \citealt{kalberla2020a}) question the applicability of the VCA to HI and claim that the whole crop of the VCA results obtained in this direction is "fake". The Velocity Gradient Technique (VGT) is also challenged by the papers that purport that the HI channel gradients are density gradients rather than velocity gradients. Note that VCA provides a unique insight into the velocity statistics, and VGT is a very promising tool e.g. map galactic magnetic fields both at high latitudes (see e.g.\citealt{YL17a,LY18a}) and within the galactic disk \citep{2019ApJ...874...25G}. {Nevertheless, it can be argued that even VCA or VGT show correspondences to the theoretical expectation, it is entirely possible that the velocity channel to be density {  dominated} as both VCA and VGT {  rely} on the fundamental properties of velocity channels as formulated by \citetalias{LP00}. This paper is in the exact timing to give a formal response to those papers \citep{susan19,2019ApJ...886L..13P,kalberla2019,kalberla2020a,kalberla2020b} and to discuss how the two methods (VCA: \S \ref{sec:VCA}, VGT: \S \ref{subsec:VGT}) will be impacted in the extreme cases when $P_d/P_v>1$. }

The points in \cite{susan19} {were} addressed in \cite{reply19} and this paper {also} provides the further theoretical, {numerical and observational} foundations for the theory describing the statistics of the velocity caustics. As we understand the current situation, the debates are focused on whether \citetalias{LP00} theory and the VCA technique adequately reflects the statistics of the multi-phase neutral hydrogen. The nature of the neutral Hydrogen velocity channel is highly related to the underlying model of turbulent HI. It is a well-known fact that neutral hydrogen has at least two stable, pressure-balanced phases (See \citealt{Draine2011PhysicsMedium,MO07}), which are called cold Neutral media (CNM, $T\sim 200K$) and warm Neutral media (WNM, $T>5250K$) (See also \citealt{2017NJPh...19f5003K}). In between the temperature ranges are the thermodynamically unstable phases. Under the differentiation of phases, WNM is believed to be subsonic, while that of CNM is believed to have sonic Mach number $M_s=v/c_s\sim 2-3$. Moreover, both phases are believed to be sub-Alfvenic \citep{C10}. In the theory of interstellar turbulence with the Alfvenic Mach number small ($M_A<1$), the statistical properties of supersonic turbulence is very different from that of subsonic turbulence (See \citealt{GS95,2007ApJ...658..423K}). Nevertheless, observationally supersonic turbulence velocity channels display different properties to that of subsonic turbulence. How do supersonic CNM and subsonic WNM behave geometrically under the framework of interstellar turbulence and their respective spatial behavior in velocity channels become a crucial question that we would like to understand.

{For example, t}he application of VDA to HVC (\S \ref{sec:observations}) allows observers to extract the velocity caustics with high accuracy in observed neutral hydrogen velocity channels and further enables observers to characterize the importance of turbulence velocity statistics in observations. Notice that HVC has a well-studied core-envelope structure. We can see that the density structure does indeed dominates the center part of the velocity spectral line, which we found to be originated by CNM density fluctuations. Yet as we move to the wing channels or focus only on small scale structures, the velocity caustics dominates over density fluctuations. Nevertheless, the WNM caustics dominates over density fluctuations for nearly all channels, which means the contributions of velocity caustics cannot be ignored for every channels. 

{Similar situation also happens for the GALFA data that we test within \S \ref{subsec:RHT}. By analyzing GALFA data, we demonstrate regions where the intensity fluctuations are clearly dominated by velocity fluctuations (See left panel of Fig.\ref{fig:GALFA00400_1035_VCA}), and we also showed that even using the so-called "cold neutral media tracing tools" like USM or RHT, we can obtain {more than} 50\% filamentary structures that are caustics (See Fig.\ref{fig:stail_pdpv}).} In fact, there is no channel that has non-zero intensity contains zero caustics contributions. Even at the peak of the velocity spectral line, the caustics contribution is still about 33\% of the total fluctuations of the velocity channel when including the large scale contribution. If we discuss only small scales ($k>5$) at $\Delta v=2.94km/s$ the percentage should be raised to 87\%. The velocity caustics map contains a unique spatial structure that reflects the properties of turbulence (See Fig.\ref{fig:HVC_decomposition}). The rich physics that is contained in velocity caustics should therefore be studied extensively.

Readers should keep in mind that the existence of velocity caustics in a specific place in the velocity channel does not exclude the possibility of cold neutral media in the same location, and vice versa. In other words, cold neutral media as a supersonic object displays both density enhancements and velocity caustics in approximately the same {spatial} location of the map, and the fluctuations that we see from velocity channels are the sum of these two (See, e.g., Fig.\ref{fig:pd2pv_multiphase_XNM}). The {dominance of density fluctuations in HI channel maps are first proposed in \cite{susan19}. In fact, as we note before, \cite{susan19} believes that there is no measurable contribution to the HI intensity from velocity caustics at all. } This argument was adopted in some of the follow-up works that correlate the locations of unsharp masked channel structures to some other CNM diagnostics (e.g., probing using NaI equivalent width \citealt{2019ApJ...886L..13P}, or CNM probed by multi-Gaussian decomposition \citealt{kalberla2019}). {Nevertheless, as we see from \S \ref{subsec:RHT}, more than half of the "filamentary structure found by unsharp masks" should be associated with velocity caustics. Moreover, those "strong density fluctuations" are originated from the few smallest $K$ from the $P_d/P_v$ diagram. Most of the small scale $P_d/P_v$ is smaller than 1, suggesting a velocity dominance in the observed HI 21cm channel maps. The works that are based on the unsharp mask algorithm should be seriously revisited.}

Finally, we should mention that some authors were so worried about the applicability of the VCA to HI that, in {an arXiv preprint} with a telling title {\bf "Are observed HI filaments turbulent fraud or density structures? Velocity caustics, facts and fakes"} \citep{kalberla2020a}, that they {give the reader the impression that some published results analyzing HI and supporting the predictions of \citetalias{LP00} theory were fake ones and no one could reproduce them, apart from the authors of the \citetalias{LP00} technique. } We believe that the analysis provided in the present paper will help the reader to make a fact-based judgment whether the confirmations of the \citetalias{LP00} predictions were real or {"fake"}.

\section{Impact to various field of astrophysical study} 
\label{sec:discussions}

{Aside from the previous studies, the development of the VDA is expected to impact the general spectroscopic and polarimetric studies. On one hand, the VDA method allows simple yet efficient ways in extracting the caustics map from spectroscopic PPV data, meaning that one can now study the velocity statistics and dynamics based on the currently available PPV data using the framework developed in this paper, which we shall discuss in \S \ref{subsec:spectroscopic}). On the other hand, the method can be easily migrated to other studies like polarimetric maps, which is also suffered from density collusion. We shall discuss the possible ways of recovering the density-free physical quantities in \S \ref{subsec:polarimetric}. Moreover, we shall also discuss how the current paper could be used in conjunction with the Differential Measure Analysis method \citep{2020arXiv200207996L} to find the magnetic field strength on the sky \S \ref{subsec:DMA}. We would also answer the questions that we asked in \S \ref{sec:intro} in \S \ref{subsec:answers}. }

\subsection{Importance to general spectroscopic studies in interstellar media}
\label{subsec:spectroscopic}
Due to the fact that turbulence is ubiquitous in numerous astrophysical processes, the development of the current paper suggests that {\bf every piece of spectroscopic PPV data could be revisited by extracting the corresponding velocity caustics structure}, practically giving a second life to all spectroscopic data for astronomers to explore with. This paper demonstrates not only the key role that velocity caustics play for the formation of channel intensity fluctuations within the spectroscopic PPV cubes, but, most importantly, it provides the algorithm of extracting velocity caustics from observation (\S \ref{sec:la},\S \ref{sec:observations},\S \ref{sec:VCA}). These caustics structures are expected to carry different morphology compared to the channel map itself, straightly follows the theory of PPV statistics \citepalias{LP00} and also the properties that we discuss in the current paper (\S \ref{sec:test},\S\ref{sec:one_sigma},see also Appendices \ref{sc:SVDA},\ref{secap:vda_emissions_absorptions}). 

The velocity caustics is the central quantity in studying the turbulent fluctuations in the quantitative theory of PPV statistics established in \citetalias{LP00} (See also \citealt{KLP16}). This theory formulated how different statistical theory of MHD turbulence (e.g., \citealt{GS95,LV99,2019tuma.book.....B}, and the references therein) could impact the spectrum and anisotropy revealed through spectroscopic observations. Nevertheless,{since the density field can be non-Kolmogorov in term of the power spectrum in a number of physical conditions} (see \citealt{2005ApJ...624L..93B,2007ApJ...658..423K}), the interpretation of the PPV data based on {the prediction derived from pure velocity statistics assumption} could be ambiguous. \citetalias{LP00} provided a way of statistical determining the spectral indexes of the velocity and density spectra, but this study provides the way of separating the actual PPV contributions arising from density and velocity fluctuations. Our present work effectively opens {a new direction in} studying turbulence by applying the VCA \citepalias{LP00} and the VCS \citep{LP06} as well as the \citetalias{LP00}-based ways of turbulence anisotropy studies (see \citealt{LP12,KLP16,KLP17a}) to the PPV data with {velocity caustics extracted} (See \S \ref{sec:VCA} for our suggested procedure). 

The velocity caustics map also has its great importance for techniques that are based on statistics of velocity channels, as velocity caustics statistics is connected to a number of studies that allows us to infer the turbulence properties (\S \ref{subsec:filaments}), the direction and magnitude of the magnetic field (\S \ref{subsec:VGT}) and the nature of the filamentary structures in observational data (\S \ref{subsec:RHT},\S \ref{subsec:nature}). The new set of tools that we developed in this paper allows astronomers to explore the velocity dynamics directly without density contamination. Similarly, the studies of the turbulence anisotropy \citep{2003ApJ...592L..37L,EL05} and {   the decomposition of observational data into the contributions arising from Alfven, fast and slow modes ( \citealt{2020NatAs.tmp..174Z}, see also \citealt{KLP16,KLP17a})} can be done much more precisely and reliably with the PPV data not contaminated by the density fluctuations. We note that the VDA provides a good synergy both with the techniques that use channel map data, different moments of channel maps (e.g., see our construction of constant velocity centroid in Appendix \ref{secap:centroid} that shows a significant promise compared to the traditional velocity centroids, see \citealt{2003ApJ...592L..37L,EL05}) as well as more sophisticated constructions. The caustics map can be combined with other tools developed by the community that applies to channel maps, e.g., wavelet transforms \citep{2010ApJ...720..742K} or the Principal Component Analysis \cite{2008ApJ...680..420H} with new insight as the caustics are density-independent and straightly follows PPV statistics from \citetalias{LP00}. Naturally, the VDA also opens new possibilities for developing new tools to directly study the properties of the velocity field of astrophysical turbulence.

\subsection{Implications of the VDA {to polarization studies}}
\label{subsec:polarimetric}

The separation of the contributions arising from velocity and density fluctuations is important for other branches of research. Take the problem of ground-state atomic alignment (GSA) {as an example} (\citealt{2006ApJ...653.1292Y,2007ApJ...657..618Y,2008ApJ...677.1401Y}). {The observational output of the ground state alignment would be the} Stokes parameter maps at a range of velocities $v$ with velocity width $\Delta v$ (hereafter GSA-PPV). {The mathematical structure of the Stokes parameter in ground state alignment problems are rather similar to that of spectroscopic maps that we study in the current paper, that the resultant observational parameter is a density-weighted average of velocity (for Doppler-shifted lines) or magnetic field information (for ground state alignments along the line of sight. As a result, the GSA-PPV} would contain "magnetic field caustics" that arises from the fluctuations of the magnetic field lines sampled in a narrow velocity range. However, Stokes parameters are produced with a weighting related to the tracers' density. For the case of ground-state alignment, that would be the atoms like [CII]. Therefore, a method similar to VDA is required in analyzing the fluctuations of GSA-PPV if we need to study only the statistics of the magnetic field in atomic alignment measurements.

{Similar technique can also be used to dust and synchrotron polarization map where both of them {  suffer} from density contamination. Using dust polarization as an example, the Stokes parameters (in the complex form $P=Q+iU$) is a density-weighted sum of double angle along the line of sight:
\begin{equation}
  P({\bf X}) \propto \epsilon\int dz \rho_{dust} e^{2i\theta({\bf X},z)} \sin^2\gamma
\end{equation}
where $\rho_{dust}$ is the dust density, $\theta$ is the 3D planar angle, and $\gamma$ is the inclination angle. Observationally we obtain both $P$ and column dust intensity $I=\int dz \rho_{dust}$. Similar to the spectroscopic counterpart, the function $P$ contains both density and magnetic field fluctuations as discussed in \cite{LP12} (under a more general framework, though). It is rather natural to consider the linear combinations of $P$ and $I$ to extract both density and magnetic fluctuations from dust emission maps. We shall discuss the fundamentals of the method and how to extend the decomposition to synchrotron polarization with strong Faraday rotation in the upcoming paper. 
}

\subsection{Studying magnetic field strength}
\label{subsec:DMA}
{Aside from the tracing of magnetic field using VGT (See \S \ref{subsec:VGT} for a dedicated discussion on how VDA could fundamentally change VGT), the VDA technique can also help to estimate the magnetic field strength based on either the Davis-Chandrasekhar-Fermi (DCF) technique \citep{1951PhRv...81..890D,CF53} or the recent synergy of gradient dispersion \citep{LYH18} and Differential Measure Approach (DMA, \citealt{2020arXiv200207996L}).

In \cite{2020arXiv200207996L} we derived the relation between the mean squared magnetic field strength and the structure functions of velocity centroid and polarization angles. The mean squared planer magnetic field strength, which is based on MHD theory of turbulence, is given by:
\begin{equation}
  \overline{B_\perp^2} = f^2 4 \pi \bar{\rho}\frac{\mathcal{D}_n^v}{D_n^\phi}
  \label{eq:Dc_over_Dtheta_rho_alt}
\end{equation}
where $\bar{\rho}$ is the mean density of the system, and the $D_n^v,D_n^\phi$ are the $n^{th}$ order structure functions for velocity centroid and polarization angle respectively. The $f^2$ factor is theoretically deduced and numerically computed \cite{2020arXiv200207996L} to be $\sim 1-2$ in most of the cases. The method of Eq.\ref{eq:Dc_over_Dtheta_rho_alt} is shown to have high accuracy even in supersonic and mildly super-Alfvenic cases in \cite{2020arXiv200207996L}, which allows observers to estimate the magnetic field strength readily with theoretical support. 

However, the method of Eq.\ref{eq:Dc_over_Dtheta_rho_alt} has a pronounced deficiency: Both $V$ and $\phi$ are implicit functions of density, even though they do not carry any units of density. For instance, in observations we can only observe the normalized velocity centroid $V_c \sim \int \rho v/\int \rho$ and the Stokes polarization angle $\phi \sim \tan_2^{-1}(\int \rho \sin 2\theta/\int \rho \cos 2\theta)$ where $\rho$ is the density. Therefore, it is obvious to ask whether we have ways to remove the contribution of density fluctuations in observations for both centroid and polarization angles. In \S \ref{secap:centroid} we discuss a method of obtaining the constant density centroid based on VDA, while the constant density Stokes parameter could be possibly extracted using the idea from \S \ref{subsec:polarimetric}. The ability of the VDA to separate the velocity and density information opens new avenues for precision magnetic field studies using the spectroscopic information. We intend to explore those in future papers.
}

\subsection{Our questions in the introduction that are answered in the present study}
\label{subsec:answers}

Based on this VDA algorithm and the multi-Gaussian decomposition algorithm \citep{2007A&A...466..555H}, we answered the questions that are asked in \S \ref{sec:intro}:
\begin{enumerate}
   \item Is the concept of density/velocity fluctuations pixel-based, or is it only valid in a statistical sense?\\
  {\bf A}: The concept of density/velocity fluctuation is a statistical concept. We cannot see any velocity statistics if we do not consider the statistical effect of caustics. (See \S \ref{sec:la},Fig.\ref{fig:sim_illustration})
  \item What is the role of velocity caustics in channel maps when the CNM dominates the emission?\\
  {\bf A}: The relative fluctuations from velocity caustics are not insignificant (Fig.\ref{fig:HVC_decomposition}, Fig.\ref{fig:HVC_Gaussian}, Fig.\ref{fig:GALFA00400_1035_VCA})
   \item What is the relative importance of velocity and density fluctuations in a spectral line's central and wing channels?  \\
  {\bf A}: Even in the case of subsonic media, most of the channels are actually velocity-dominated in small scales. An example would be the {  High Velocity Cloud} (\S \ref{sec:observations}) where its CNM density contribution is approximately the same as of the velocity counterpart (Fig.\ref{fig:HVC_Gaussian}). However, in small scales the density fluctuations in velocity channels are approximately $1/2$ of that of velocity. (Fig.\ref{fig:pd2pv_HVC}).
\end{enumerate}

\section{Summary} \label{sec:conclusion}

This paper {develops a set of self-consistent, comprehensive tools in extracting the velocity caustics from PPV cubes} by combining analytical and numerical approaches {and has been thoughtfully tested in observations}. {Our numerical part} includes isothermal and multiphase numerical simulations. We developed a new algorithm for isolating the velocity caustics, explored the difference of the PPV statistics of the central and wing velocity channels, and applied our new approaches to HI 21 cm GALFA data. Our results are summarized as:
\begin{enumerate}
  \item We developed a {  new} algorithm termed Velocity Decomposition Algorithm (VDA) for extracting the velocity information from observational spectroscopic PPV cube. The algorithm provides an excellent decomposition of velocity and density contributions for sub-sonic turbulence, and it also shows significant promise for supersonic turbulence (See also Appendix \ref{sc:SVDA}).
  \item Our numerical study demonstrates that in thin velocity channels the contributions from velocity caustics dominate at small scales the density fluctuations in the intensity fluctuations of the wing channel at small scales. 
  \item We also applied our approach to multiphase simulations with realistic CNM and WNM mass fractions and demonstrated that the ratio of the spectral energies associated with density and velocity fluctuation, i.e. $P_d/P_v$, is less than unity when the channel width is thin regardless of whether the channel is at the center or at the wing (Fig.\ref{fig:pd2pv_multiphase}). In particular, the cold phase media which weighs the most in terms of spectral studies and exhibit velocity-dominant features in small scales as predicted in \citetalias{LP00}.
  \item We demonstrated that {the contribution of velocity caustics fluctuations is maximized in the wings and exhibit a double peak pattern for isolated emission regions. The distance of the peaks as a function of velocity is approximately $\Delta v_{effective}$ (See Fig.\ref{fig:betatoppv}, Fig.\ref{fig:one_sigma}). This distance is unchanged in the presence of the galactic rotation curve (Fig.\ref{fig:betatoppv_shear}).}
  \item We test our method with a selected high-velocity cloud (\S \ref{sec:observations}). This high-velocity low-metallicity cloud is isolated and has simple core-envelope geometry, making it a good testing ground for our approach. We determine that the small-scale fluctuations in the high-velocity cloud PPV data are dominated by velocity caustics regardless of the cloud's multiphase nature (Fig.\ref{fig:pd2pv_multiphase_XNM}). {Moreover, we see the double peak behavior as we predicted from \S \ref{fig:one_sigma}}.
  \item Using the VDA and the $1-\sigma$ (\S \ref{sec:one_sigma}) condition we propose a modification of the VCA technique that is advantageous when the observational data is significantly thermally broadened (\S \ref{sec:VCA}). The modification above extends the applicability of the VCA technique and increases its accuracy in recovering the 3D velocity spectral indices. With VDA, we can focus our analysis on the statistics of pure velocity caustics. That means we can employ the \citetalias{LP00}'s theory directly without worrying about the density contributions.
  \item We show that the possible "cold neutral media" location traced by RHT/USM can be {associated} with the velocity caustics of cold neutral media (\S \ref{subsec:filaments},\S \ref{subsec:RHT}). More importantly, our study confirms {that the velocity caustics are also filamentary and ubiquitous under the RHT algorithm} in HI channel maps (Fig.\ref{fig:RHT}) and establishes the foundations for further advancing the ways of studying turbulence and magnetic fields using spectroscopic data.
  \item The VDA method has significant implications to both the Gradient Technique (GT) (\S \ref{subsec:VGT}) and also the other structure identification algorithms like RHT (\S \ref{subsec:RHT}). Since the velocity caustics have well-predicted properties from the theory of MHD turbulence \citep{GS95} and the PPV statistics \citepalias{LP00}, the gradients or "fibres" of the caustics are expected to trace magnetic field and shocks much better than any density-weighted variants.
  \item {Most importantly, the VDA could derive a completely new set of unexplored velocity caustics data (\S \ref{subsec:spectroscopic}) from every spectroscopic data set. Moreover, the VDA allows one to check with a well-established theory of PPV statistics \citepalias{LP00} and apply the methods that are derived from \citetalias{LP00}. The potential of the VDA in studying MHD turbulence in observations should not be underestimated.}
\end{enumerate}

\noindent {\bf Acknowledgment} {  We thank the anomynous referee for providing extensive comments and suggestions to our paper.} We acknowledge our discussions with Dmitri Pogosyan that significantly advanced our understanding of the statistical theory of the fluctuations in the channel maps, in particular, in relation to the 1-$\sigma$ condition (\S \ref{sec:one_sigma}). We thank Alexey Chepurnov, Chris Mckee, and Snezana Stanimirovic for the elucidating exchanges on the velocity caustics they provided. We thank Alexei Kritsuk for his inspirational numerical studies in multi-phase media and Chi Yan Law for sharing his expertise in studying molecular cloud spectral line data. K.H.Y thank Huirong Yan for elucidating the discussion of apply VDA in different MHD modes. We thank Victor Lazarian and Yiming(Amy) Qin for the improvement of the language. K.H.Y, K.W.H. and A.L. acknowledge the support the NSF AST 1816234 and NASA TCAN 144AAG1967 have provided. Flariton Institute is supported by the Simons Foundation. The numerical part of the research used resources of both Center for High Throughput Computing (CHTC) at the University of Wisconsin and National Energy Research Scientific Computing Center (NERSC), a U.S. Department of Energy Office of Science User Facility operated under Contract No. DE-AC02-05CH11231, as allocated by TCAN 144AAG1967. {This work mostly employs the software Julia (https://julialang.org/), and we sincerely thank the development team.}

\appendix
\section{The correlation of density and velocity in the case of subsonic media}
\label{sec:apb}

To derive the orthogonality condition of VDA (See \S \ref{sec:la}, also \S \ref{eq:SVDA}) from the \citetalias{LP00}'s point of view, we need to start with the basic thermal broadening equation in subsonic turbulence in the velocity space form (Eq.\ref{eq:rhov_PPV}) and perform some simple analysis in the case of very thin channels ($\Delta v < \delta v_R=D({\bf X},0) < c_s$). Then from Eq.\ref{eq:rho_PPV} we can simply write

\begin{equation}
  \begin{aligned}
  p_d &\propto \delta \rho e^{-\zeta_1^2/2c_s^2}\\
  p_v &\propto \langle \rho \rangle e^{-\zeta_2^2/2c_s^2}
  \end{aligned}
\end{equation}
where $\zeta_{1,2}$ are two values in $(v_0-\Delta v/2, v_0+\Delta v/2)$ selected according to mean value theorem. We can already make some observation here. First of all, if $c_s$ is large and we know that from MHD theory $\langle \delta \rho \cdot v\rangle=0$, $\langle \delta \rho\rangle=0$, the correlation term can be separated and proved to be zero:
\begin{equation}
  \langle p_d p_v\rangle \sim \langle \rho \rangle \langle \delta \rho \rangle \langle e^{-\zeta_1^2/2c_s^2}e^{-\zeta_2^2/2c_s^2}\rangle =0
\end{equation}
The problem here is that, the orthogonality of the inner product above relies on two factors: (1) $\langle \delta \rho \cdot v\rangle=0$ , (2) $\langle \delta \rho\rangle=0$. The first one is not true when we have large $M_s$ (See the correction of VDA in \S \ref{sc:SVDA}), and the second one is not true if we are having insufficient statistical sampling (See the discussion of statistical sampling from \cite{YL17a}).

\begin{figure*}[ht]
  \centering
  \includegraphics[width=0.49\textwidth]{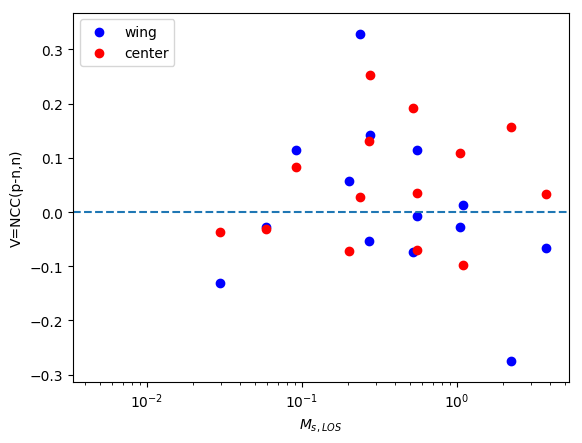}
  \caption{\label{fig:V} The correlation of $\langle p_d p_v\rangle$ as a function of $M_{s,LOS} = v_{LOS}/c_s$ for center channel $(v=v_{peak})$ and wing channel $(v=v_{peak}-\delta v_{los})$ using the simulation set "beta" from \cite{LY18a}.
 }
\end{figure*}

To proceed, we have to employ the numerical simulations and see how $\langle p_d p_v \rangle$ varies as a function of $M_s$. To start with we have to compute the channel map $p$ from Eq.\ref{eq:rho_PPV} and the {true velocity caustics} map $n$ by setting $\rho=\text{const}$ in Eq.\ref{eq:rho_PPV}. We then compute the term $NCC(p-n,n)=\langle (p-n)n\rangle$ as from definition in \S \ref{sec:la} that $p_d = p-n$ and $p_n = n$, and NCC is a normalized correlation (See Eq.\ref{eq:NCC}). Fig.\ref{fig:V} shows the relation of $\langle p_d p_v \rangle$ as a function of the line of sight sonic Mach number $M_{s,LOS} = v_{LOS}/c_s$. We compute $\langle p_d p_v \rangle$ for both the center and the wing channels. From Fig.\ref{fig:V} we see that the term $\langle p_d p_v \rangle$ is generally non-zero especially when $M_s\gg 1$, regardless of whether one picks the center or the wing. However, it is worth noting that despite the term $\langle p_d p_v \rangle$ is non-zero in supersonic turbulence, we still have surprisingly good extraction result in the main text (See Fig.\ref{fig:illus}) for the velocity caustics.

\section{Kernel of velocity channels}

\label{secap:kernel}
\begin{figure*}[ht]
  \centering
  \includegraphics[width=0.98\textwidth]{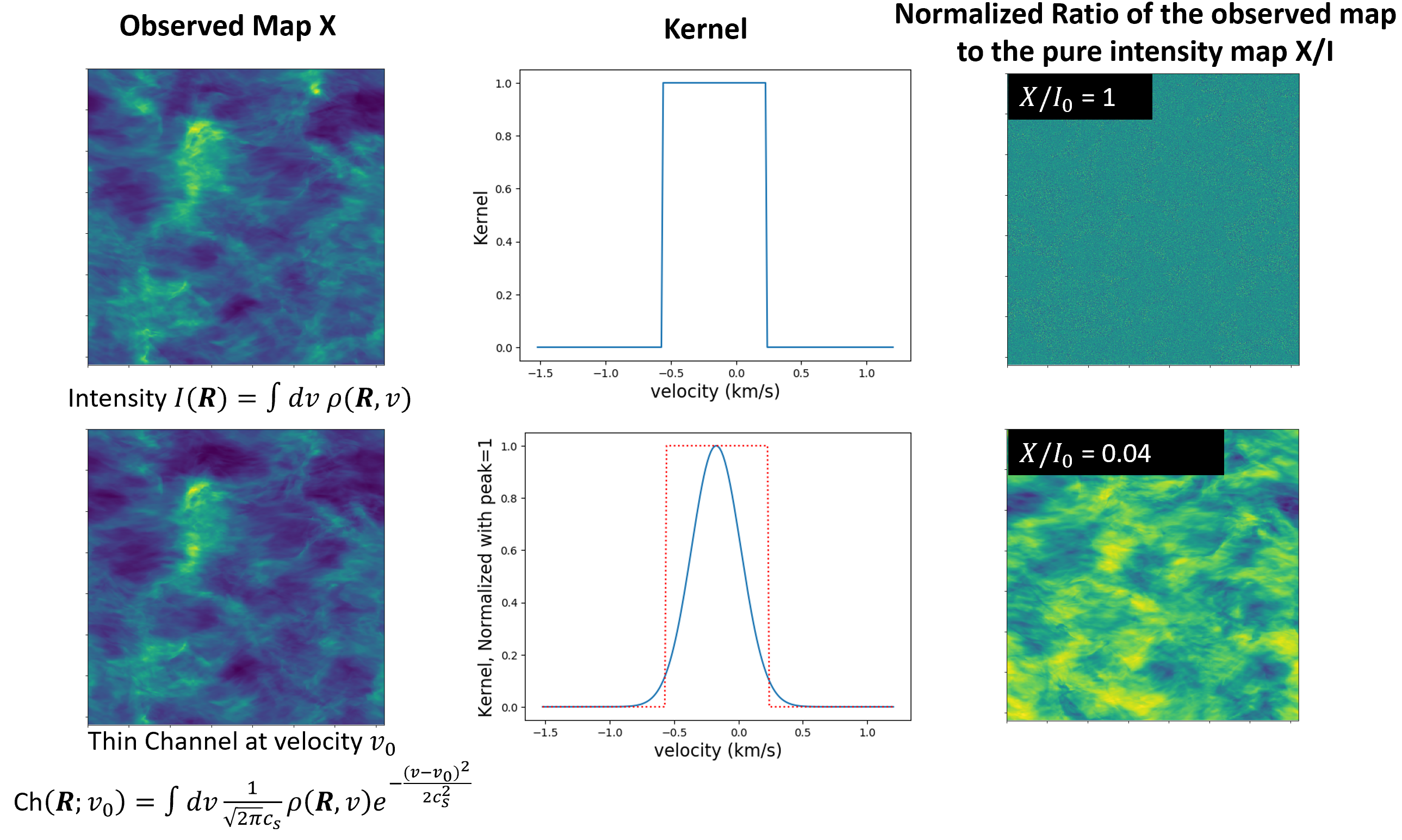}
  \caption{\label{fig:illus_b} An illustration on how the intensity map kernel should be different from that of the channel map kernel in a realistic MHD simulation. From the left column: The intensity and the thin channel map from the simulation; the middle column: the respective shape of the kernel. We can see that the thin channel kernel is intrinsically different from the full intensity kernel; the right column: the ratio of intensities between the observed map and the total intensity map.
 }
\end{figure*}

A fundamental concept throughout this paper is the {\it effective shape of the observational kernel} in the velocity space (See Fig.\ref{fig:illus_b}). This discussion of the observational kernel shape is crucial in understanding the differences between the physical scenarios in \citetalias{LP00} and \cite{susan19}, and the physical reality. In simple words, the effective kernel shape describes the relation of the observed maps and the unbroadened velocity channels, according to Eq.\ref{eq:rhov_PPV}. For instance, the kernel to produce the column density map is a squared top-hat over all channels, while the kernel to produce the thick velocity channel map is a Gaussian function with large upper and lower limits. Considering the kernel's shape is crucial in determining what is contained in a certain observed map. ({See also Fig.\ref{fig:cartoons}}).

\section{What gradients are we exactly computing in velocity channels?}

\label{secap:origin}
There were a number of concerns on whether the gradients of velocity channels are actually "velocity gradients" in some of {the recent publications} (See e.g., \cite{susan19} for their discussion section). We shall make a simple deviation in this section to show that the gradients of velocities purely define the gradients directions of velocity channels within the velocity channel width {\it as long as the channel is thin}. We shall start with Eq.\ref{eq:rhov_PPV} and explicitly write out the spatial and spectral dependencies of the channel map by assuming the x-axis is the LOS direction:
\begin{equation}
\begin{aligned}
p({\bf X}=(y,z),v_x({\bf X});\Delta v) \propto \int_{v_x-\Delta v/2}^{v_x+\Delta v/2} dv f(v) W(v) \exp\left(-\frac{(v-v_{center})^2}{2c_s^2}\right)
\end{aligned}
\end{equation}
where $f(v)$ is the probability distribution function (PDF) for velocity $v$, and $W(v)$ is the thermal kernel (See \S\ref{secap:kernel}).

We would consider the isothermal case but the result can be easily generalized to the non-isothermal cases. The spatial gradient can be derived by the Fundamental theorem of Calculus setting $v_{center}=0$ without loss of generality, also we will put $W(v) = 1$ for simplicity:
\begin{equation}
\begin{aligned}
\nabla_{\bf X} p({\bf X},v_x({\bf X}),\Delta v) &\propto \nabla_{\bf X} v_x \Big[f(v_x({\bf X})+\Delta v/2)\exp\left(-\frac{(v_x({\bf X})+\Delta v/2)^2}{2c_s^2}\right)-f(v_x({\bf X})-\Delta v/2)\exp\left(-\frac{(v_x({\bf X})-\Delta v/2)^2}{2c_s^2}\right)\Big]
\end{aligned}
\end{equation}

which we can see that the gradients of a velocity channel at velocity $v_x$ is actually given by the gradients of the velocity itself with a complicated weighting parameter. There is an ambiguity in {the above equation} since the $\nabla v_x$ here actually represents the gradients of velocity {\bf within the velocity range $v_x\pm \Delta v/2$ only}. To proceed we assume $\Delta v \ll c_s$ (which is an extra condition for "thin" channel for this paper, which we shall call it thermally thin in the moment) and thus an expansion is doable:

\begin{equation}
\begin{aligned}
\nabla_{\bf X} p({\bf X},v_x({\bf X}),\Delta v) &\propto \nabla_{\bf X} v_x\Big|_{v_x\in[v_x-\Delta v/2,v_x+\Delta v/2]} \exp\Big(-\frac{v_x^2({\bf X})}{2c_s^2}\Big) \left[\frac{\partial f}{\partial v}\Big|_{v=v_x({\bf X})}\Delta v-f(v_x({\bf X}))\Big(-\frac{v_x\Delta v}{c_s^2}\Big)\right]
\end{aligned}
\end{equation}

One can see that the formulation above actually works even for $f(v)$ being the {\it true density} PDF. As a well known fact that if we assume $f$ to be a Gaussian : $f\sim \exp(-v^2/2\delta^2)$ with $\delta$ to be the linewidth (usually $\delta \sim \sqrt{\delta v^2+2c_s^2}$), then the above expression gives

\begin{equation}
\begin{aligned}
\nabla_{\bf X} p({\bf X},v_x({\bf X}),\Delta v) &\propto v_x\Delta v\nabla_{\bf X} v_x\Big|_{v_x\in[v_x-\Delta v/2,v_x+\Delta v/2]} \exp\Big(-\frac{v_x^2({\bf X})}{2c_s^2}\Big) f(v_x({\bf X}))\left[\frac{1}{c_s^2}+\frac{1}{\delta^2}\right]
\end{aligned}
\end{equation}

The key here is that, the gradients of velocity channels are actually defined by the map that contains the velocity pixel values between $[v_0-\Delta v/2,v_0+\Delta v/2]$ multiplied with two exponent factors. Observe that the gradient operator can actually be regrouped:

\begin{equation}
\begin{aligned}
\nabla_{\bf X} p({\bf X},v_x({\bf X}),\Delta v) &\propto \Delta v\left(1+\frac{c_s^2}{\delta v^2+2c_s^2}\right) f(v_x({\bf X})) {\nabla_{\bf X} \exp\Big(-\frac{v_x^2({\bf X})}{2c_s^2}\Big)}
\end{aligned}
\label{eq:eq24}
\end{equation}
The important thing is that it is last factor $\nabla_{\bf X} \exp\Big(-\frac{v_x^2({\bf X})}{2c_s^2}\Big)$ that gives the direction of gradients, and Eq.\ref{eq:eq24} works for all $v$.
The result here indicates that the orientation of gradients in thermally thin velocity channels is purely defined by the gradients of the velocity pixel values between $[v_0-\Delta v/2,v_0+\Delta v/2]$, with the amplitude of the gradients be related to the PDF $f$ and also a number of extra constant factors.

\section{Constant density Velocity Centroid}

\label{secap:centroid}

{\citeauthor{2003ApJ...592L..37L}(\citeyear{2003ApJ...592L..37L}, see also \citealt{EL05,2015ApJ...814...77E})} discussed the properties of velocity centroid and the gradients of which are {extensively} applied in observations (See \cite{YL17a} and works from the same authors). However, the realistic velocity centroid computed by the following formula
\begin{equation}
  C({\bf X}) =\frac{\int dz \rho({\bf X},z)v({\bf X},z)}{\int dz \rho({\bf X},z)}
  \label{eq:centroid}
\end{equation}
has a density contribution both in the numerator and the denominator. Statistical analysis on velocity centroid, e.g., \cite{KLP16}, {often needs to assume that density is constant in order to study the properties of the velocity field in such observables.} The weighting of density in velocity centroid also makes it a less favorable observable than thin velocity channels in terms of tracing magnetic fields with gradients, since the latter is contains a higher {portion} of velocity contributions\footnote{This statement is given quantitatively from our knowledge according to the alignment measure of the gradients of centroid and channel when compared to the local B-field, see \cite{LY18a}.}.

With the availability of velocity caustics $p_v$ using VDA, the current study allows one to construct the so-called velocity-weighted caustics {\it in sub-sonic turbulence} by the following formula:
\begin{equation}
  V_{n}({\bf X}) = \int dv v p_v({\bf X},v) 
  \label{eq:C_VDA}
\end{equation}

\begin{figure*}[th]
  \centering
  \includegraphics[width=0.67\textwidth]{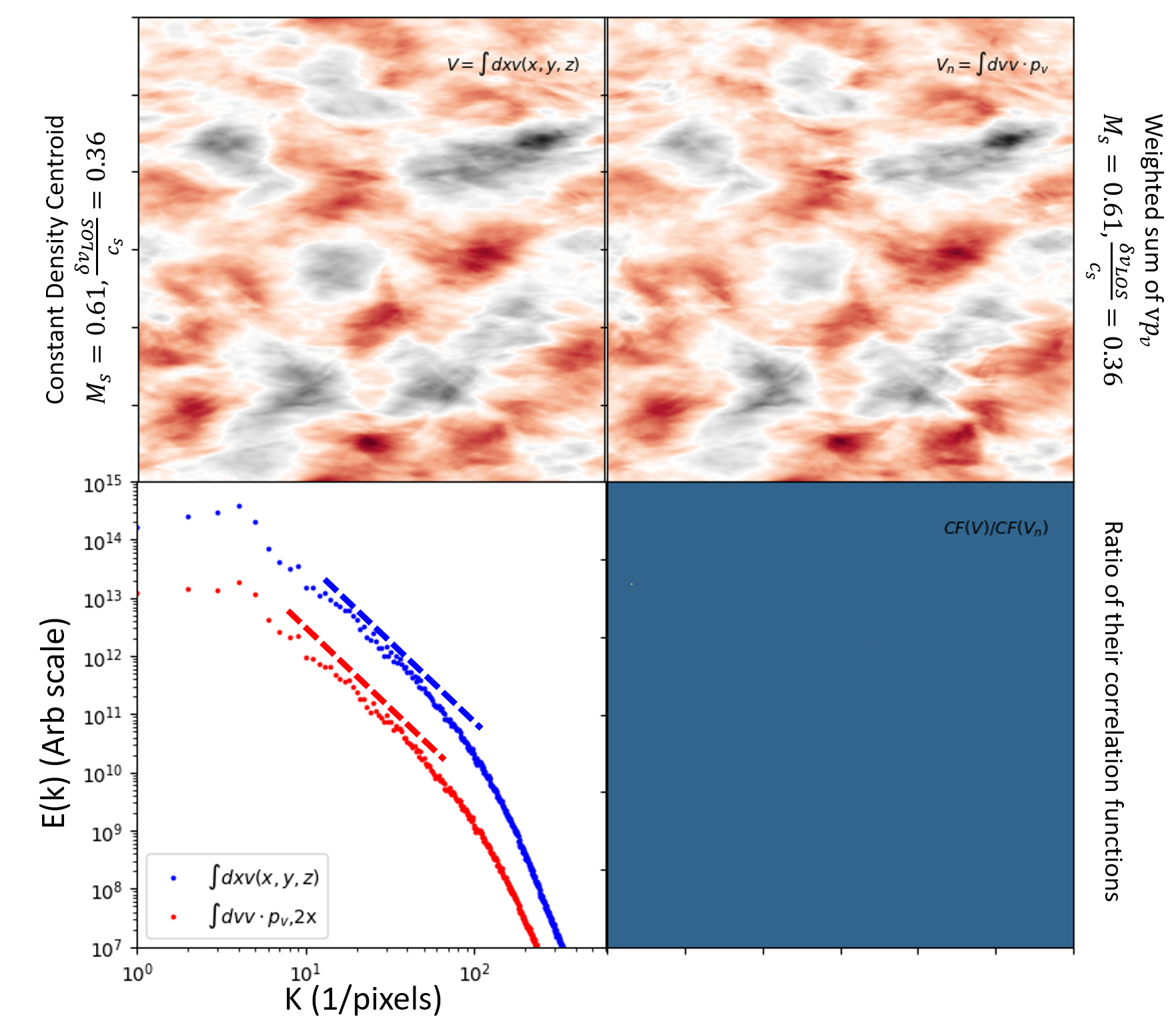}
  \caption{\label{fig:centroid} A set of figures comparing the structure and statistics of the true constant density centroid and the observationally available velocity-weighted caustics. Top row: structures of the true constant density centroid (top left) and velocity-weighted caustics (Top right). Lower left: The spectra of the true constant density centroid $V$ (blue) and the velocity-weighted caustics $V_n$ (red, shifted downward with a constant factor for better visualization). The dash lines are drawn for assisting the readers to locate the slopes of the inertial range of the spectra, which these lines have the same slope. Lower right: {The ratio of the two maps $V/V_n$ which} reflects the spatial dependence of $R$ (Eq.\ref{eq:R}). {As we can see from this panel $V/V_n$} is obviously a constant map.
 }
\end{figure*}

Notice that in the case of constant density, the centroid is simply the sum of the LOS velocity:
\begin{equation}
  V = \frac{1}{L_z}\int dz v({\bf X},z )
  \label{eq:C_real}
\end{equation}
where $L_z$ is the line of sight depth. We show the structures of the two {quantities from Eqs.\ref{eq:C_VDA},\ref{eq:C_real}} in the top row of Fig.\ref{fig:centroid} and it is obvious that they look almost exactly the same. To further test whether their statistical properties are the same, we plot the spectra of the two maps in the lower left panel of Fig.\ref{fig:centroid} and their ratio of correlation functions in the lower right panel of Fig.\ref{fig:centroid} computed by:
\begin{equation}
  R({\bf K}) = \frac{|\mathcal{F}\{V\}|^2}{|\mathcal{F}\{V_n\}|^2}
  \label{eq:R}
\end{equation}
The expression $R$ is a function of position, and will only be a constant if the two maps are exactly the same. It is obvious from the lower left corner of Fig.\ref{fig:centroid} that the spectra of both Eq.\ref{eq:C_VDA} and \ref{eq:C_real} are having similar structures, with the inertial ranges of them (the dash lines) carrying the same slope. As for the correlation function ratio (lower right corner of Fig.\ref{fig:centroid}) the value is basically a constant, indicating that the structures of the $V$ and $V_n$ maps are basically the same.

\section{Supersonic Velocity Decomposition Algorithm}
\label{sc:SVDA}

To obtain a unified decomposition method for both subsonic and supersonic turbulence, one has to make changes to the properties that we listed in \S \ref{sec:la} and proceed with the supersonic turbulence. {Notice that the modification in this section can also be used in some other cases where turbulence is involved, for example, situations where we have strong shocks or self-gravity. }

\begin{figure}[th]
  \centering
  \includegraphics[width=0.49\textwidth]{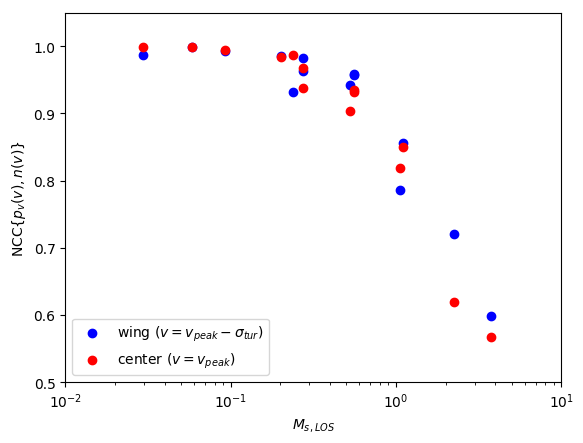}
  \caption{\label{fig:SVDA} A figure showing the variation of the NCC (Eq.\ref{eq:NCC}) between the decomposed $\hat{p}_v$ and the pure velocity caustics $n$, the latter being obtained by setting $\rho=\text{const}$ in Eq.\ref{eq:rho_PPV}, as a function of the ratio of line-of-sight velocity to sonic Mach number $M_{s,LOS}=v_{LOS}/c_s$ using the simulations "beta" listed in \cite{LY18a}.}
\end{figure}

Before that let's first recall the three properties (\S \ref{sec:la}):
\begin{enumerate}
  \item $\langle p_d p_v \rangle=0$;
  \item $p_v=0$ when $\Delta v\rightarrow \infty$;
  \item $p_d \sim I$ when $c_s \gg \delta v_{LOS}$
\end{enumerate}

Even with the simple recipe in \S \ref{sec:la}, we can see from Fig.\ref{fig:SVDA} that the decomposed $p_v$ correctly traces the true velocity caustics $n$ produced by synthesizing the velocity channel with a constant density term in terms of the NCC value (Eq.\ref{eq:NCC}). In fact even when $M_{s,LOS} = \delta v_{los}/c_s \sim 5$ the NCC value between $p_v$ and $n$ is still above 0.5 for both center and the wing channels, meaning the our approach in \S \ref{sec:la} actually works very well even in supersonic media. To further improve the method, we have to recall the properties of supersonic turbulence. 

\subsection{Vector space concept in channel maps}
\label{subsecap:vs}
Here we are using the "Vector Space" language. That means, the velocity channels can be written as a sum of:
\begin{equation}
  p(v) = A \hat{p}_d + B\hat{p}_v
  \label{eq:as}
\end{equation}
where $\hat{p}_{d,v} = p_{d,v}/\sigma_{p_{d,v}}$. Property (1) gives orthogonality. Property (2) means sum of $B(v)\hat{p}_v(v)$ along v is zero. Property (3) tells us what is $\hat{p}_d$ for all v, and it is a constant of v. To proceed we have to formally introduce the vector space concept {for velocity channel decomposition. For a } certain {statistical area} $A$ {and given} a function $x({\bf X},v)$ where ${\bf X}\in A \subset \mathcal{R}^2$, we can define the following operations in $(A, \cdot =\langle ...\rangle)$:
\begin{enumerate}
\item The "amplitude" of $x({\bf X})$ in a certain area $A$ is $\sigma_x$.
\item The "unit vector" of $x({\bf X})$ is:
$$
\hat{x} = \frac{x-\langle x \rangle}{\sigma_x}
$$

\item The inner product of two "vectors" are:
$$
x \cdot y = \langle (x - \langle x\rangle)(y - \langle y\rangle)\rangle
$$
\end{enumerate}

\subsection{Modification on property 1: {Correction when $\langle p_d p_v\rangle\neq 0$}}
\label{subsecap:nonortho}
From the discussion in \S\ref{sec:apb}, property (1) is not correct in the case of supersonic media. In fact $\langle p_d p_v\rangle$ will deviate from zero for a magnitude of $\pm 0.15$ but numerical experiment suggests that the number is never going over $\pm 0.3$ (See Fig.\ref{fig:V}.) As a result, we can write 
\begin{equation}
\hat{p}_d = U\hat{z} + V \hat{p}_v
\end{equation}
where $\langle \hat{z} \hat{p}_v\rangle=0$ and $U,V$ are some constants that relies on numerical calculations to estimate. Notice that $U^2+V^2=1$, that means we only need to estimate $V$. Incidentally, the constant $V$ is given by Fig.\ref{fig:V} above.

In matrix form we can write
\begin{equation}
\begin{aligned}
p(v) &= \left[\begin{matrix}
U & 0 \\ V & 1
\end{matrix}\right]\left[\begin{matrix}
A \\ B
\end{matrix}\right]\\
&=UA\hat{z} + (VA+B)\hat{p}_v\\
\end{aligned}
\end{equation}

Then with some algebra we can obtain the exact form of $p_v$ and $p_d$ for the case when $\langle p_v p_d\rangle\neq0$:
\begin{equation}
\begin{aligned}
p_d &= \Big\langle (p-\langle p\rangle)\hat{z}\Big\rangle (\hat{z} + \frac{V}{U} \hat{p}_v)\\
p_v &= \Big\langle (p-\langle p\rangle)(\hat{p}_v-\frac{V}{U}\hat{z})\Big\rangle \hat{p}_v
\end{aligned}
\label{eq:SVDA}
\end{equation}
{where the value of V is given by Fig.\ref{fig:SVDA}.}

\subsection{Can we modify property 2?}
{Readers might wonder whether we can modify property 2 in our deviation. Unfortunately, property 2 is subjected to the deviation of \citetalias{LP00} that the caustics will not appear when we {consider} the thick channel. We shall not discuss the possible modification of property 2 in the current paper. }

\subsection{Modification on property 3: {The velocity channel differential}}
\label{secap:log}

Here we describe a brand new method based on VDA in supersonic turbulence, which gives surprisingly good in wing and center channels. {Moreover, the method is more accurate than the method that we developed in the main text (\S \ref{sec:la}), with a cost that we do not obtain the velocity caustics directly from observations plus being more noise dependent}. To start with, we shall consider the observationally available function:
\begin{equation}
  V({\bf X},v,\Delta v) = -c_s^2\frac{\partial \log p ({\bf X},v,\Delta v)}{\partial v}
  \label{eq:partial_centroid}
\end{equation}
Notice that when $\Delta v\rightarrow \infty$, from Eq.\ref{eq:rhov_PPV} 
\begin{equation}
\begin{aligned}
   V({\bf X},v,\Delta v\rightarrow\infty) &= -c_s^2\frac{\int dv' \rho(v') \frac{-(v'-v)}{c_s^2}\exp\left(-\frac{(v'-v)^2}{2c_s^2}\right)}{\int dv' \rho(v') \exp\left(-\frac{(v'-v)^2}{2c_s^2}\right)}\\
   &= C({\bf X})
\end{aligned}
\end{equation}
which is basically the normalized velocity centroid (Eq.\ref{eq:centroid}). We shall name Eq.\ref{eq:partial_centroid} to be the {\it partial velocity centroid} (See also the discussion of the centroid at \S\ref{secap:centroid}) at velocity $v$ and velocity channel width $\Delta v$. Readers should notice that Eq.\ref{eq:partial_centroid} is simply the centroid function that integrates within the range $[v-\Delta v/2,v+\Delta v/2]$.

\begin{figure*}[bth]
  \centering
  \includegraphics[width=0.67\textwidth]{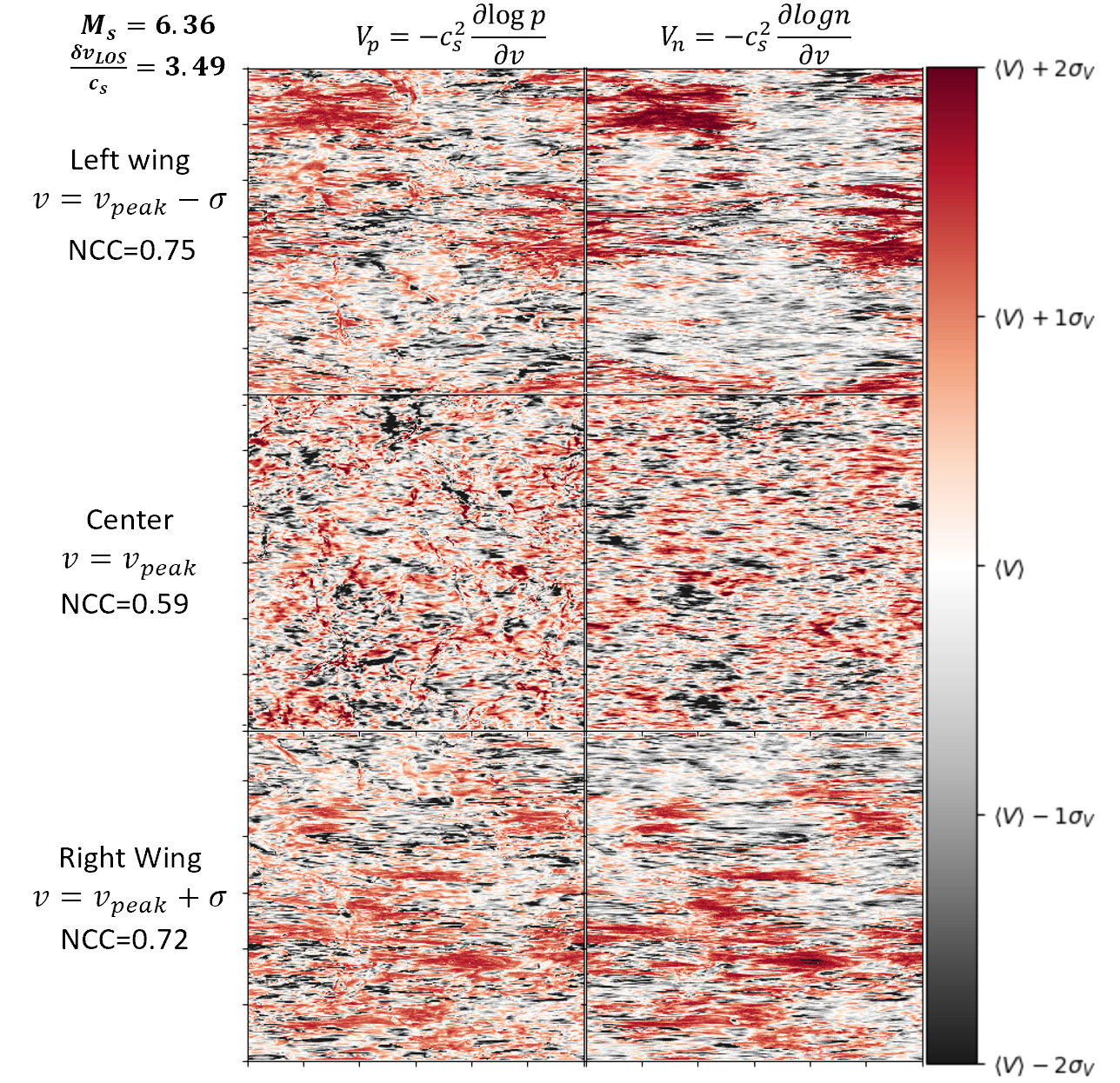}
  \caption{\label{fig:SDAillus} A set of figures showing how $V_p$ and $V_n$ behave in the wing channel (top and bottom rows) and the center channel (middle two) in a supersonic simulation h0-1200 ($M_s=6.36$ and $\delta v_{LOS}/c_s=3.49$). The NCC between $V_p$ and $V_n$ are $0.75$ and $0.72$ for left and right wings respectively, while that for the center is $0.58$.}
\end{figure*}

\begin{figure*}[bth]
  \centering
  \includegraphics[width=0.49\textwidth]{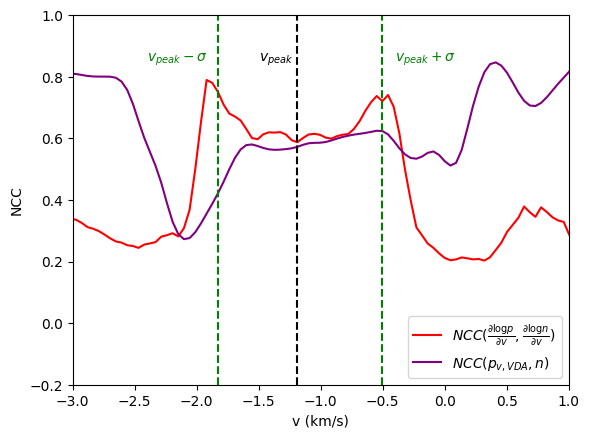}
  \includegraphics[width=0.49\textwidth]{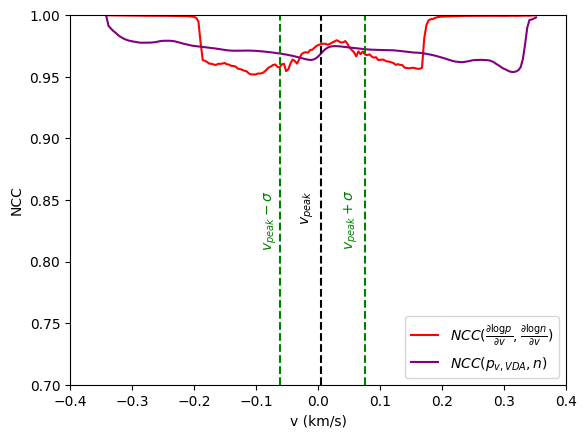}
  \caption{\label{fig:SDA_v2} The NCC variation curves of the pairs $V_p=\frac{\partial \log p}{\partial v}$, $V_n=\frac{\partial \log n}{\partial v}$ and $p_v,n$ as a function of $v$ for the supersonic case h0-1200 (left, $M_s=6.36$ and $\delta v_{LOS}/c_s=3.49$) and subsonic case "e5r3" (right, $M_s=0.61$ and $\delta v_{LOS}/c_s=0.36$). Notice we do not use the correction formulae (Eq.\ref{eq:SVDA}) but simply stick with the formulate in \S \ref{sec:la}.}
\end{figure*}

This special $V$ function has a unique behavior in the tracing of $p_v$ in supersonic media. Notice that if we assume $\Delta v \ll c_s \ll \delta v$ (the latter inequality comes from $M_s>1$) then the partial centroid (Eq.\ref{eq:partial_centroid}) is simply the partial velocity projection along the line of sight regardless of what density weights we are using:
\begin{equation}
  V({\bf X},v_0,\Delta v\ll c_s\ll\delta v) \sim \sum_{v\in[v_0-\Delta v/2,v_0+\Delta v/2]} v
\end{equation}
Therefore, we can compute $V$ from both with density weighting (denoted as $V_p$) and without density weighting (denoted as $V_n$) and use NCC (Eq.\ref{eq:NCC}) to see whether they are alike. Fig.\ref{fig:SDAillus} shows how $V_p$ and $V_n$ look in the wing and the center channel for the simulation h0-1200 ({See Tab\ref{tab:sim}}) which has a total $M_s=6.36$ and $\delta v_{LOS}/c_s=3.49$. We can see that they look pretty much alike. Notice the $V_n$ is purely velocity dominant. This means the $V_p$ terms is also very much dominated by velocity fluctuations. Furthermore, we can see from the left of Fig.\ref{fig:SDA_v2} that the NCC of the partial centroids to the caustics counterpart is generally higher than that of the $p_v$ to the caustics. Moreover, at $v=v_{peak}\pm\sigma$ ($\sigma=\delta v_{LOS})$ the differences of the NCC between the partial centroid to $p_v$ is the largest. This shows that the partial centroid method is a more robust method than the VDA method that we developed in the {main text in supersonic turbulence. However, note that the supersonic variant that we developed here is very sensitive to noise. The observation application of the supersonic method will be tested elsewhere.}

How can we understand the results physically? Notice that technically we can always write the partial centroid as the linear combination of $p_d$ and $p_v$ in different $v$ using the differential approximation and the knowledge that $V_p\approx V_n$::
\begin{equation}
  \begin{aligned}
  V_p({\bf X},v) \approx V_n({\bf X},v)&\approx \frac{c_s^2}{p_v({\bf X},v)\Delta v} \Big[p_v({\bf X},v) - p_v({\bf X},v+\Delta v)\Big]\\  
  \rightarrow p_v(v+\Delta v)&\approx p_v({\bf X},v)\left[1-\frac{\Delta v}{c_s^2}V_p({\bf X},v)\right]
  \end{aligned}
\end{equation}
We can see that as long as we have an initial guess of $p_v$ at some velocity, we can extrapolate $p_v$ for different velocities. Notice that in the extreme wing channels, the logarithm of velocity fluctuations are purely velocity like, then we can successively produce the eigenmaps of $p_v(v)$ and use Eq.\ref{eq:SVDA} to compute also $p_d$ out. Readers might wonder whether the parameter $V_p$ could be used in subsonic turbulence. We show the same NCC dependence curve for subsonic turbulence in the right of Fig.\ref{fig:SDA_v2}. One can see that the performance of $V_p$ is comparable to that of $p_v$ in tracing the corresponding velocity caustics structures. This indicates the partial centroid $V$ is a versatile parameter in studying caustics in both subsonic and supersonic turbulence.

As a remark, incidentally, we also report that the logarithm of the channel map $\log(p)$ itself also performs very well in the wings in tracing its velocity caustics equivalent $\log(n)$. There is a theoretical and numerical finding (e.g., \citealt{2007ApJ...658..423K}) that using the logarithm of channel maps suppress the high-density fluctuations, and $\log\rho$ is also a natural parameter in analyzing the linear waves in MHD turbulence (See \citealt{2003matu.book.....B}). However, we found that a good correlation usually only occurs when $|v-v_{peak}|>1.5\sigma$. Nevertheless, while we can postulate that the channel map or the $V$ parameter (Eq.\ref{eq:partial_centroid}) is the linear combination of density and velocity contributions, the logarithm of velocity channels do not. Therefore, based on the formulation of this work, there still needs a way to understand why the logarithm of channels could trace those far wing caustics in observations. Incidentally, the $V\propto -d(\log p)/dv$ parameter (Eq.\ref{eq:partial_centroid} traces the caustics counterpart ($V_n \propto -d(\log n)/dv$) very well in the ranges of velocities $|v-v_{peak}|<1.5\sigma$ (See Fig.\ref{fig:SDA_v2}) exactly opposite to the ranges of applicability of the $\log p$ method, which means they can be complementary to each other in terms of observational applications.

\section{The Velocity Decomposition Algorithm in heavy molecular tracers lines}
\label{secap:vda_emissions_absorptions}
In the main text, we assume the observed velocity channel's thermal kernel is related to the system's intrinsic temperature. This is true for neutral hydrogen, since the emission agent is the major species of the turbulence system. However, in realistic observation, the thermal speed of the observed species is different from the ambient thermal speed of the turbulence system. For example, suppose we have turbulent partially ionized gases observed in channel maps, where we recognize that the neutrals are lighter (e.g. $H_2$) while the ions are heavier ($HCO^+$). Under these circumstances, the effective sonic speed for the emission lines of these two species would be inversely related to the mean molecular weight of the observed species since $c_s^2 = dP/d\rho = \gamma P/\rho \propto \mu^{-1}$ where $\mu$ is the mean molecular weight. For example, if the line of sight sonic Mach number of a certain turbulence system is {$v_{los}/c_s\sim 0.5$ } and we have the mean molecular weight of ions to be 10 times that of neutrals, then the effective line of sight sonic Mach number of the ions would be then {$v_{los}/c_s \sim 5.0$}. This means that while the neutrals' velocity channels are suffered from strong thermal broadening, the thermal broadening effect can be mostly ignored in the ions' channel map. 

\begin{figure*}[bth]
  \centering
  \includegraphics[width=0.99\textwidth]{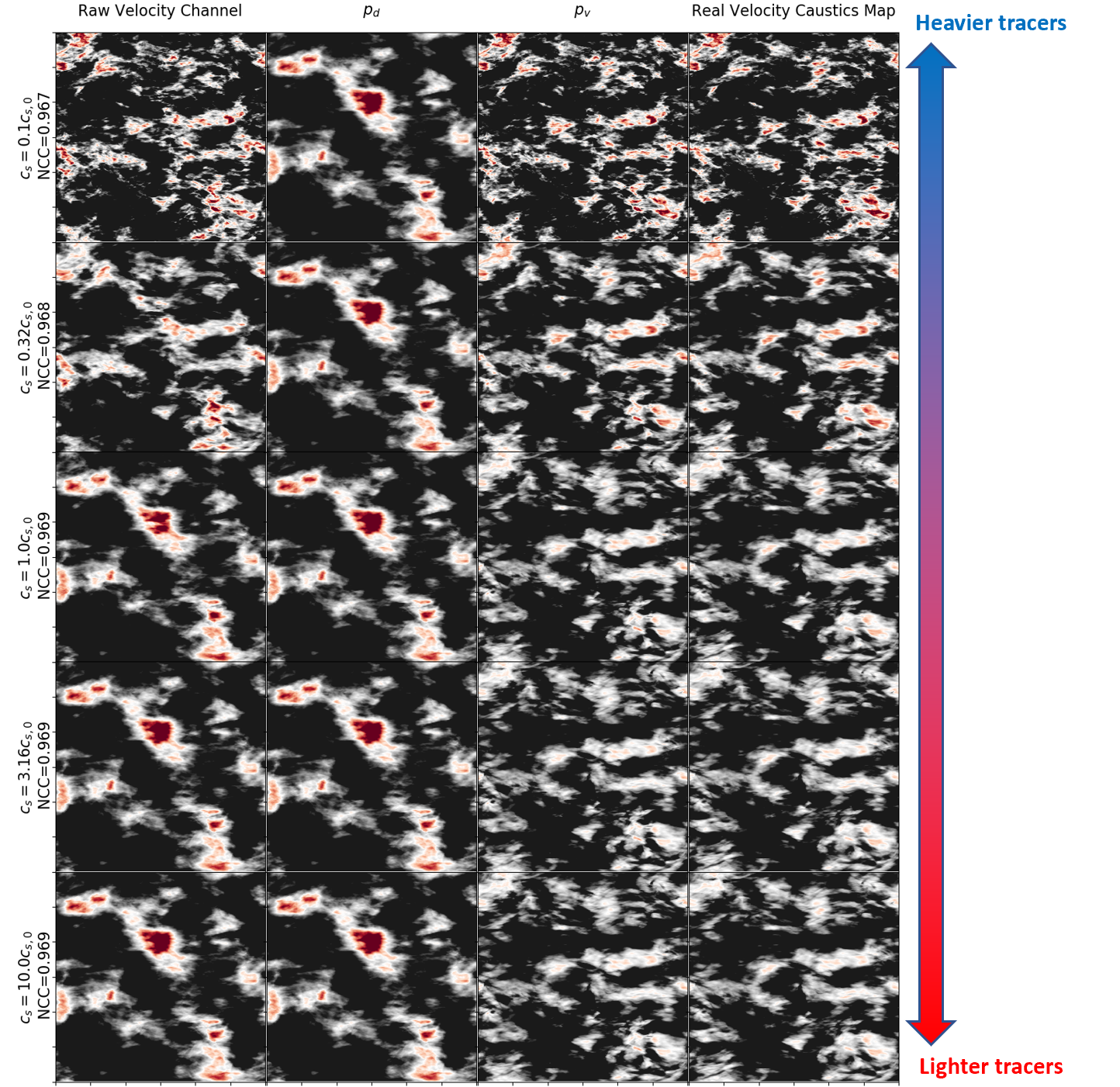}
  \caption{\label{fig:emission_illus} A set of figures showing how the structure of the channel maps from the $v=0$ channel of the simulation "e5r3" ($M_s=0.61, v_{los}/c_{s,0}$) will change as a function of the effective thermal speed $c_s$. Here we represent $c_s$ as multiples of $c_{s,0}$, the intrinsic thermal speed of the simulation as we use in the main text. From the left column: Raw velocity channel $p$, decomposed $\rho_d$, decomposed $\rho_v$ and the real velocity channel map $n$. The corresponding NCC value between $\rho_v$ and $n$ for each row is labeled in the left. The color bar is drawn between $[p,p+3\sigma_p]$. }
\end{figure*}

From our discussion in the main text, it is rather evident that the line of sight sonic Mach number $v_{los}/c_s$ (not the intrinsic 3D Mach number, see \S \ref{sec:theory}) is a critical parameter affecting the performance of VDA. In this section, we would like to briefly study how the change of the effective thermal kernel width, which is inversely proportional to the molecular weight, would affect the performance of VDA. Fig.\ref{fig:emission_illus} shows how the change of the effective thermal kernel width will alter the performance of VDA in the center channel of "e5r3" in the main text. 

For the sake of completeness, we consider both heavier and lighter tracers, which corresponds to a decrease and increase of $c_s$, respectively. We can see from Fig.\ref{fig:emission_illus} that when the molecular tracers are heavier than the fluid in bulk (i.e., $c_s/c_{s,0}<1$, the structure of the raw velocity channel will be more fragmented and vice versa for the lighter tracers. 

In Fig.\ref{fig:emission_illus} we clearly see that the structure of observed maps gets more and more closer to the maps of pure velocity caustics as the thermal broadening of the species decreases. At the same time, it resembles more the density map as the thermal broadening related increases. This is a direct consequence of the \citetalias{LP00} theory, which predicts that the thermal width is similar to the width of the velocity channels. The theory also predicts that the effect of velocity caustics decreases with the increase of the velocity channel width (See Eq.\ref{eq:effective_dv}). We note that the role of the heavy species with lower thermal broadening can also be played by clumps of the colder gas that is being moved by, the hotter gas in the multi-phase medium. This was the model of the multi-phase HI that was adopted in \citetalias{LP00} to justify the application of the theory to the cold fraction of interstellar HI. Our multi-phase numerical simulations presented in the main body of the paper support this model of HI.

One might wonder why there are still density fluctuations in the heavy tracer limits as the latter should correspond to the limiting case discussed in \citetalias{LP00}. It is true that in the case of $c_s\rightarrow 0$, there should be only velocity caustics fluctuations in velocity channels, but since realistic observations always have $c_s>0$, the density fluctuations would still exist despite being small in magnitude. Not to mention, the relative fluctuation of density and velocity fluctuations is always a function of velocity channel position (See \S \ref{sec:la}, \ref{sec:test}). Neither velocity nor density fluctuations are negligible in real spectroscopic data. Yet, with our development of the VDA, it is always possible to separate the density and velocity fluctuations in observations and thus quantify their relative importance. To better quantify the performance of VDA under this scenario, we use the NCC (Eq.\ref{eq:NCC}) again as in the main text. 

The corresponding value of $NCC(n,p_v)$ for each row is listed on the left-hand side of the row. We observe that while the raw velocity channel structure is changing as a function of $c_s$, the decomposed $p_v$ is almost an exact match with the real velocity caustics map. From previous sections, we know that the intrinsic line of sight sonic Mach number is an essential parameter in affecting the performance of VDA. However, there is no drop of performance from the change of the effective thermal kernel width due to the change of the molecular species. Therefore, we conclude that the VDA technique can apply to molecular tracer emission lines.

It is also worth noting that the caustics map structure is still a function of thermal kernel width, as we discussed in the main text. Readers might wonder what physically the caustics represent when $c_s$ changes. Indeed, they are still pure velocity structures, but being integrated for a different effective channel width $\Delta v_{eff}^2 \sim \delta v^2 + 2c_s^2$ (See Eq.\ref{eq:thin_criterion}). A larger $c_s$ would represent a larger collection of velocity structures along the line of sight. 

\begin{thebibliography}{}
\providecommand\natexlab[1]{#1}
\providecommand\JournalTitle[1]{#1}

\bibitem[{Armstrong {et~al.}(1995)Armstrong, Rickett, \& Spangler}]{Armstrong1995ElectronMedium}
Armstrong, J.~W., Rickett, B.~J., \& Spangler, S.~R. 1995,\href{http://dx.doi.org/10.1086/175515}{\JournalTitle{The Astrophysical Journal}, 443, 209}

\bibitem[Beresnyak et al.(2005)]{2005ApJ...624L..93B} Beresnyak, A., Lazarian, A., \& Cho, J.\ 2005, \apjl, 624, L93 

\bibitem[Beresnyak \& Lazarian(2019)]{2019tuma.book.....B} Beresnyak, A., \& Lazarian, A.\ 2019, Turbulence in Magnetohydrodynamics


\bibitem[Biskamp(2003)]{2003matu.book.....B} Biskamp, D.\ 2003, Magnetohydrodynamic Turbulence, by Dieter Biskamp, pp.~310.~ISBN 0521810116.~Cambridge, UK: Cambridge University Press, September 2003., 310 


\bibitem[Burton et al.(2001)]{2001A&A...369..616B} Burton, W.~B., Braun, R., \& Chengalur, J.~N.\ 2001, \aap, 369, 616

\bibitem[Burkhart et al.(2013)]{2013ApJ...770..141B} Burkhart, B., Lazarian, A., Goodman, A., et al.\ 2013, \apj, 770, 141



\bibitem[Chandrasekhar \& Fermi(1953)]{CF53} Chandrasekhar, S., \& Fermi, E.\ 1953, \apj, 118, 113 


\bibitem[Chepurnov \& Lazarian(2009)]{2009ApJ...693.1074C} Chepurnov, A. \& Lazarian, A.\ 2009, \apj, 693, 1074. doi:10.1088/0004-637X/693/2/1074

\bibitem[{Chepurnov \& Lazarian(2010)}]{Chepurnov2010ExtendingData}
Chepurnov, A., \& Lazarian, A. 2010, \href{http://dx.doi.org/10.1088/0004-637X/710/1/853}{\JournalTitle{The
 Astrophysical Journal, Volume 710, Issue 1, pp. 853-858 (2010).}, 710, 853}
 
 \bibitem[Chepurnov et al.(2015)]{2015ApJ...810...33C} Chepurnov, A., Burkhart, B., Lazarian, A., et al.\ 2015, \apj, 810, 33

 \bibitem[{Cho \& Lazarian(2002)}]{CL02}
Cho, J., \& Lazarian, A. 2002, \href{http://dx.doi.org/10.1103/PhysRevLett.88.245001}{\JournalTitle{Physical
 Review Letters, vol. 88, Issue 24, id. 245001}, 88}
 
\bibitem[{Cho \& Lazarian(2003)}]{CL03} \mnras, \ 2003, 345, 325
 

\bibitem[Clark et al.(2014)]{Clark14} Clark, S.~E., Peek, J.~E.~G., \& Putman, M.~E.\ 2014, \apj, 789, 82 

\bibitem[Clark et al.(2015)]{Clark15} Clark, S.~E., Hill, J.~C., Peek, J.~E.~G., Putman, M.~E., \& Babler, B.~L.\ 2015, Physical Review Letters, 115, 241302 

\bibitem[Clark et al.(2019)]{susan19} Clark, S.~E., Peek, J.~E.~G., \& Miville-Deschenes, M.-A.\ 2019, \apj, 874,171, arXiv:1902.01409

\bibitem[Clarke et al.(2018)]{2018MNRAS.479.1722C} Clarke, S.~D., Whitworth, A.~P., Spowage, R.~L., et al.\ 2018, \mnras, 479, 1722


\bibitem[Crutcher et al.(2010)]{C10} Crutcher, R.~M., Hakobian, N., \& Troland, T.~H.\ 2010, \mnras, 402, L64 


\bibitem[Davis(1951)]{1951PhRv...81..890D} Davis, L.\ 1951, Physical Review, 81, 890. doi:10.1103/PhysRev.81.890.2
 { 
\bibitem[Dib \& Burkert(2005)]{2005ApJ...630..238D} Dib, S. \& Burkert, A.\ 2005, \apj, 630, 238. doi:10.1086/431785

\bibitem[Dib et al.(2006)]{2006ApJ...638..797D} Dib, S., Bell, E., \& Burkert, A.\ 2006, \apj, 638, 797. doi:10.1086/498857

\bibitem[Dib et al.(2007)]{2007ApJ...661..262D} Dib, S., Kim, J., V{\'a}zquez-Semadeni, E., et al.\ 2007, \apj, 661, 262. doi:10.1086/513708

\bibitem[Dib et al.(2008)]{2008ApJ...678L.105D} Dib, S., Brandenburg, A., Kim, J., et al.\ 2008, \apjl, 678, L105. doi:10.1086/588608
}

\bibitem[{Draine(2011)}]{Draine2011PhysicsMedium}
Draine, B.~T. 2011, {Physics of the interstellar and intergalactic medium}
 (Princeton University Press), 540
 
\bibitem[Esquivel \& Lazarian(2005)]{EL05} Esquivel, A. \& Lazarian, A.\ 2005, \apj, 631, 320. doi:10.1086/432458

\bibitem[Esquivel et al.(2015)]{2015ApJ...814...77E} Esquivel, A., Lazarian, A., \& Pogosyan, D.\ 2015, \apj, 814, 77. doi:10.1088/0004-637X/814/1/77

\bibitem[{Goldreich \& Sridhar (1995)}]{GS95} Goldreich, P.~;Sridhar, S. 1995,  \href{http://dx.doi.org/10.1086/174600}{\JournalTitle{The Astronomical Journal}, 438, 763} 


\bibitem[Gonz{\'a}lez-Casanova, \& Lazarian(2019)]{2019ApJ...874...25G} Gonz{\'a}lez-Casanova, D.~F., \& Lazarian, A.\ 2019, \apj, 874, 25.


\bibitem[Hayes et al.(2006)]{2006ApJS..165..188H} Hayes, J.~C., Norman, M.~L., Fiedler, R.~A., et al.\ 2006, \apjs, 165, 188. doi:10.1086/504594


\bibitem[Haud \& Kalberla(2007)]{2007A&A...466..555H} Haud, U., \& Kalberla, P.~M.~W.\ 2007, \aap, 466, 555
 

\bibitem[Heyer et al.(2008)]{2008ApJ...680..420H} Heyer, M., Gong, H., Ostriker, E., \& Brunt, C.\ 2008, \apj, 680, 420-427 


\bibitem[Heiles \& Troland (2003)]{2003ApJ...586..1067H} Heiles, C., Troland, T.~H., \ 2003, \apj, 586, 1067 


\bibitem[Hopkins et al.(2012)]{2012MNRAS.421.3488H} Hopkins, P.~F., Quataert, E., \& Murray, N.\ 2012, \mnras, 421, 3488. doi:10.1111/j.1365-2966.2012.20578.x

\bibitem[Hsieh et al.(2019)]{2019ApJ...873...16H} Hsieh, C.-. han ., Hu, Y., Lai, S.-P., et al.\ 2019, \apj, 873, 16.

\bibitem[Hu et al.(2018)]{PCA} Hu, Y., Yuen, K.~H., \& Lazarian, A.\ 2018, \mnras, 480, 1333.

\bibitem[Hu et al.(2019a)]{survey} Hu, Y., Yuen, K. H., Lazarian V., et al. \ 2019, Nature Astronomy

\bibitem[Hu et al.(2019b)]{velac} Hu, Y., Yuen, K.~H., Lazarian, A., et al.\ 2019, \apj , arXiv:1904.04391.

\bibitem[Hu et al.(2019c)]{IGVHRO} Hu, Y., Yuen, K.~H., \& Lazarian, A.\ 2019, \apj, arXiv:1908.09488v1

\bibitem[Inoue \& Inutsuka(2009)]{2009ApJ...704..161I} Inoue, T. \& Inutsuka, S.-. ichiro .\ 2009, \apj, 704, 161. doi:10.1088/0004-637X/704/1/161

\bibitem[Inoue \& Inutsuka(2016)]{2016ApJ...833...10I} Inoue, T. \& Inutsuka, S.-. ichiro .\ 2016, \apj, 833, 10. doi:10.3847/0004-637X/833/1/10 

\bibitem[Kalberla \& Haud(2019)]{kalberla2019} Kalberla, P.~M.~W., \& Haud, U.\ 2019, \aap, 627, A112 


\bibitem[Kalberla \& Haud(2020)]{kalberla2020a} Kalberla, P.~M.~W., \& Haud, U.\ 2020, arXiv e-prints, arXiv:2003.01454
\bibitem[Kalberla et al.(2020)]{kalberla2020b} Kalberla, P.~M.~W., Kerp, J., \& Haud, U.\ 2020, arXiv e-prints, arXiv:2004.14630

\bibitem[Kandel et al.(2016)]{KLP16} Kandel, D., Lazarian, A., \& Pogosyan, D.\ 2016, \mnras, 461, 1227 
\bibitem[Kandel et al.(2017a)]{KLP17a} Kandel, D., Lazarian, A., \& Pogosyan, D.\ 2017, \mnras, 464, 3617 
\bibitem[Kandel et al.(2017b)]{KLP17b} Kandel, D.,
Lazarian, A., \& Pogosyan, D.\ 2017, \mnras, 470, 3103 


\bibitem[Kritsuk et al.(2017)]{2017NJPh...19f5003K} Kritsuk, A.~G., Ustyugov, S.~D., \& Norman, M.~L.\ 2017, New Journal of Physics, 19, 065003 



\bibitem[Kowal et al.(2007)]{2007ApJ...658..423K} Kowal, G., Lazarian, A., \& Beresnyak, A.\ 2007, \apj, 658, 423 


\bibitem[Kowal \& Lazarian(2010)]{2010ApJ...720..742K} Kowal, G. \& Lazarian, A.\ 2010, \apj, 720, 742. doi:10.1088/0004-637X/720/1/742

\bibitem[Koyama \& Inutsuka (2002)]{Koyama02} Koyama, H., \& Inutsuka, S.\ 2000, \apj, 532, 980

\bibitem[Lazarian (2009)]{2009fohl.book..357L} Lazarian, A.\ 2009, From the Outer Heliosphere to the Local Bubble, 357. doi:10.1007/978-1-4419-0247-4-29


\bibitem[Lazarian \& Esquivel(2003)]{2003ApJ...592L..37L} Lazarian, A., \& Esquivel, A.\ 2003, \apjl, 592, L37 
 
\bibitem[Lazarian \& Pogosyan(2000)]{LP00} Lazarian, A., \& Pogosyan, D.\ 2000, \apj, 537, 720 

\bibitem[Lazarian \& Pogosyan(2004)]{LP04} Lazarian, A., \& Pogosyan, D.\ 2004, \apj, 616, 943 

\bibitem[Lazarian \& Pogosyan(2006)]{LP06} Lazarian, A., \& Pogosyan, D.\ 2006, \apj, 652, 1348 

\bibitem[Lazarian \& Pogosyan(2008)]{LP08} Lazarian, A., \& Pogosyan, D.\ 2008, \apj, 686, 350


\bibitem[Lazarian \& Pogosyan(2012)]{LP12} Lazarian, A., \& Pogosyan, D.\ 2012, \apj


\bibitem[{Lazarian \& Vishniac(1999)}]{LV99} Lazarian, A., \& Vishniac, E.~T. 1999, \href{http://dx.doi.org/10.1086/307233}{\JournalTitle{The Astrophysical Journal, Volume 517, Issue 2, pp. 700-718.}, 517, 700}

\bibitem[Lazarian \& Yuen(2018a)]{LY18a} Lazarian, A., \& Yuen, K.~H.\ 2018, \apj, 853, 96 

\bibitem[Lazarian \& Yuen(2018b)]{LY18b} Lazarian, A., \& Yuen, K.~H.\ 2018, \apj, 865, 59 


\bibitem[Lazarian et al.(2017)]{Letal17} Lazarian, A., Yuen, K.~H., Lee, H., \& Cho, J.\ 2017, \apj, 842, 30 

\bibitem[Lazarian et al.(2018)]{LYH18} Lazarian, A., Yuen, K.~H., Ho, K.~W., et al.\ 2018, \apj, 865, 46.

\bibitem[Lazarian et al.(2020)]{2020arXiv200207996L} Lazarian, A., Yuen, K.~H., \& Pogosyan, D.\ 2020, arXiv:2002.07996



\bibitem[Li(2018)]{2018MNRAS.477.4951L} Li, G.-X.\ 2018, \mnras, 477, 4951. doi:10.1093/mnras/sty657


\bibitem[Lu et al.(2020)]{2020MNRAS.496.2868L} Lu, Z., Lazarian, A., \& Pogosyan, D.\ 2020, \mnras, 496, 2868. doi:10.1093/mnras/staa1570

\bibitem[McKee(1990)]{1990ASPC...12....3M} McKee, C.~F.\ 1990, The Evolution of the Interstellar Medium, 12, 3

\bibitem[McKee \& Ostriker(1977)]{1977ApJ...218..148M} McKee, C.~F. \& Ostriker, J.~P.\ 1977, \apj, 218, 148. doi:10.1086/155667


\bibitem[McKee \& Ostriker(2007)]{MO07} McKee, C.~F., \& Ostriker, E.~C.\ 2007, \araa, 45, 565 


\bibitem[Padoan et al.(2006)]{2006ApJ...653L.125P} Padoan, P., Juvela, M., Kritsuk, A., \& Norman, M.~L.\ 2006, \apjl, 653, L125 

\bibitem[Peek et al.(2018)]{stail} Peek, J.~E.~G., Babler, B.~L., Zheng, Y., et al.\ 2018, \apjs, 234, 2

\bibitem[Peek \& Clark(2019)]{2019ApJ...886L..13P} Peek, J.~E.~G., \& Clark, S.~E.\ 2019, \apjl, 886, L13


\bibitem[Seifried et al.(2020)]{2020MNRAS.497.4196S} Seifried, D., Walch, S., Weis, M., et al.\ 2020, \mnras, 497, 4196. doi:10.1093/mnras/staa2231

\bibitem[Stanimirovi{\'c} \& Lazarian(2001)]{2001ApJ...551L..53S} Stanimirovi{\'c}, S. \& Lazarian, A.\ 2001, \apjl, 551, L53. doi:10.1086/319837

\bibitem[Stanimirovi{\'c} et al.(2006)]{2006ApJ...653.1210S} Stanimirovi{\'c}, S., Putman, M., Heiles, C., et al.\ 2006, \apj, 653, 1210

\bibitem[Stone et al.(1998)]{1998ApJ...508L..99S} Stone, J.~M., Ostriker, E.~C., \& Gammie, C.~F.\ 1998, \apjl, 508, L99. doi:10.1086/311718

\bibitem[Stone et al.(2020)]{2020ApJS..249....4S} Stone, J.~M., Tomida, K., White, C.~J., et al.\ 2020, \apjs, 249, 4. doi:10.3847/1538-4365/ab929b

\bibitem[White et al.(2016)]{2016ApJS..225...22W} White, C.~J., Stone, J.~M., \& Gammie, C.~F.\ 2016, \apjs, 225, 22. doi:10.3847/0067-0049/225/2/22


\bibitem[Xu et al.(2019)]{2019ApJ...878..157X} Xu, S., Ji, S., \& Lazarian, A.\ 2019, \apj, 878, 157 
\bibitem[Yan \& Lazarian(2006)]{2006ApJ...653.1292Y} Yan, H. \& Lazarian, A.\ 2006, \apj, 653, 1292. doi:10.1086/508704
\bibitem[Yan \& Lazarian(2007)]{2007ApJ...657..618Y} Yan, H. \& Lazarian, A.\ 2007, \apj, 657, 618. doi:10.1086/510847
\bibitem[Yan \& Lazarian(2008)]{2008ApJ...677.1401Y} Yan, H. \& Lazarian, A.\ 2008, \apj, 677, 1401. doi:10.1086/533410

\bibitem[Yuen \& Lazarian(2017a)]{YL17a} Yuen, K.~H., \& Lazarian, A.\ 2017, \apjl, 837, L24 

\bibitem[Yuen \& Lazarian(2017b)]{YL17b} Yuen, K.~H., \& Lazarian, A.\ 2017, arXiv:1703.03026 

\bibitem[Yuen et al.(2018)]{2018ApJ...865...54Y} Yuen, K.~H., Chen, J., Hu, Y., et al.\ 2018, \apj, 865, 54 

\bibitem[Yuen et al.(2019)]{reply19} Yuen, K.~H., Hu, Y., Lazarian, A., \& Pogosyan, D.\ 2019, arXiv:1904.03173 

\bibitem[Yuen \& Lazarian (2020a)]{GA} Yuen, K.~H. \& Lazarian, A., 2020, \apj, 898,65
\bibitem[Yuen \& Lazarian (2020b)]{curvature} Yuen, K.~H. \& Lazarian, A., 2020, \apj, 898,66




\bibitem[Zhang et al.(2020)]{2020NatAs.tmp..174Z} Zhang, H., Chepurnov, A., Yan, H., et al.\ 2020, Nature Astronomy. doi:10.1038/s41550-020-1093-4

\end{thebibliography}
\end{document}